\documentclass[12pt]{article}
\usepackage{graphicx}
\usepackage{amsmath}
\usepackage{amsfonts}
\usepackage{amssymb}
\newtheorem{theorem}{Theorem}[section]

\newtheorem{lemma}[theorem]{Lemma}

\newtheorem{corollary}[theorem]{Corollary}
\newtheorem{remark}[theorem]{Remark}
\newenvironment{proof}[1][Proof]{\textsc{#1.} }{\ \rule{0.5em}{0.5em}}
\numberwithin{equation}{section}

\begin{document}

\title{Static vacuum solutions
from convergent null data expansions at space-like infinity}

\author{Helmut Friedrich\\ 
Max-Planck-Institut f\"ur Gravitationsphysik\\
Am M\"uhlenberg 1\\
14476 Golm, Germany}

\maketitle

\begin{abstract}
{\footnotesize
We study formal expansions of asymptotically flat solutions to the static
vacuum field equations which are determined by minimal sets of freely
specifyable data referred to as `null data'. These are given by sequences of
symmetric trace free tensors  at space-like infinity of increasing order.
They are $1:1$ related to the sequences of Geroch multipoles. 
Necessary and sufficient growth estimates on the null data are obtained for 
the formal expansions to be absolutely convergent. This provides a complete
characterization of all asymptotically flat solutions to the static vacuum field
equations. }
\end{abstract}


{\footnotesize

\section{Introduction}

\vspace{.5cm}

In this article will be given a 
characterization of the asymptotically flat, static solutions to
Einstein's vacu\-um field equations $Ric[\tilde{g}] = 0$. We
thus consider Lorentz metrics which take in coordinates
suitably adapted to a hypersurface orthogo\-nal, time-like Killing field
$K$ the form
\begin{equation}
\label{staticlineel}
\tilde{g} = v^2\,d\,t^2 + \tilde{h},\quad \quad
v = v(x^c),\quad \quad
\tilde{h} =  \tilde{h}_{ab}(x^c)\,dx^a\,dx^b, 
\end{equation}
where $\tilde{h}$ denotes a negative definite metric on the time slices 
$\tilde{S}_c = \{t = c = const.\}$ and the Killing field is given by 
$K = \partial_t$. In this representation Einstein's vacuum field equations
reduce to the {\it static vacuum field equations}
\begin{equation}
\label{statEinstvac}
R_{ab}[\tilde{h}] = \frac{1}{v}\,\tilde{D}_a\,\tilde{D}_b\,v,
\quad\quad
\Delta_{\tilde{h}}\,v = 0
\quad \mbox{on}\quad \tilde{S} \equiv \tilde{S}_0. 
\end{equation}
It will be assumed that
$\tilde{S}$ is diffeomorphic to the complement of a closed ball $B_R(0)$ in
$\mathbb{R}^3$ with a diffeomorphism whose components define coordinates
$x^a$, $a = 1, 2, 3$, on $\tilde{S}$ in which the asymptotic flatness 
condition\footnote{The terms $O_k(|x|^{-(1 + \epsilon)})$ behave like 
$O(|x|^{-(1 + \epsilon + j)})$ under differentiations of order $j \le
k$. }
\begin{equation}
\label{falloff}
\tilde{h}_{ac} = \left(1 + \frac{2\,m}{|x|}\right)\,\delta_{ac} 
+ O_k(|x|^{-(1 + \epsilon)}),
\quad \quad
v = 1 - \frac{m}{|x|} + O_k(|x|^{-(1 + \epsilon)})
\quad \quad \mbox{as} \quad |x| \rightarrow \infty,
\end{equation}
is realized with some $\epsilon > 0$ and $k \ge 2$, where $|\,.\,|$
denotes the standard Euclidean norm. 

Solutions to equations (\ref{statEinstvac})
satisfying the fall-off conditions (\ref{falloff}) have been
characterized  by Reula (\cite{reula:1989}) and Miao (\cite{miao:2003})
in terms of boundary value problems for the static field equations
where the data are prescribed on the sphere $\partial \tilde{S}$, which
encompasses the asymptotic end. 

Our interest in static solutions comes, however, from
the observation that for vacuum solutions arising from asymptotically
flat, time symmetric initial data asymptotic smoothness at null
infinity appears to be related to asymptotic
staticity of the data at space-like infinity
(\cite{friedrich:cargese}, \cite{valientekroon:2004a}). To analyse
this situtation we wish to control
the static vacuum solutions in terms of quantities defined at space-like
infinity.  

Another reason for giving such a characterization results from the work
by  Corvino (\cite{corvino}, \cite{corvino:2005}),
Corvino and Schoen (\cite{corvino:schoen}), and Chru\'sciel and Delay
(\cite{chrusciel:delay:2002}, \cite{chrusciel:delay:2003}). These
authors deform given asymptotically flat vacuum data outside prescribed
compact sets to vacuum data which are {\it exactly static or
stationary near} or {\it asymptotically static or stationary at}
space-like infinity and use such data to discuss the existence of
null geodesically complete solutions which have a smooth asymptotic
structure at null infinity. To assess the scope of these results it is
desirable to have a complete description of the asymptotically flat
static vacuum solutions in terms of asymptotic quantities. 

A characterization of this type has been suggested by Geroch by giving
a definition of multipole moments for static solutions 
(\cite{geroch:1970}). He assumes the metric
$\tilde{h}$ to admit a smooth
conformal extension in the following sense.
With an additional point $i$, which is to represent space-like
infinity, the set $S = \tilde{S} \cup \{i\}$ is assumed to acquire
a smooth differential structure which induces on $\tilde{S}$ the given
one, which makes $S$ diffeomorphic to an open ball in $\mathbb{R}^3$ with the
center representing $i$, and which admits a function 
$\Omega \in C^2(S) \cap C^{\infty}(\tilde{S})$
with the properties
\begin{equation}
\label{Opos}
\Omega > 0 \quad \mbox{on} \quad \tilde{S},
\end{equation}
\begin{equation}
\label{regres}
h_{ab} = \Omega^2\,\tilde{h}_{ab} \quad \mbox{extends to a smooth
negative definite metric on $S$},
\end{equation}
\begin{equation}
\label{Oati}
\Omega = 0,\,\,\,\,
D_a\Omega = 0,\,\,\,\,
D_a D_b \Omega = - 2\,h_{ab} 
\quad \mbox{at} \quad i,
\end{equation}
where $D$ denotes the covariant derivative operator defined by $h$.
We note that these conditions are preserved under rescalings 
$h \rightarrow \vartheta^4 \,h$, $\Omega \rightarrow \vartheta^2
\,\Omega$ with smooth positive functions
$\vartheta$ satisfying $\vartheta(i) = 1$. 

With these assumptions Geroch defines a sequence of tensor fields 
$P$, $P_a$, $P_{a_2a_1}$, $\ldots$ near $i$ by 
setting\footnote{We depart from the convention of \cite{geroch:1970}
by changing the sign of $P$.}
\[
P = \Omega^{-1/2}\,(1 - v),\quad P_a = D_a\,P,\quad
P_{a_2 a_1} = 
{\cal C}(D_{a_2}P_{a_1}
- \frac{1}{2}\,P\,R_{a_2 a_1}),
\]
\[
P_{a_{p + 1} \ldots a_1} = 
{\cal C}(D_{a_{p + 1}}P_{a_p \ldots a_1}
- c_p\,P_{a_{p + 1} \ldots a_3}\,R_{a_2 a_1}),
\quad \mbox{with} \quad c_p = \frac{p\,(2\,p - 1)}{2}, \quad
p = 2, 3, \,\ldots \,,
\]
where $R_{ab}$ denotes the Ricci tensor of
$h_{ab}$ and 
${\cal C}$ the projector onto the symmetric, trace free part of the
respective tensor fields. The multipole moments are then defined as the
tensors 
\[
\nu = P(i), \quad \quad 
\nu_{a_p \ldots a_1} = P_{a_p \ldots a_1}(i),
\quad p = 1, 2, 3, \ldots \,,
\]
at $i$. Setting aside the monopole $\nu$, we will denote the remaining
series of multipoles by 
\begin{equation} \label{multpoleseries}
{\cal D}_{mp} =
\{\nu_{a_1},
\,\nu_{a_2 a_1},\, 
\nu_{a_3 a_2 a_1}, \,\ldots \}.
\end{equation}

The problem of characterizing solutions to a quasi-linear,
gauge-elliptic system of equations of the type (\ref{statEinstvac})
by a minimal set of data given at an ideal point representing space-like
infinity is unusual and certainly quite different from a standard
boundary value problem for (\ref{statEinstvac}). There are available some
results which go into this direction but little has been done on the
general question of existence.

M\"uller zum Hagen has shown that solutions  $v$, $\tilde{h}_{ab}$ to
(\ref{statEinstvac}) are real analytic in $\tilde{h}$-harmonic
coordinates (\cite{mueller zum hagen}). 
The question to what extent the multipoles introduced above determine
the metric $h_{ab}$ and the function $v$ raises the question whether
this metric is real analytic even at $i$ in suitable coordinates and
conformal scalings. Beig and Simon (\cite{beig:simon}) have shown (under
assumptions which have been relaxed later by Kennefick and O'Murchadha
\cite{kennefick:o'murchadha}) that the rescaled metric does
indeed extend in a suitable gauge as a real analytic metric to $i$ if
it is assumed that the ADM mass satisfies 
\begin{equation}
\label{nonvanmass}
m \neq 0.
\end{equation}
We shall assume this result in the following and shall not go through
the argument again, though  its structural basis  will be pointed out in passing. Beig and Simon also provide an argument which esssentially
shows that given sets of multipoles determine unique formal
expansions of `formal solutions' to the static vacuum field equations. 

For axisymmetric static vacuum solutions, which are 
special in admitting explicit descriptions (\cite{Weyl:1917}),
the question under which assumptions a sequences of multipoles does
indeed determine a converging expansion of a static solution
has been studied by B\"ackdahl and Herberthson
(\cite{baeckdahl:herberthson}). For the general case, for which the
freedom to prescribe data is much larger, this problem has never been
analysed. For this reason the results referred to above remained
essentially of heuristic value. 

It is the purpose of this article to derive, under the
assumption (\ref{nonvanmass}), necessary and sufficient
conditions for certain minimal sets of asymptotic data, denoted
collectively by ${\cal D}_{n}$ and referred to as {\it null data},  to
determine (unique) real analytic solutions and thus to provide a complete
characterization of all possible asymptotically flat solutions to the
static vacuum field equations. The behaviour of these solutions in the
large  will not be studied here. We shall only be interested in
what could be called `germs of static solutions at space-like
infinity', for which $S$ may comprise only a neighbourhood of the point
$i$ which is quite small in terms of
$h$ (in terms of
$\tilde{h}$ they cover still infinite domains extending to space-like
infinity). 

While the multipoles above are defined for any conformal
gauge, it will be concenient for our analysis to remove the conformal 
gauge freedom. As shown below, the metric 
$h = \Omega^2\,\tilde{h}$ defined with the {\it preferred
gauge}
\[ 
\Omega = \left(\frac{1 - v}{m}\right)^2,
\]
on a suitable neighbourhood $\tilde{S}$ of space-like infinity, can be
extended with (\ref{Opos}) - (\ref{Oati}) in suitable coordinates to a real
analytic metric at $i$. The metric so obtained satisfies
$R[h] = 0$ on $S$.
In this gauge we get with the notation above
\begin{equation}
\label{Amultmomingauge}
P = m,\quad P_a = 0\quad
P_{a_2 a_1} = - \frac{m}{2}\,s_{a_2 a_1},
\end{equation}
\begin{equation}
\label{Bmultmomingauge}
P_{a_{p + 1} \ldots a_1} = 
{\cal C}(D_{a_{p + 1}}P_{a_p \ldots a_1}
- c_p\,P_{a_{p + 1} \ldots a_3}\,s_{a_2 a_1}),
\quad p = 2, 3,\, \ldots ,
\end{equation}
where $s_{ab}$
denotes the trace free part of the Ricci tensor of $h$.
In the given gauge we consider now the set
\[
{\cal D}_n = \{s_{a_2 a_1}(i),\,\,
{\cal  C}(D_{a_3}s_{a_2 a_1})(i),\,\,
{\cal  C}(D_{a_4}D_{a_3}s_{a_2 a_1})(i),\,\,
{\cal  C}(D_{a_5}D_{a_4}D_{a_3}s_{a_2 a_1})(i),\,\,
\ldots\ldots\,\,\}.
\] 
Given $m \neq 0$ and the sequence 
${\cal D}_n$ associated with $h$, one calculate the multipoles 
${\cal D}_{mp}$ of $h$ and vice versa. The sets 
${\cal D}_n$ and ${\cal D}_{mp}$ thus carry the same information, but 
${\cal D}_n$ is easier to work with because the expressions are
linear in the curvature.

Let now $c_{\bf a}$, $ {\bf a} = 1, 2, 3$, be an $h$-orthonormal frame
field near
$i$ which is $h$-parallely propagate along the geodesics through $i$
and denote the covariant derivative in the direction of $c_{\bf a}$
by $D_{\bf a}$. We express the tensors in
${\cal D}_n$ in terms of this frame and write 
\begin{equation}
\label{canulldata}
{\cal D}^*_n = \{s_{{\bf a}_2 {\bf a}_1}(i),\,\,
{\cal  C}(D_{{\bf a}_3}s_{{\bf a}_2 {\bf a}_1})(i),\,\,
{\cal  C}(D_{{\bf a}_4}D_{{\bf a}_3}s_{{\bf a}_2 {\bf a}_1})(i),\,\,
{\cal  C}(D_{{\bf a}_5}D_{{\bf a}_4}D_{{\bf a}_3}s_{{\bf a}_2
{\bf a}_1})(i),\,\,
\ldots\,\,\}.
\end{equation}
We note that these tensors are defined uniquely up to a rigid rotation
$c_{\bf a} \rightarrow s^{\bf c}\,_{\bf a}\,c_{\bf c}$ with
$(s^{\bf c}\,_{\bf a}) \in O(3, \mathbb{R})$. These data will be referred to as the
{\it null data of $h$ in the frame $c_{\bf a}$}.

It will be shown that if these data are derived from an real
analytic metric $h$ near $i$ there exist constants
$M, \,r > 0$ so that the components of these tensors satisfy the {\it
Cauchy estimates}
\[
|{\cal  C}(D_{{\bf a}_p}\,\ldots\,D_{{\bf a}_1}s_{\bf b\,c})(i)|
\le
\frac{M\,p!}{r^p},\,\,\,\,\,\,
\quad {\bf a}_p, \ldots , {\bf a}_1, {\bf b}, {\bf c} = 1, 2, 3,
\quad \quad p = 0, 1, 2, \ldots\,\,.
\] 
Conversely, we get the following existence result.
\begin{theorem}
\label{mainres}
Suppose $m \neq 0$ and 
\begin{equation}
\label{absnulldata}
\hat{{\cal D}}_{n} =
\{\psi_{{\bf a}_2 {\bf a}_1},\,\,\psi_{{\bf a}_3 {\bf
a}_2{\bf a}_1},
\,\,\psi_{{\bf a}_4 {\bf a}_3 {\bf a}_2 {\bf a}_1}, 
\,\,\ldots\,\,\},
\end{equation}
is a infinite sequence of symmetric, trace free tensors given in an
orthonormal frame at the origin of a 3-dimensional Euclidean space.
If there exist constants $M, r > 0$ such that the
components of these tensors satisfy the estimates
\[
|\psi_{{\bf a}_p \,\,\ldots \,\,{\bf a}_1 \bf b\,c}| \le
\frac{M\,p!}{r^p},
\,\,\,\,\,\,
\quad {\bf a}_p, \ldots , {\bf a}_1, {\bf b, c} = 1, 2, 3,
\quad \quad p = 0, 1, 2, \ldots\,,
\] 
then there exists a germ $(\tilde{h}, v)$ of an analytic, asymptotically
flat, static vacuum solution at space-like infinity with ADM mass $m$,
unique up to isometries, so that the null data implied by 
$h = \left(\frac{m}{1 - v}\right)^4\tilde{h}$
in a suitable frame $c_{\bf a}$  as described above satisfy
\[
{\cal  C}(D_{{\bf a}_q}\,\ldots\,D_{{\bf a}_3}s_{{\bf a}_2
{\bf a}_1})(i) =
\psi_{{\bf a}_q \,\,\ldots \,\,{\bf a}_1},
\quad \quad q = 2, 3, 4, \ldots\,\,.
\] 
\end{theorem}

A series of data of the form (\ref{absnulldata}) (not
necessarily satisfying any estimates) will in the following be
referred to as {\it abstract null data}. 
The type of estimate imposed here on the abstract
null data does not depend on the orthonormal frame in which they are
given (cf. the discussion leading to (\ref{tbasest})).  Since these
estimates are necessary as well as sufficient, all possible germs of
asymptotically flat static vacuum solutions at space-like infinity are
characterized by this result. 

The proof of the result above will be given in terms of the conformal metric $h_{ab}$. For this purpose equations (\ref{statEinstvac}) are reexpressed in  
chapter \ref{secconfstatvacequ} as `conformal static vacuum field equations' for $h_{ab}$ and fields derived from $h_{ab}$ and $v$. In chapter \ref{exactsetarg} it is shown by a direct argument that in a certain setting a set of abstract null data defines the expansion coefficients of a formal expansion of a solution to these equations uniquely. Showing the convergence of the series so obtained appears difficult, however. 
Using the analyticity of the solutions to the conformal static vacuum field equations 
at the point $i$, we study in  chapter \ref{charivp} their analytic extensions  into the complex domain.  Denote by ${\cal N}_i$ the `cone' with vertex at $i$ generated by the complex null geodesics through the point $i$.  
The null data are then represented by a function on  ${\cal N}_i$, the component of the Ricci tensor obtained by contracting it with the null vector tangent to  
${\cal N}_i$.  In this setting the original problem assumes the form of a characteristic initial value problem with data prescribed on  ${\cal N}_i$.

We wish to obtain the equations in a form which allows us to derive from prescribed estimates on the null data appropriate estimates on the expansion coefficients. This requires a choice of gauge which is suitably adapted to  ${\cal N}_i$. Because of the vertex, any such gauge will necessarily be singular at a certain  subset of the manifold. The manifold $\hat{S}$ considered in chapter \ref{charivp} organizes the singularity in a geometric way. In chapter \ref{confstatvaconShat} the conformal static vaccum field equations are considered on $\hat{S}$, and it is shown how to determine a formal solution to the complete set of conformal field equations from a  given set of abstract null data.  The convergence of the series so obtained is shown in chapter \ref{converge}.
Making use of the Lemmas proven in the previous chapters, this result is translated 
in chapter \ref{analyticati}  into a gauge which is regular near $i$ and allows us to prove Theorem \ref{mainres}. 
A translation of the estimates on the null data
into equivalent estimates on the multipoles and a generalization of the
present result to stationary solutions will be discussed elsewhere.

\section{The static field equations in the conformal
setting}
\label{secconfstatvacequ}

The existence problem will be analysed completely in terms of
the conformally rescaled metric. We begin by describing the conformal
gauge and then express the static field equations in terms of the
conformal fields. This discussion follows essentially that of 
\cite{friedrich:static} and \cite{friedrich:cargese}.

\subsection{The choice of the conformal gauge}

Consider a situation as described by conditions (\ref{Opos}) - (\ref{Oati}).
If the  metric $\tilde{h}$ is asymptotically flat and has vanishing Ricci
scalar $R[\tilde{h}]$ on $\tilde{S}$ the  function $\Omega$
satisfies (cf. \cite{friedrich:cargese})
\[
(\Delta_{h} - \frac{1}{8}R[h])\,(\Omega^{- 1/2}) = 0
\quad \mbox{on} \quad \tilde{S} 
\quad \mbox{and} \quad
r\,\,\Omega^{- 1/2} \rightarrow 1 \quad \mbox{as} \quad
r \rightarrow 0,
\] 
where $r$ denotes the $h$-distance from $i$. Sufficiently close to $i$
one obtains the representation
\[
\Omega^{- 1/2} = \zeta^{- 1/2} + W,
\]
with smooth functions $\zeta$ and $W$ satisfying 
\begin{equation}
\label{Wequ}
(\Delta_{h} - \frac{1}{8}R[h])\,W = 0, 
\end{equation}
and 
\begin{equation}
\label{sigmaati}
\zeta(i) = 0,\,\,\,\,
D_a\zeta(i) = 0,\,\,\,\,
D_a D_b \zeta(i) = - 2\,h_{ab}.
\end{equation}
The functions $\zeta$ and $W$ are
real analytic if the metric $h$ is real analytic. 
In \cite{beig:simon} Beig and Simon consider
static vacuum metrics in the form
\[
\tilde{g} = e^{2\,U}\,dt^2 
+ e^{- 2\,U}\,\hat{h}_{a b}\,dx^{a}\,dx^{b},
\]
related to (\ref{staticlineel}) by $v = e^U$ and 
$\hat{h}_{ab} = v^2\,\tilde{h}_{ab}$, and show that 
the function $\omega = (U/m)^2$ and the metric 
\begin{equation}
\label{BSscaling}
h'_{ab} = \omega^2\,\hat{h}_{ab} = 
\Omega^{'2}\,\tilde{h}_{ab} 
\quad \mbox{with} \quad
\Omega' = \omega\,e^{U},
\end{equation}
extend in $h'$-harmonic coordinates near $i$ to real analytic fields at
$i$ so that $\Omega'$ satisfies requirements (\ref{Opos}) - (\ref{Oati})
with the $h'$-covariant derivative operator $D'$.

It follows that  
$\Omega^{'- 1/2} = \zeta^{'- 1/2} + W'$ 
with 
$\zeta' = \frac{\omega}{\cosh^2(U/2)}$ and
$W' = \frac{m}{2}\,\frac{\sinh(U/2)}{U/2}$
(\cite{friedrich:static}).
Assume $S$ to be chosen so that $U \neq 0$ on $\tilde{S}$.
Rescaling with $\vartheta = W'/W'(i) > 0$ on $S$ gives
\[
h = \vartheta^4\,h' = \Omega^2\,\tilde{h}
\quad \mbox{with} \quad 
\Omega = \vartheta^2\,\Omega',
\]
where the conformal factor can be written 
\begin{equation}
\label{preferredgauge} 
\Omega = \left(\frac{1 - v}{m}\right)^2
\quad \mbox{on} \quad S.
\end{equation}
Because of equations (\ref{Wequ}) the metric $h$ 
has then vanishing Ricci scalar 
\begin{equation}
\label{Ricvan}
R[h] = 0
\quad \mbox{on} \quad S,
\end{equation}
and it follows that  
\begin{equation}
\label{fundsol}
\Omega^{- 1/2} = \zeta^{- 1/2} + W,
\end{equation}
where
\begin{equation}
\label{sigma}
W = \frac{m}{2}, \quad \quad
\zeta = \frac{1}{\mu} \left(\frac{1 - v}{1 + v}\right)^2 
\quad \mbox{with} \quad
\mu = \frac{m^2}{4}.
\end{equation}
The fields $h$ and $\zeta$ are real analytic on $S$ and the functions $W$
and $\zeta$ satisfy (\ref{Wequ}), (\ref{sigmaati}). In the
following the gauge (\ref{preferredgauge}) and thus (\ref{Ricvan}) -
(\ref{sigma}) will be assumed.

\subsection{The conformal static vacuum field equations}
\label{confstatvacequ}

The function $\zeta$ satisfies on $S$ the equation
\begin{equation}
\label{zetafundsol}
\Delta_h\,(\zeta^{-1/2}) = 4\,\pi\,\delta_i,
\end{equation}
where $\delta_i$ denotes the Dirac distribution with weight $1$ at $i$.
This equation implies
\begin{equation}
\label{Ban2fequ}
2\,\zeta\,s =
D_a\zeta\,D^a\zeta
\quad \mbox{on} \quad S \quad
\mbox{with} \quad s = \frac{1}{3}\,\Delta_h\,\zeta,
\end{equation}
which, together with (\ref{sigmaati}), implies in turn the equation above.
The function $\zeta^{-1/2}$ can be characterized as a fundamental solution
of $\Delta_h$ with pole at $i$ so that $\zeta$ is real analytic on $S$
and satisfies (\ref{sigmaati}). It is uniquely determined by $h$ because
the expansion coefficients of $\zeta$ in $h$-normal coordinates centered
at $i$ are recursively determined by (\ref{sigmaati}), (\ref{Ban2fequ}).

We derive now a representation of the static vacuum field equations 
(\ref{statEinstvac}) in terms of the conformal metric $h$ and fields
derived from it. With (\ref{Ricvan}) follows
\begin{equation}
\label{hequ}
R_{ab}[h] = s_{ab},
\end{equation}
where $s_{ab}$ is a trace free symmetric tensor field.
The first of equations (\ref{statEinstvac}) implies in the gauge
(\ref{preferredgauge}) 
\begin{equation}
\label{Bn1fequ}
0 = \Sigma_{ab} \equiv D_a\,D_b\,\zeta - s\,h_{ab}
+ \zeta\,(1 - \mu\,\zeta)\,s_{ab}, 
\end{equation}
with $s$ as in (\ref{Ban2fequ}). 
With the Bianchi identity $D^a s_{ab} = 0$ the integrability conditions
\[
0 = \frac{1}{2}\,\,D^c\,\Sigma_{ca}, \quad \quad
0 = 
\frac{1}{\zeta}\left(D_{[c}\,\Sigma_{a]b} +
\frac{1}{2}\,\,D^d\,\Sigma_{d[c}\,h_{a]b}\right)
\]
for the overdetermined system (\ref{Bn1fequ}) take the form
\begin{equation}
\label{B1intco}
0 = S_a \equiv
D_a\,s + (1 - \mu\,\zeta)\,s_{ab}\,D^b\,\zeta,
\end{equation}
and 
\begin{equation}
\label{Bcott}
0 = H_{cab} \equiv
(1 - \mu\,\zeta)\,D_{[c}s_{a]b} 
- \mu\,(
2\,D_{[c}\zeta\,s_{a]b} + D^d\,\zeta\,s_{d[c}\,h_{a]b}).
\end{equation}
We note that this can be read as an expression of the Cotton tensor
$B_{bca} = D_{[c}R_{a]b} - \frac{1}{4}\,D_{[c}R\,h_{a]b}$ in terms of the
undifferentiated curvature. Its dualized version reads by (\ref{Bcott})
\begin{equation}
\label{BtwoindCott}
B_{ab} =
\frac{1}{2}\,B_{acd}\,\epsilon_b\,^{cd}
= \frac{\mu}{1 - \mu\,\zeta}(
s_{da}\,\epsilon_{b}\,^{cd}D_c \zeta 
- \frac{1}{2}\,s_{de}\,\epsilon_{ba}\,^d D^e\zeta).
\end{equation}

Equations (\ref{hequ}), (\ref{Bn1fequ}), (\ref{B1intco}),
(\ref{Bcott}) together with conditions (\ref{sigmaati}), which imply 
\begin{equation}
\label{sati}
s(i) = - 2,
\end{equation}
will be referred to as the {\it conformal static vacuum field equations}
for the unknown fields 
\begin{equation}
\label{unknowns}
h,\,\,\,\zeta,\,\,s,\,\,s_{ab}.
\end{equation}

The second of equations (\ref{statEinstvac}) implies that  
$R[\tilde{h}] = 0$ and can thus also be read as the conformally covariant
Laplace equation for $v$. With the conformal covariance of the latter and
(\ref{preferredgauge}), (\ref{Ricvan}),  (\ref{sigma}), its conformal version reduces to (\ref{zetafundsol}).
The identity
\[
D_a(2\,\zeta\,s - D_c\,\zeta\,D^c\,\sigma) = 
2\,\zeta\,S_a
- 2\,\Sigma_{ac}\,D^c\zeta,
\]
shows that (\ref{Ban2fequ}), whence (\ref{zetafundsol}), is a
consequence of equations (\ref{sigmaati}) and (\ref{Bn1fequ}).
It follows that for given $m \neq 0$, which defines $W$ and $\mu$, a
solution of the conformal static vacuum field equations provides a unique
solution to the static vacuum field equations (\ref{statEinstvac}).

The system (\ref{hequ}), (\ref{Bn1fequ}), (\ref{B1intco}),
(\ref{Bcott}) represents a quasi-linear, overdetermined, gauge-elliptic
system of PDE's. The Ricci operator becomes
elliptic in a suitable gauge and the elliptic character of the remaining
equations can be seen by taking the trace of (\ref{Bn1fequ}), by
contracting (\ref{B1intco}) with $D^a$, and by contracting (\ref{Bcott}) with
$D^c$ and using the Bianchi identity and (\ref{Bn1fequ}) again so that in all three cases one obtains an equation with the Laplacian acting on the respective unknown.
By deducing from the fall-off behaviour of the physical solution at
space-like infinity a certain minimal smoothness of the conformal fields
at $i$ and invoking a general theorem of Morrey (\cite{morrey}) on
elliptic systems of this type, Beig and Simon (\cite{beig:simon})
concluded that the solutions are in fact real analytic at $i$.
To avoid introducing additional constraints by taking derivatives, we
shall deal with the system of first order above.

\section{The exact sets of equations argument}
\label{exactsetarg}

Constructing solutions from minimal sets of data prescribed at $i$ poses quite  an unusual problem for a system of the type of the static
conformal field equations. To see how it might be done, we
study expansions of the fields in normal coordinates.

For convenience assume in the following $S$ to coincide with a convex
$h$-normal neighbourhood of $i$. Let $c_{\bf a}$, ${\bf a} = 1, 2, 3$, be
an $h$-orthonormal frame field on $S$ which is parallely transported along
the $h$-geodesics through $i$ and let $x^a$ denote normal coordinates
centered at $i$ so that $c^b\,_{\bf a} \equiv \,<dx^b, c_{\bf a}>\, =
\delta^b\,_{\bf a}$ at $i$. We refer to such a frame as {\it
normal frame centered at $i$}. Its dual frame will be
denoted by $\chi^{\bf c} =
\chi^{\bf c}\,_b\,dx^b$.

At the point with coordinates $x^a$ the coefficients of
the frame then satisfy
\[
c^{b}\,_{\bf a}\,x^a = \delta^{b}\,_{\bf a}\,x^a,
\,\,\,\,\,\,\,\,\,\,
x_b\,c^{b}\,_{\bf a} = x_b\,\delta^b\,_{\bf a},
\]
(where we set $x_a = x^b\,\delta_{ba}$ and assume, as in the following,
that the summation rule does not distinguish between bold face and
other indices). Equivalently, the coefficients of the dual frame satisfy 
\begin{equation}
\label{normalformchar}
\chi^{\bf a}\,_b\,x^b = \delta^{\bf a}\,_b\,x^b ,
\,\,\,\,\,\,\,\,\,\,
x_a\,\chi^{\bf a}\,_b = x_a\,\delta^{\bf a}\,_b, 
\end{equation}
which implies with the coordinate expression
$h_{ab} = - \delta_{\bf ac}\,\chi^{\bf a}\,_b\,\chi^{\bf
c}\,_d$ of the metric the well known characterization $x^a\,h_{ab} = -
x^a\,\delta_{ab}$ of the $x^a$ as $h$-normal coordinates centered at $i$.
In the following all tensor fields, except the frame
field $c_{\bf a}$ and the coframe field $\chi^{\bf c}$, will be
expressed in terms of this frame field, so that the metric is given by 
$h_{\bf ab} \equiv h(c_{\bf a}, c_{\bf c})  = - \delta_{\bf ab}$.
With $D_{\bf a} \equiv D_{c_{\bf a}}$ the connection
coefficients with respect to $c_{\bf a}$ are defined by
$D_{\bf a}\,c_{\bf c} = \Gamma_{\bf a}\,^{\bf b}\,_{\bf c}\,c_{\bf b}$. 

An analytic tensor field $T_{{\bf a}_1 \ldots {\bf
a}_k}$ on $S$ has in the normal coordinates $x^a$ a {\it normal expansion}
at
$i$, which can be written (cf.
\cite{friedrich:i-null})
\begin{equation}
\label{normalexp}
T_{{\bf a}_1 \ldots {\bf a}_k}(x) = 
\sum_{p \ge 0} \frac{1}{p\,!}\,x^{c_p} \ldots x^{c_1}\,
D_{{\bf c}_p} \ldots D_{{\bf c}_1}\,
T_{{\bf a}_1 \ldots {\bf a}_k}(i).
\end{equation} 
(This is a convenient short version of the correct expression; more
precisely, the
$x^a$ should be replaced here by the components of the vector field $X$
which has in normal coordinates the expansion $X(x) = x^b\,\delta^{\bf
a}\,_b\,c_{\bf a}$ and which is characterized by the conditions 
$D_V V = 0$, $\,V(i) = 0$.)
In the following will be shown how normal expansions can be obtained for
solutions
\begin{equation}
\label{Bunknowns}
h_{\bf ab},\,\,\zeta,\,\,s,\,\, s_{\bf ab},
\end{equation}
to the conformal static vacuum field equations.
In $3$ dimensions the curvature tensor satisfies
\[
R_{abcd}[h] = 2 \{h_{a[c} L_{d]b} + h_{b[d} L_{c]a}\} 
\quad \mbox{with} \quad 
L_{ab}[h] = R_{ab}[h] - \frac{1}{4}\,R[h]\,h_{ab},
\]
and can be expressed because of (\ref{Ricvan})
completely in terms of $s_{\bf ab}$. Once the
latter is known, the connection coefficients $\Gamma_{\bf a}\,^{\bf
b}\,_{\bf c}$  and the coefficients of the 1-forms $\chi^{\bf a}$ 
can thus be obtained, order by order, from the structural equations in
polar coordinates (cf. \cite{dieudonne:IV}),
\[
\frac{d}{d\,s}\left(s\,\chi^{\bf a}\,_{b}(s\,x^f)\right)
= \delta^{\bf a}\,_{b} 
+ \Gamma_{\bf c}\,^{\bf a}\,_{\bf d}(s\,x^f)
\,s\,\chi^{\bf c}\,_{b}(s\,x^f)\,\,x^d,
\]
\[
\frac{d}{d\,s}\left(\Gamma_{\bf a}\,^{\bf c}\,_{\bf e}(s\,x^f)
\,s\,\chi^{\bf a}\,_{b}(s\,x^f)\right)
= R^{\bf c}\,_{\bf e d a}(s\,x^f)\,x^d
\,\,s\,\chi^{\bf a}\,_{b}(s\,x^f),
\]
where $s$ denotes an affine parameter along the $h$-geodesics through
$i$ with unit tangent vectors which vanishes at $i$, so that 
$s^2 = \delta_{ab}\,x^a\,x^b$.

By formally taking covariant derivatives, the expansion coefficients of
$\zeta$ and $s$ up to order $m + 2$ resp. $m + 1$ can be obtained from
equations (\ref{Bn1fequ}) and (\ref{B1intco}) once  $s_{\bf ab}$ is known up
to order $m$. Calculating the expansion coefficients for $s_{\bf ab}$ by
means of equation (\ref{Bcott}) leads, however, to some complicated
algebra. It turns out that the latter simplifies considerably in the
space spinor formalism. 

To achieve the transition to the space-spinor formalism we introduce
the constant van der Waerden symbols
\[
\alpha^{AB}\,_{a}, \quad
\alpha^{a}\,_{AB},
\quad
a = 1, 2, 3,\,\,\,\,A, B = 0, 1,
\] 
which map one-index objects onto two-index objects
which are symmetric in the two indices. 
If the latter are
read as matrices, the symbols are given by 
\[
\xi^{a} \rightarrow \xi^{AB} = \alpha^{AB}\,_{a}\,\xi^{a} =
\frac{1}{\sqrt{2}}
\left( \begin{array}{cc}
- \xi^1 - i \xi^2 & \xi^3 \\
\xi^3 & \xi^1 - i \xi^2
\end{array} \right),
\]
\[
\xi_{a} \rightarrow \xi_{AB} = \xi_{a} \, \alpha^{a}\,_{AB} =
\frac{1}{\sqrt{2}} \left( \begin{array}{cc}
- \xi_1 + i \xi_2 & \xi_3 \\
\xi_3 & \xi_1 + i \xi_2
\end{array} \right).
\]

With the summation rule also applying to capital indices one gets 
\[
\delta^{c}\,_{a} = \alpha^{c}\,_{AB}\,\,\alpha^{AB}\,_{a},
\quad\quad
- \,\delta_{ab}\,\alpha^{a}\,_{AB}\,\alpha^{b}\,_{CD}
= - \epsilon_{A(C}\,\epsilon_{D)B} \equiv h_{ABCD},  
\]
\[
a, b = 1,2,3,\,\,\,
A,B, C, D = 0, 1,
\]
where the constant $\epsilon$-spinor is antisymmetric, 
$\epsilon_{AB} = - \epsilon_{BA}$, and satisfies $\epsilon_{01} = 1$.
It is used to pull indices according to the rules
$\iota_B = \iota^A\,\epsilon_{AB}$, 
$\iota^A = \epsilon^{AB}\,\iota_B$, so that $\epsilon_A\,^B$ corresponds
to the Kronecker delta. We shall denote the `scalar product'
$\kappa_A\,\iota^A$ of two spinors $\kappa^A$ and $\iota^A$ 
occasionally also by $\epsilon(\kappa, \iota)$. It is important
here to observe the order in which the spinors occur.

Given the van der Waerden symbols, we associate with an tensor field
$T^{{\bf a}_1 \ldots {\bf a}_p}\,_{{\bf b}_1 \ldots {\bf b}_q}$
{\it given in the frame $c_{\bf a}$} the space spinor field
\[
T^{A_1B_1 \ldots A_pB_p}\,_{C_1D_1 \ldots C_q D_q}
=
T^{{\bf a}_1 \ldots {\bf a}_p}\,_{{\bf b}_1 \ldots {\bf b}_q}
\alpha^{A_1B_1}\,_{a_1} \ldots \ldots
\alpha^{b_q}\,_{C_q D_q}
\]
\[
= 
T^{(A_1B_1) \ldots (A_pB_p)}\,_{(C_1D_1) \ldots (C_q D_q)}.
\]
In the following we shall employ tensor or spinor notation as it appears
convenient.
With the spinor field 
\[
\tau^{AA'} =  \epsilon_0\,^A\,\epsilon_{0'}\,^{A'}
+ \epsilon_1\,^A\,\epsilon_{1'}\,^{A'},
\]
and the notation 
\[
\xi^+_{A B \ldots H} = 
\tau_A\,^{A'}\,\tau_B\,^{B'}\, \ldots \tau_H\,^{H'}\,
\bar{\xi}_{A' B' \ldots H'},
\]
where the bar denotes complex conjugation, one finds that a space spinor
field
\[
T_{A_1B_1 \ldots A_pB_p} = 
T_{(A_1B_1) \ldots (A_pB_p)},
\]
arises from a real tensor field
$T_{{\bf a}_1 \ldots {\bf a}_p}$ if and only if it satisfies the reality
condition
\begin{equation}
\label{reality}
T_{A_1B_1 \ldots A_pB_p} = 
(- 1)^p\,T^+_{A_1B_1 \ldots A_pB_p}.
\end{equation} 
It follows in particular  
\[
\xi_{AB}\,\xi^{AB} = 2\,(\xi_{00}\,\xi_{11} - \xi_{01}\,\xi_{01})
= 2\,\det(\xi_{AB})  = - \delta_{ab}\,\xi^a\,\xi^b,
\] 
and we can have $\xi_{AB}\,\xi^{AB} = 0$ for vectors $\xi^{AB} \neq 0$ only
if $\xi^a$ is complex. Since $\xi^{AB} = \xi^{(AB)}$, the relations  
$\xi_{AB}\,\xi^{AB} = 0$, $\xi^{AB} \neq 0$ imply by the
equation above that  
$\xi^{AB} = \kappa^A\,\kappa^B$ for some $\kappa^A \neq 0$. This fact
will allow us to interprete the data (\ref{canulldata}) as `null data'.

Any spinor field $T_{ABC \ldots GH}$, symmetric or not, admits a
decomposition into products of totally symmetric spinor fields and epsilon
spinors which can be written schematically in the form  
(cf. \cite{penrose:rindler:I})
\begin{equation}
\label{spinordecomp}
T_{ABC \ldots GH} = T_{(ABC \ldots GH)} +
\sum\,\epsilon's \times symmetrized \,\,contractions\,\,of\,\,
T.
\end{equation}

Later on it will be important for us that spinor fields 
$T_{A_1B_1 \ldots A_pB_p}$ arising from tensor fields $T_{{\bf a}_1
\ldots {\bf a}_p}$ satisfy 
\[
T_{(A_1B_1 \ldots A_pB_p)} =
{\cal C}(T_{{\bf a}_1 \ldots {\bf a}_p})\,
\alpha^{a_1}\,_{A_1B_1} \,\ldots\,\alpha^{a_p}\,_{A_pB_p},
\] 
i.e. the projectors ${\cal C}$ onto the trace free symmetric
part of tensors is represented in the space spinor notation simply by
symmetrization. If convenient we shall denote the latter also  by the
symbol $sym$.

To discuss vector analysis in terms of spinors, a complex
frame field and its dual 1-form field are defined by
\[
c_{AB} = \alpha^{a}\,_{AB}\,c_{\bf a},\quad \quad
\chi^{AB} = \alpha^{AB}\,_{a}\,\chi^{\bf a},
\]
so that $h(c_{AB}, c_{AB}) = h_{ABCD}$. If the derivative of a
function $f$ in the  direction of $c_{AB}$ is denoted by 
$c_{AB}(f) = f_{,a}\,c^a\,_{AB}$ and 
the spinor connection coefficients are defined by 
\[
\Gamma_{AB}\,^C\,_D = \frac{1}{2}\,\Gamma_{\bf a}\,^{\bf b}\,_{\bf c}\,
\alpha^{a}\,_{AB}\,\alpha^{CH}\,_b\,\alpha^{c}\,_{DH},
\quad \mbox{so that} \quad
\Gamma_{ABCD} = \Gamma_{(AB)(CD)},
\]
the covariant derivative of a spinor field $\iota^A$ is
given by
\[
D_{AB}\,\iota^C = e^a\,_{AB}(\iota^C)
+ \Gamma_{AB}\,^C\,_B\,\iota^B.
\]
If it is required to satisfies the Leibniz rule with respect to tensor
products, it follows that covariant derivatives in the $c_{\bf a}$-frame
formalism translate under contractions with the van der Waerden symbols
into spinor covariant derivatives and vice versa.

The commutator of covariant spinor derivatives satisfies
\begin{equation}
\label{covdercomm}
(D_{CD}\,D_{EF} - D_{EF}\,D_{CD})\,\iota^A =
R^A\,_{BCDEF}\,\iota^B,
\end{equation}
with the curvature spinor
\[
R_{ABCDEF} 
= \frac{1}{2}\left\{
(s_{ABCE} - \frac{R[h]}{6}\,h_{ABCE})\,\epsilon_{DF}
+ (s_{ABDF} - \frac{R[h]}{6}\,h_{ABDF})\,\epsilon_{CE}
\right\},
\]
where $R[h]$ is the Ricci scalar and 
$s_{ABCD} = s_{\bf ab}\,\alpha^{a}\,_{AB}\,\alpha^{b}\,_{CD}$
represents the trace free part of the Ricci tensor of $h$, which is
completely symmetric, $s_{ABCD} = s_{(ABCD)}$. The gauge condition
(\ref{Ricvan}) implies
\begin{equation}
\label{curveingauge}
R_{ABCDEF} 
= \frac{1}{2}\left(
s_{ABCE}\,\epsilon_{DF} + s_{ABDF}\,\epsilon_{CE}
\right).
\end{equation}

In the space-spinor formalism equations (\ref{Bcott}) acquire the concise
form
\begin{equation}
\label{spsptwoindCott}
D_A\,^E s_{BCDE} =
\frac{2\,\mu}{1 - \mu\,\zeta}\,s_{E(BCD}\,D_{A)}\,^E\zeta.
\end{equation}
Applying to this equation and to the spinor versions of equations 
(\ref{Bn1fequ}) and (\ref{B1intco}) the theory of `exact sets of fields'
discussed in \cite{penrose:rindler:I}, we get the following result.

\begin{lemma}
\label{realformalexpansion}
Let there be given a sequence 
\[
\hat{{\cal D}}_{n} =
\{\psi_{A_2B_2A_1B_1},\,\,\psi_{A_3B_3A_2B_2A_1B_1},\,\,
\psi_{A_4B_4A_3B_3A_2B_2A_1B_1},\,\,\ldots\},
\]
of totally symmetric spinors satisfying the reality condition
\ref{reality}.  Assume that there exists a solution $h$, $\zeta$, $s$,
$s_{ABCD}$ to the conformal static field equations (\ref{sigmaati}),
(\ref{hequ}), (\ref{Bn1fequ}), (\ref{B1intco}), (\ref{Bcott}) so that 
the spinors given by $\hat{{\cal D}}_{n}$ coincide with the null
data ${\cal D}^*_{n}$ given by (\ref{canulldata}) of the metric $h$ in
terms of an $h$-orthonormal normal frame centered at $i$, i.e. 
\begin{equation}
\label{psipDps}
\psi_{A_pB_p \ldots A_3B_3A_2B_2A_1B_1} =
D_{(A_p B_p}\,\ldots\,D_{A_3 B_3}\,s_{A_2B_2A_1B_1)}(i),
\quad p \ge 2.
\end{equation}

Then the coefficients of the normal expansions (\ref{normalexp}) of
the fields (\ref{unknowns}), in particular of  
\begin{equation}
\label{snormalexp}
s_{ABCD}(x) = 
\sum_{p \ge 0} \frac{1}{p!}\,x^{A_pB_p} \ldots x^{A_1B_1}\,
D_{A_pB_p} \ldots D_{A_1B_1}\,
s_{ABCD}(i),
\end{equation} 
with $x^{AB} = \alpha^{AB}\,_a\,x^a$, are uniquely determined by the data
$\hat{{\cal D}}_n$ and satisfy the reality conditions.
\end{lemma}

\begin{proof}
It holds $s_{ABCD}(i) = \psi_{ABCD}$ by assumption and the expansion
coefficients for $\zeta$, $s$ of lowest order are given by 
(\ref{sigmaati}), (\ref{sati}). The induction steps for $\zeta$ and $s$ being
obvious by (\ref{Bn1fequ}) and
(\ref{B1intco}), we only need to consider $s_{ABCD}$ and
(\ref{spsptwoindCott}). Assume $m \ge 0$. If spinors
$D_{A_p B_p}\,\ldots\,D_{A_1 B_1}\,s_{CDEF}(i)$, $p \le m$, have
been obtained which satisfy
(\ref{psipDps}) and, up to that order, equation (\ref{spsptwoindCott}), the
totally symmetric part of 
\[
D_{A_{m+1} B_{m+1}}\, \ldots \,D_{A_1 B_1}\,s_{CDEF}(i),
\] is given by
the prescribed data while its contractions,
which define the remaining terms in the decomposition
corresponding to (\ref{spinordecomp}), are determined as follows. Observing
the symmetries involved, essentially two cases can occcur: 

i) If one of the indices $B_j$ is contracted with $F$,
say, the operator $D_{A_j B_j}$ can be commuted with other
covariant derivatives, generating by (\ref{covdercomm}),
(\ref{curveingauge}) only terms of lower order,  
until it applies directly to $s_{CDEF}$.  
Equation (\ref{spsptwoindCott}) then shows how to express the resulting 
term by quantities of lower order. 

ii) If the index $B_j$ is contracted with $B_k$, $k \neq j$, the
operators $D_{A_j B_j}$ and $D_{A_k B_k}$ can be commuted with other
covariant derivatives, until the operator 
$D_{A_j H}\,D_{A_k}\,^H$ applies directly to $s_{CDEF}$. If the
corresponding term is symmetrized in $A_j$ and $A_k$ the general identity
\[
D_{H(A}\,D^H\,_{B)}\,s_{CDEF} = - 2\,s_{H(CDE}\,s_{F)AB}\,^H,
\]
implied by (\ref{covdercomm}), (\ref{curveingauge}) shows that this term is
in fact of lower order. If a contraction of $A_j$ and $A_k$ is involved,
the general identity
\[
D_{AB}\,D^{AB}\,s_{CDEF} = - 2\,D_F\,^G \,D_G\,^H\, s_{CDEH}
+ 3\,s_{GH(CD}\,s_{E)F}\,^{GH},
\] 
shows together with (\ref{spsptwoindCott}) that the corresponding term can
again be expressed in terms of quantities of lower order, showing that 
$D_{A_{m+1} B_{m + 1}}\,\ldots\,D_{A_{1} B_{1}}\,s_{CDEF}(i)$ is
determined by our data and terms of order $\le m$. That the expansion
coefficients satisfy the reality condition is a consequence of the
formalism and the fact that they are satisfied by the data 
$\hat{{\cal D}}_n$. \end{proof}

To achieve our goal, we have to show the convergence of the formal
series determined in Lemma \ref{realformalexpansion}. This requires us to
impose estimates on the free coefficients given by ${\cal D}_n$. We get
the following result.

\begin{lemma} 
\label{necdatest}
A necessary condition for the formal series (\ref{snormalexp}) determined
in Lemma \ref{realformalexpansion} to be absolutely convergent near the
origin is that the data given by $\hat{{\cal D}}_n$ satisfy estimates of
the type
\begin{equation}
\label{symmest}
|\psi_{A_pB_p \ldots A_1 B_1 CDEF}| \le \frac{p!\,M}{r^p},
\quad p = 0, 1, 2, \ldots\,,
\end{equation}
with some constants $M, r > 0$.
\end{lemma} 

\begin{proof}
If $f$ is a real analytic  function defined on some neighbourhood of the
origin in $\mathbb{R}^n$, it can be analytically extended to a function
which is defined, holomorphic, and bounded on  a polydisc
$P(0, r) = \{x \in \mathbb{C}^n|\,\,|x^j| < r,\, 1 \le j \le n\}$ with some 
$r > 0$. Its Taylor expansion  
$f = \sum_{|\alpha| \ge 0}\,\frac{1}{\alpha\,!}\,
\partial^{\alpha}f(0)\,x^{\alpha}$ is absolutly convergent on 
$P(0, r)$ with
$\sup_{x \in P(0, r)}|f(x)| \le M < \infty$ so that its derivatives satisfy the estimates
\begin{equation}
\label{basicest}
|\partial^{\alpha}f(0)| \le \frac{\alpha\, !\,\,M}{r^{|\alpha|}}
\le \frac{|\alpha|\, !\,\,M}{r^{|\alpha|}}.
\end{equation}
The first of these estimates are known as Cauchy inequalities.
Here $\alpha \in \mathbb{N}^n$ denotes a multi-index and we use the notation
$|\alpha| = \alpha_1 + \ldots + \alpha_n$,
$\,\,\alpha \,! = \alpha_1 ! \cdot \ldots \cdot \alpha_n !$,
$\,\,\partial^{\alpha} = 
\partial^{\alpha_1}_1 \cdot \ldots \cdot \partial^{\alpha_n}_n$, and
$x^{\alpha} = (x^1)^{\alpha_1} \cdot \ldots \cdot (x^n)^{\alpha_n}$.

If the series (\ref{snormalexp}) and thus 
\begin{equation}
\label{Csnormalexp}
s_{{\bf ab}}(x) = 
\sum_{p \ge 0} \frac{1}{p!}\,x^{c_p} \ldots x^{c_1}\,
D_{{\bf c}_p} \ldots D_{{\bf c}_1}\,
s_{{\bf ab}}(i),
\end{equation} 
is absolutely convergent near the origin, there exist therefore by the second of
the estimates (\ref{basicest}) constants $M_*, r_* > 0$ with
\[
|D_{{\bf c}_p} \ldots D_{{\bf c}_1}\,s_{{\bf ab}}(i)| \le
\frac{p\,!\,M_*}{r_*^p}, \quad \quad
{\bf c}_p, \,\ldots \,, {\bf c}_1, {\bf a}, {\bf b} = 1, 2, 3,
\quad \quad p = 0, 1, 2, \ldots\,\,.
\]
Observing the transition rule from tensor to spinor quantities, one gets
from this the estimates
\begin{equation}
\label{genCauchyest}
|D_{A_p B_p}\,\ldots\,D_{A_1 B_1}\,s_{CDEF}(i)| \le
\frac{p\,!\,M}{r^p}, \quad \quad A_p, B_p, \ldots E, F = 0, 1,
\quad \quad p = 0, 1, 2, \ldots\,\,
\end{equation}
with $M = 9\,c^2\,M_*$ and $r = r_*/3\,c$, where 
$c = \max_{a = 1, 2, 3;\,A, B = 0, 1}|\alpha^a\,_{AB}|$.
To derive from these estimates the estimates (\ref{symmest}) we 
consider instead of (\ref{spinordecomp}) directly
the symmetrization operator to get
\[
|\psi_{A_pB_p \ldots A_1 B_1 CDEF}| =
|D_{(A_p B_p}\,\ldots\,D_{A_1 B_1}\,s_{CDEF)}(i)| \le
\]
\[
\frac{1}{(2 p + 4)!}\sum_{\pi \in {\cal S}_{2 p + 4}}
|D_{\pi(A_p B_p}\,\ldots\,D_{A_1 B_1}\,s_{CDEF)}(i)|
\le \frac{p!\,M}{r^p},
\]
where ${\cal S}_{m}$ denotes the group of permutations of $m$ elements.
\end{proof}

We note for later use that  if the derivatives of a smooth function $f$ satisfy 
estimates of the type (\ref{basicest}) with some constants $M, r > 0$ then the 
function $f$ is real
analytic near the origin because its Taylor series is majorized by 
\begin{equation}
\label{Amajorant}
\sum_{\alpha}\,M\,r^{- |\alpha|}\,x^{\alpha}
= \frac{M\,r^n}{(r - x^1) \cdot \ldots \cdot (r - x^n)}, \quad\quad
|x^a| < 1,
\end{equation}
and
\begin{equation}
\label{Bmajorant}
\sum_{\alpha}\,\frac{|\alpha|\,!}{\alpha\,!} \,
M\,r^{- |\alpha|}\,x^{\alpha}
= \frac{M\,r}{(r - x^1 - \ldots - x^n)}, \quad\quad
\sum_{j = 1}^n\,|x^j| < 1.
\end{equation}

\subsection{Relations between null data and multipoles}

We express the relation between the sequences 
${\cal D}^*_n$ of null data and the sequences
${\cal D}^*_{mp}$ of multipoles of $h$ (in the same normal frame
centered at $i$) in terms of space-spinor notation.

\begin{lemma}
\label{mpnrel}
The spinor fields $P_{A_pB_p \,\ldots\, A_1B_1}$ near $i$, given by 
(\ref{Amultmomingauge}), (\ref{Bmultmomingauge}), are of the form
\begin{equation}
\label{PsFrel}
P_{A_pB_p \,\ldots\, A_1B_1} = 
- \frac{m}{2}\left\{D_{(A_p B_p}\,\ldots\,D_{A_3 B_3}\,s_{A_2 B_2A_1 B_1)}
+ F_{A_pB_p \,\ldots\, A_1B_1}\right\},
\end{equation}
with symmetric spinor-valued functions 
\[
F_{p} \equiv  F_{A_pB_p \,\ldots\, A_1B_1}
= F_{A_pB_p \,\ldots\, A_1B_1}[\{
D_{(A_q B_q}\,\ldots\,D_{A_3 B_3}\,s_{A_2 B_2A_1 B_1)}\}_{q \le p - 2}],
\quad p \ge 2,
\]
which satisfy
\[
F_{A_2B_2A_1B_1} = 0,\,\,\,\,F_{A_3B_3A_2B_2 A_1B_1} = 0,
\] 
and which are real linear combinations of symmetrized tensor products
of 
\[
s_{A_2 B_2A_1 B_1},\, D_{(A_3 B_3}\,s_{A_2 B_2A_1 B_1)}, \,\ldots \,
, D_{(A_{p - 2}B_{p - 2}}\,\ldots\, D_{A_3 B_3}\,s_{A_2 B_2A_1 B_1)},
\]
for $p \ge 4$.
\end{lemma}

\begin{proof}
The first two results on $F$ follow by direct calculations from
(\ref{Amultmomingauge}), (\ref{Bmultmomingauge}). Inserting (\ref{PsFrel}) into
the recursion relation (\ref{Bmultmomingauge}) gives for $p \ge 3$ the
recursion relations
\begin{equation}
\label{Frecurs}
F_{A_{p + 1}B_{p + 1} \,\ldots\, A_1B_1}
= D_{(A_{p + 1} B_{p + 1}}\,F_{A_pB_p \,\ldots\, A_1B_1)}
\end{equation}
\[
- c_p\left\{
s_{(A_{p + 1} B_{p + 1} A_p B_p}\,\,
D_{A_{p - 1} B_{p - 1}}\,\ldots\,s_{A_2 B_2A_1 B_1)}
+
s_{(A_{p + 1} B_{p + 1}}\,\,F_{A_{p - 1}B_{p - 1}\,\ldots\,
A_1B_1)}\right\}.
\]
With the induction hypothesis which assumes the properties of
the $F$'s stated above for $F_{A_{q}B_{q} \,\ldots\, A_1B_1}$,
$q \le p$, the relations (\ref{Frecurs}) imply these properties 
for $F_{A_{p + 1}B_{p + 1} \,\ldots\, A_1B_1}$.
\end{proof}

A further calculation gives 
\[
F_4 = - c_3\,s_{(A_4 B_4 A_3 B_3}\,s_{A_2 B_2 A_1 B_1)},\,\,\,\,\,
F_5 = - (2\,c_3 + c_4)\,
s_{(A_5 B_5 A_4 B_4}\,D_{A_3 B_3}s_{A_2 B_2 A_1 B_1)},
\]
and by induction the recursion law above implies the general
expressions
\[
F_{2p} =
\alpha_{2p}\,sym(s \otimes D^{2p - 4}s) + \,\ldots\,
+ \omega_{2p}\,sym(\otimes^ps),\quad p \ge 3,
\]
\[
F_{2p + 1} =
\alpha_{2p+1}\,sym(s \otimes D^{2p - 3}s) + \,\ldots\,
+ \omega_{2p+1}\,sym(\otimes^{p-1}s \otimes Ds), \quad p \ge 3,
\]
with real coefficients $\alpha_{2p}, \alpha_{2p+1}, \ldots , \omega_{2p},
\omega_{2p + 1}$. The first terms on the right hand sides denote the
term with the highest power of $D$ occuring in the respective expression.
The sum of the powers of $D$ occuring in each term is even in the case
of $F_{2p}$ and odd in the case of $F_{2p+1}$. 
The sum of the powers of $D$ occuring in each of the terms indicated by
dots lies between $2$ and $2\,p - 4$ in the case of 
$F_{2p}$ and between $3$ and $2\,p - 3$ in the case of $F_{2p+1}$. The
coefficients indicated above are determined by
\[
\alpha_6 = - (2\,c_3 + c_4 + c_5), \quad
\alpha_7 = - (2\,c_3 + c_4 + c_5 + c_6), \quad
\omega_5 = - (2\,c_3 + c_4), \quad
\omega_6 = c_3\,c_5,
\]
and, for $p \ge 3$, by
\[
\alpha_{2p + 1} = \alpha_{2p} - c_{2p}, \quad
\alpha_{2p + 2} = \alpha_{2p+1} - c_{2p+1}, \quad
\] 
\[
\omega_{2p + 1} = p\,\omega_{2p} - c_{2p}\,\omega_{2p - 1}, \quad
\omega_{2p + 2} = - c_{2p + 1}\,\omega_{2p}, \quad
\] 
which implies in particular
\begin{equation}
\label{omega2p}
\omega_{2p} = (-1)^{p+1}\,\Pi_{l = 1}^{p - 1} \,c_{2l + 1}, 
\quad p \ge 3.
\end{equation}

Restricting the relation (\ref{PsFrel}) to $i$ defines with the
identification (\ref{psipDps}) a non-linear map which can be read as a map 
\[
\Psi: \{\, \hat{{\cal D}}_n\, \} \rightarrow \{\,
\hat{{\cal D}}_{mp}\,\},
\]
of the set of {\it abstract null data} into the set of {\it abstract
multipoles} (i.e. sequences of symmetric spinors not necessarily
derived from a metric) satisfying
\begin{equation}
\label{absnullabsmpmap}
\nu_{A_pB_p \,\ldots\, A_1B_1} = 
- \frac{m}{2}\left(\psi_{A_p B_p \,\ldots\,A_1 B_1}
+ F_{A_pB_p \,\ldots\, A_1B_1}[\{
\psi_{A_q B_q\,\ldots\,A_1 B_1}\}_{q \le p - 2}]
\right), \quad  p \ge 2.
\end{equation}

\begin{corollary}
\label{mpnmap}
For given $m$ the map $\Psi$ which maps sequences 
$\hat{{\cal D}}_n$ of abstract null data onto sequences 
$\hat{{\cal D}}_{mp}$
of abstract multipoles is bijective. 
\end{corollary}

\begin{proof}
An inverse of $\Psi$ can be constructed because $F_2 = 0$, $F_3 = 0$, and
the $F_p$ depend only on the $\psi_{A_q B_q\,\ldots\,A_1 B_1}$ with 
$q \le p - 2$. The relations (\ref{absnullabsmpmap}) therefore
determine for a given sequence $\hat{{\cal D}}_{mp}$ recursively a
unique sequence $\hat{{\cal D}}_n$.
\end{proof}

It follows that for a given metric $h$ the sequences of multipoles and the
sequences of null data in a given standard frame carry the same
information on $h$. The relation is not simple, however. It can
happen that a sequence $\hat{{\cal D}}_n$ with only a finite number of
non-vanishing members is mapped onto an sequence $\hat{{\cal
D}}_{mp}$ with an infinite number of
non-vanishing members and vice versa. For instance, the relations
given above show that the sequence
$\hat{{\cal D}}_{n} = \{\psi_2,\,\,0,\,\,0,\,\,
0, \ldots\}$ with $\psi_2 \equiv \psi_{A_2B_2A_1B_1} \neq 0$
is mapped onto the sequence
$\hat{{\cal D}}_{mp} =
\{\nu_2,\,\,0,\,\,\nu_4,\,\,0,\,\,\nu_6,\,\,\ldots\}$ 
with $\nu_q = \nu_{A_qB_q \,\ldots\,A_1B_1}$, where
\[
\nu_2 = \psi_2, \quad \quad
\nu_{2p} = (-1)^{p+1}\,(\Pi_{l = 1}^{p - 1} \,c_{2l + 1})\,
sym(\otimes^{p}\psi_2) \neq 0, \quad p \ge 2.
\]

\section{The characteristic initial value problem}
\label{charivp}

To complete the analysis one would have to show that the estimates
(\ref{symmest}) imply estimates of the type
(\ref{genCauchyest}) for the coefficients of (\ref{snormalexp}).  
The induction argument used in the proof of
Lemma \ref{realformalexpansion} leads, however, to complicated algebraic
considerations.  The commutation of covariant derivatives
generates with the subsequent derivative operations more and more
non-linear terms of lower order. Formalising this procedure 
to derive estimates does not look very attractive. 
To arrive at a formulation of our question which looks more similar
to a boundary value problem to which Cauchy-Kowalevskaya
type arguments apply, we make use of the inherent geometric nature
of the problem and the geometric meaning of the null data.

The fields $h$, $\zeta$, $s$, $s_{ABCD}$ are necessarily real
analytic in the normal coordinates $x^a$ and a standard frame $c_{AB}$
centered at $i$. They can thus be extended near $i$ by
analyticity into the complex domain and considered as holomorphic
fields on a complex analytic manifold $S_c$. Choosing $S_c$ to be a
sufficiently small neighbourhood of $i$, we can assume the extended
coordinates, again denoted by
$x^a$, to define a holomorphic coordinate system on $S_c$ which
identifies the latter with an open neighbourhood of the origin in $\mathbb{C}^3$.
The original manifold $S$ is then a real, $3$-dimensional, real analytic
submanifold of the real, $6$-dimensional, real analytic manifold
underlying $S_c$. 
If $\alpha^a$, $\beta^a$, $a = 1, 2, 3$,  define real local coordinates on
the real $6$-dimensional manifold underlying $S_c$ so that
the holomorphic coordinates $x^a$ can be written
$x^a = \alpha^a + i\,\beta^a$, we use the standard notation  
$\partial_{x^a} = \frac{1}{2}(\partial_{\alpha^a} -
i\,\partial_{\beta^a})$ and
$\partial_{\bar{x}^a} = \frac{1}{2}(\partial_{\alpha^a} +
i\,\partial_{\beta^a})$. The assumption that the complex-valued function 
$f = f(x^a)$ be holomorphic is then equivalent to the requirement
that $\partial_{\bar{x}^a}f = 0$ so that we will only have to deal with
the operators $\partial_{x^a}$. 
Under the analytic extension the main differential
geometric concepts and formulas remain valid. The coordinates $x^a$ and
the extended frame, again denoted by
$c_{AB}$, satisfy the same defining equations and the 
extended fields, denoted again by $h$, $\zeta$, $s$, $s_{ABCD}$,
satisfy the conformal static vacuum field equations as before.

The analytic function $\Gamma = \delta_{ab}\,x^a\,x^b$ on $S$ extends
to a holomorphic function on $S_c$ which satsifies again the
eikonal equation $h^{ab}\,D_a\Gamma\,D_b\Gamma = - 4\,\Gamma$. On $S$
it vanishes only at $i$, but the set 
\[
{\cal N}_i = \{p \in S_c|\,\,\Gamma(p) = 0\},
\]
is an irreducible analytical set (cf. \cite{range}) such that  
${\cal N}_i \setminus \{i\}$ is $2$-dimensional complex submanifold of
$S_c$. It is the cone swept out by the complex null geodesics through $i$
and we will refer to it shortly as the {\it null cone at $i$}.  
While some of the following considerations may be reminiscent of
considerations concerning cones swept out by real null geodesics
through given points of $4$-dimensional Lorentz spaces, there are basic
differences. In the present case there do not exist splittings into
future and past cones. The set ${\cal N}_i \setminus \{i\}$ is
connected and its set of  of complex null generators is diffeomorph to
$P^1(\mathbb{C}) \sim S^2$. If ${\cal N}_i \setminus \{i\}$ is considered as a
$4$-dimensional submanifold of the $6$-dimensional real manifold
underlying $S_c$, the set of real null generators is not simply connected
but diffeomorphic to $SO(3, \mathbb{R})$.

The set ${\cal N}_i$ will be important for geometrizing our problem.
Let $u \rightarrow x^a(u)$ be a null geodesic through $i$ so that 
$x^a(0) = 0$. Its tangent vector is then of the form 
$\dot{x}^{AB} = \iota^A\,\iota^B$ with a spinor field
$\iota^A = \iota^A(u)$ satisfying $D_{\dot{x}}\iota^A = 0$ along the geodesic . Then
\begin{equation}
\label{nulldatum}
s_0(u) = \iota^A\,\iota^B\,\iota^C\,\iota^D\,s_{ABCD}(x(u)),
\end{equation}
is an analytic function of $u$ with Taylor expansion
\[
s_0 = \sum_{p = 0}^{\infty} \frac{1}{p!}\,u^p\,
\frac{d^p}{du^p}\,s_0(0),
\]
where
\[
\frac{d^p}{du^p}\,s_0(0)
= \iota^{A_p}\,\iota^{B_p} \dots
\iota^C\,\iota^D\,D_{A_pB_p} \ldots D_{A_1 B_1} s_{ABCD}(i)
\]
\[
= \iota^{A_p}\,\iota^{B_p} \dots
\iota^C\,\iota^D\,D_{(A_pB_p} \ldots D_{A_1 B_1} s_{ABCD)}(i).
\]
Knowing these expansion coefficients for initial null vectors 
$\iota^A\,\iota^B$ covering an open subset of the null directions at $i$
is equivalent to knowing the null data ${\cal D}^*_n$ of the metric
$h$. 

Our problem can thus be formulated as the boundary value
problem for the conformal static vacuum equations with 
data given by the function (\ref{nulldatum}) on ${\cal N}_i$, where
the $\iota^A \iota^B$ are parallely propagated null vectors tangent to
${\cal N}_i$. The set ${\cal N}_i$ can be regarded as a (complex)
characteristic of the (extended) operator $\Delta_h$ and also to the
conformal static equations.  Therefore we shall refer to this problem as
the {\it characteristic initial value problem for the conformal static
vacuum field equations with data on the null cone at space-like
infinity}.

The conformal static vacuum field equations 
(\ref{hequ}), (\ref{Bn1fequ}), (\ref{B1intco}), (\ref{Bcott}) form a
$3$-dimensional analogue of the $4$-dimensional conformal Einstein
equations (\cite{friedrich:1981}). Characteristic initial value 
problems for these two type of systems are therefore quite similar in
character. 

The existence of analytic solutions to characteristic initial
value problems for the conformal Einstein equations has been shown in 
\cite{friedrich:1982} by using Cauchy-Kowalevskaya type arguments. 
In the present case we shall employ somewhat different techniques for the
following reason.

The remaining and in fact the main difficulty in our problem arises from
fact that ${\cal N}_i$ is not a smooth hypersurface but an
analytic set with a vertex at the point $i$. A characteristic initial
value problem for the conformal Einstein equations with data on a cone
has been studied in \cite{friedrich:pure rad} and some of the techniques
introduced there and further developed in \cite{friedrich:i-null} will be
used in the following. The method we use to derive estimates on the
expansion coefficients has apparently not been used before in the context
of Einstein's field equations.

\subsection{The geometric gauge}

To obtain a setting in which the mechanism of calculating the expansion
coefficients allows one to derive estimates on the
coefficients from the conditions imposed on the data, a gauge needs to be
chosen which is suitably adapted to the singular set ${\cal N}_i$.  The
coordinates and the frame field will then necessarily be singular and
the frame will no longer define a smooth lift to the bundle of frames
but a subset which becomes tangent to the fibres over some points. The
setting described in the following will organize this situation
in a geometric way and provide control on 
the singularity and the smoothness of the fields. 

Let $SU(2)$ be the group of complex  $2 \times 2$ matrices 
$(s^{A}\,_{B})_{A, B = 0,1}$ satisfying 
\begin{equation}
\label{SU2}
\epsilon_{AB}\,s^{A}\,_{C}\,s^{B}\,_{D} = \epsilon_{CD}, 
\quad
\tau_{AB'} \,s^{A}\,_{C}\,\bar{s}^{B'}\,_{D'} = 
\tau_{CD'},
\end{equation}
where $s^{B}\,_{D} \rightarrow \bar{s}^{B'}\,_{D'}$
denotes complex conjugation. The map 
\begin{equation}
\label{SU2SO3}
SU(2) \ni s^{A}\,_{B} 
\rightarrow s^{(A}\,_{(C}\,s^{B)}\,_{D)}
\rightarrow s^a\,_b = \alpha^a\,_{AB}\,s^{A}\,_{C}\,s^{B}\,_{D}
\,\alpha^{CD}\,_b \in SO(3, R),
\end{equation}
realizes the $2:1$ covering homomorphism of $SU(2)$ onto the group
$SO(3, \mathbb{R})$. 
Under holomorphic extension the map above extends to a
$2:1$ covering homomorphism of the group $SL(2, \mathbb{C})$ onto the group 
$SO(3, \mathbb{C})$, where $SL(2, \mathbb{C})$ denotes the group of complex $2 \times 2$ matrices satisfying only the first of conditions (\ref{SU2}).

We will make use of the principal bundle of normalized spin
frames 
$SU(S) \stackrel{\pi}{\rightarrow} S$ with structure group $SU(2)$. A
point $\delta \in SU(S)$ is given by a pair of spinors 
$\delta = (\delta^A_{0}, \delta^A_{1})$ at a given point of $S$
which satisfies
\begin{equation}
\label{su2spinframe}
\epsilon(\delta_{A},\,\delta_{B}) = 
\epsilon_{AB}, \quad
\epsilon(\delta_{A},\,\delta^+\,_{B'}) = 
\tau_{AB'},
\end{equation}
where the lower index, which labels the members of the spin frame, 
is assumed to acquire a prime under the ``$+$''-operation. The action of
the structure group is given for $s \in SU(2)$ by  
\[
\delta \rightarrow \delta \cdot s
\quad \mbox{where} \quad
(\delta \cdot s)_{A} =  s^{B}\,_{A}\,\delta_{B}.
\]
The projection $\pi$ maps a frame $\delta$ onto its base point in $S$.
The bundle of spin frames is mapped by a $2:1$ bundle morphism 
$SU(S) \stackrel{p}{\rightarrow} SO(S)$ onto the
bundle $SO(S) \stackrel{\pi'}{\rightarrow } S$ of oriented, orthonormal
frames on $S$ so that $\pi' \circ p = \pi$. 
For any spin frame $\delta$ we can identify by (\ref{su2spinframe}) the
matrix $(\delta^A_{B})_{A, {A, B} = 0, 1}$ with an
element of the group $SU(2)$. With this reading the map $p$ will be
assumed to be realized by
\[
SU(S) \ni \delta \rightarrow p(\delta)_{AB} 
= \delta^E_{A}\,\delta^F_{B}\,c_{EF} \in SO(S),
\]
where $c_{AB}$ denotes the normal frame field on $S$ introduced before. We
refer to $p(\delta)$ as the frame associated with the spin frame
$\delta$.

Under holomorphic extension the bundle $SU(S) \rightarrow S$ is extended to
the principal bundle $SL(S_c) \stackrel{\pi}{\rightarrow} S_c$ of spin
frames  $\delta = (\delta^A_{0}, \delta^A_{1})$ at given points of $S_c$
which satisfy only the first of conditions (\ref{su2spinframe}). Its
structure group is $SL(2, \mathbb{C})$.
The  bundle $SU(S) \stackrel{\pi}{\rightarrow} S$ is embedded
into $SL(S_c) \stackrel{\pi}{\rightarrow} S_c$ as a real analytic
subbundle. The bundle morphism $p$ extends to a $2:1$ bundle morphism,
again denoted by $p$, of $SL(S_c) \stackrel{\pi}{\rightarrow} S_c$ onto
the bundle $S0(S_c) \stackrel{\pi'}{\rightarrow} S_c$ of oriented,
normalized frames of $S_c$ with structure group $SO(3, \mathbb{C})$.
We shall make use of several structures on $SM(S_c)$.

With each $\alpha \in sl(2, \mathbb{C})$, i.e. $\alpha = (\alpha^A\,_B)$ with
$\alpha_{AB} = \alpha_{BA}$, is associated a 
{\it vertical vector field $Z_{\alpha}$} tangent to the fibres, which is
given at $\delta \in SL(S_c)$ by
$Z_{\alpha}(\delta) = \frac{d}{dv} (\delta \cdot
exp(v\,\alpha))|_{v = 0}$, where $v \in  \mathbb{C}$ and 
$exp$ denotes the exponential map $sl(2, \mathbb{C}) \rightarrow 
SL(2,  \mathbb{C})$.

The $ \mathbb{C}^3$-valued {\it solder form} $\sigma^{AB} = \sigma^{(AB)}$ maps
a tangent vector $X \in T_{\delta}SL(S_c)$ onto the components
of its projection $T_{\delta}(\pi)X \in
T_{\pi(\delta)}S_c$  in the frame $p(\delta)$ associated
with $\delta$ so that 
$T_{\delta}(\pi)X  = \,<\sigma^{AB}, X>\,p(\delta)_{AB}$. It follows that
$<\sigma^{AB}, Z_{\alpha}>\, = 0$ for any vertical vector field
$Z_{\alpha}$.

The $sl(2,  \mathbb{C})$-valued {\it connection form $\omega^A\,_B$ on
$SL(S_c)$} transforms with the adjoint transformation
under the action of $SL(2,  \mathbb{C})$ and maps any vertical vector field
$Z_{\alpha}$ onto its generator so that 
$<\omega^A\,_B, Z_{\alpha}>\,= \alpha^A\,_B$.

With $x^{AB} = x^{(AB)} \in  \mathbb{C}^3$ is associated the {\it horizontal
vector field $H_x$} on $SL(S_c)$ which is horizontal in the sense that
$<\omega^A\,_B, H_x>\, = 0$ and which satisfies
$<\sigma^{AB}, H_x>\, = x^{AB}$.
Denoting by $H_{AB}$, $A, B = 0, 1$, the horizontal vector fields
satisfying $<\sigma^{AB}, H_{CD}>\, = h^{AB}\,_{CD}$, it follows that
$H_x = x^{AB}\,H_{AB}$. An integral curve of a horizontal vector field
projects onto an $h$-geodesic and represents a spin frame field which is
parallely transported along this geodesic.

A holomorphic spinor field $\psi$ on $S_c$ is represented on $SL(S_c)$ by
a  holomorphic spinor-valued function $\psi_{A_1 \ldots A_j}(\delta)$
on $SL(S_c)$, given by the components of $\psi$ in the frame $\delta$.
We shall use the notation $\psi_k = \psi_{(A_1 \ldots A_j)_k}$,
$k = 0, \ldots , j$, where  $(\ldots \ldots)_k$ denotes the
operation {\it`symmetrize and set
$k$ indices equal to $1$ the rest equal to $0$'}. If $\psi$ is symmetric,
these functions completely specify $\psi$. They are referred
to as the {\it esssential components of $\psi$}.

\subsection{The submanifold $\hat{S}$ of $SL(S_c)$}

We combine the construction of a coordinate system and a
frame field with the definition of an analytic submanifold $M$ of
$SL(S_c)$ which is obtained as follows. We choose a spin frame $\delta^*$
in the fibre of $SL(S_c)$ over $i$ which is projected by $\pi'$ onto the
frame $c_{AB}$ at considered $i$ before. The curve
\[
C \ni v \rightarrow \delta(v) = \delta^* \cdot s(v) \in SL(S_c),
\]
with
\begin{equation}
\label{thegenerator}
s(v) = exp(v\,\alpha) = \left( \begin{array}{cc}
1 & 0 \\
v & 1  \\
\end{array} 
\right), \quad
\alpha = \left( \begin{array}{cc}
0 & 0 \\
1 & 0  \\
\end{array} 
\right) \in sl(2, C),
\end{equation}
in the fibre of $SL(S_c)$ over $i$ defines a vertical, 1-dimensional,
holomorphic submanifold $I$ through $\delta^*$ on which $v$ defines a
coordinate. The associated family of frames $e_{AB} = e_{AB}(v)$ at $i$ is
given explicitly by
\[
e_{00}(v) = c_{00} + 2\,v\,c_{01} + v^2\,c_{11},\,\,\,\,\,\,\,\,
e_{01}(v) = c_{01} + v\,c_{11},\,\,\,\,\,\,\,\,
e_{11}(v) = c_{11}.
\]
The following construction is carried out in some neighbourhood of $I$. If
the latter is chosen small enough all the following statements will be
correct.

The set $I$ is moved with the flow of $H_{11}$ to obtain
a holomorphic $2$-manifold $U_0$ of $SL(S_c)$ containing $I$. The parameter
on the integral curves of $H_{11}$ which vanishes on $I$ will be denoted
by $w$ and $v$ is extended to $U_0$ by assuming it to be constant on
the integral curves of $H_{11}$. All these integral curves are mapped by
$\pi$ onto the null geodesics $\gamma(w)$ with affine parameter $w$
and tangent vector $\gamma'(0) = c_{11}$ at $\gamma(0) = i$. The parameter
$v$ specifies frame fields which are parallely propagted along $\gamma$.

The set $U_0$ is moved with the flow of $H_{00}$
to obtain a holomorphic $3$-submanifold $\hat{S}$ of $SL(S_c)$ containing
$U_0$. We denote by $u$ the parameter on the integral curves of $H_{00}$
which vanishes on $U_0$, extend $v$ and $w$ to $\hat{S}$ by assuming
them to be constant on the integral curves of $H_{00}$. The
functions $z^1 = u$, $z^2 = v$, $z^3 = w$ define 
holomorphic coordinates on $\hat{S}$.
The restriction the projection to $\hat{S}$ will be again denoted by $\pi$.

The projections of the integral curves of $H_{00}$ with a fixed value of
$w$ sweep out, together with $\gamma$, the cone ${\cal N}_{\gamma(w)}$
near $\gamma(w)$ which is generated by the null geodesics through the
point $\gamma(w)$. On the null geodesics $u$ is an affine parameter which
vanishes at $\gamma(w)$ while $v$ parametrizes the different generators.
In terms of the base space $S_c$ our gauge is based on the
nested family of cones ${\cal N}_{\gamma(w)}$ which share the
generator $\gamma$. The set $W_0 = \{w = 0\}$, which projects onto 
${\cal N}_i \setminus \gamma$, will define the intial data set for
our problem. 
The map $\pi$ induces a biholomorphic diffeomorphism  of
$\hat{S}' \equiv \hat{S} \setminus U_0$ onto $\pi(\hat{S}')$. 
The  singularity of the gauge at points of $U_0$ (resp. over $\gamma$) 
consists in $\pi$ dropping rank on $U_0$ because the curves
$w = const.$ on $U_0$ are tangent to the fibres over
$\gamma(w)$ where $\partial_v = Z_{\alpha}$. The null curve $\gamma(w)$
will be referred to as the singular generator of  
${\cal N}_i$ in the gauge determined by the spin frame $\delta^*$ resp.
the corresponding frame $c_{AB}$ at $i$.

The solder and the connection form pull back to holomorphic
$1$-forms on $\hat{S}$, which will be denoted again by $\sigma^{AB}$ and
$\omega^A\,_B$. Corresponding to the behaviour of $\pi$ the $1$-forms
$\sigma^{00}$, $\sigma^{01}$, $\sigma^{11}$ are linearly independent
on $\hat{S}'$ while the rank of this system drops to $2$ on
$U_0$ because $<\sigma^{AB}, \partial_v> \,= \,<\sigma^{AB}, Z_{\alpha}>\,=
0$. If the pull back of the curvature form 
$\Omega^A\,_B = \frac{1}{2}\,r^A\,_{BCDEF}\,\sigma^{CD} \wedge \sigma^{EF}$
to $\hat{S}$ is denoted again
by $\Omega^A\,_B$, the solder and the connection form satisfy the
structural equations 
\[
d\,\sigma^{AB} = - \omega^A\,_C \wedge \sigma^{CB}
- \omega^B\,_C \wedge \sigma^{AC},
\quad\quad\quad\quad
d\,\omega^A\,_B = - \omega^A\,_C \wedge \omega^C\,_B 
+ \Omega^A\,_B.
\]
By construction of $\hat{S}$ we have
\[
< \sigma^{AB}, \partial_v>\, = 0,\,\,\,\,\,\,
< \sigma^{AB}, \partial_w>\, = \epsilon_1\,^A\,\epsilon_1\,^B
\,\,\,\,\,\,\mbox{on}\,\,\,\,\,\,U_0,
\]

\[
< \omega^A\,_B, \partial_w>\, = 0,\,\,\,\,\,\,
< \omega^A\,_B, \partial_v>\, =
\,< \omega^A\,_B, Z_{\alpha}>\, = \epsilon_1\,^A\,\epsilon_B\,^0
\,\,\,\,\,\,\mbox{on}\,\,\,\,\,\,U_0,
\]

\[
< \sigma^{AB}, \partial_u> \,= \epsilon_0\,^A\,\epsilon_0\,^B
\,\,\, \mbox{and} \,\,\,
< \omega^A\,_B, \partial_u> \,= 0
\,\,\,\mbox{on}\,\,\,\hat{S}
\,\,\, \mbox{while} \,\,\,
< \sigma^{AB}, \partial_v>\, \neq 0
\,\,\,\mbox{on}\,\,\,\hat{S}'.
\]

To obtain more precise information on $\sigma^{AB}$ and $\omega^A\,_B$ we
note the following general properties (cf. \cite{friedrich:pure rad}
and \cite{friedrich:i-null} for more details). If, for given $x^{AB} \in
C^3$, the Lie derivative with respective $H_x$ is denoted by ${\cal
L}_x$, then
\[
{\cal L}_x \sigma^{AB} = \,2\,x^{C(A}\,\omega^{B)}\,_C,
\,\,\,\,\,\,\,\,\,
<{\cal L}_x \omega^A\,_B,\,.\,>\,\, = \,\,<\Omega^A\,_B, H_x \wedge\,.\,>.
\]

Since $0 = [\partial_u, \partial_v] = [H_{00}, \partial_v]$ on $\hat{S}$
and $\Omega^A\,_B$ is horizontal, it follows that 
\[
\partial_u<\sigma^{AB}, \partial_v> \,
= 2\,\epsilon_0\,^{(A} <\omega^{B)}\,_0, \partial_v>,
\,\,\,
\partial_u<\omega^A\,_B, \partial_v>|_{u = 0} =
\,<\Omega^A\,_B, H_x \wedge Z_{\alpha}>|_{u = 0} = 0,
\]
which gives with the relations above 
\[
<\omega^A\,_B, \partial_v>\, = \epsilon_1\,^A\,\epsilon_B\,^0 + O(u^2)
\,\,\,\mbox{whence}\,\,\,
<\omega^A\,_B, \partial_v>\, = \,2\,u\,\epsilon_0\,^{(A} \epsilon_1\,^{B)}
+ O(u^3)
\,\,\,\,\,\,\,\mbox{as}\,\,\,\,\,\,\,
u \rightarrow 0.
\]
Similarly we obtain with 
$0 = [\partial_u, \partial_w] = [H_{00}, \partial_w]$  on $\hat{S}$
\[
\partial_u<\sigma^{AB}, \partial_w>\, = 
2\,\epsilon_0\,^{(A} <\omega^{B)}\,_0, \partial_w>,
\,\,\,\,\,\,\,\,\,\,
\partial_u<\omega^A\,_B, \partial_w>|_{u = 0} =
\frac{1}{2}\,r^A\,_{B 00 11}.
\]
In terms of the coordinates $z^a$ we thus get $\sigma^{AB} =
\sigma^{AB}\,_a\,dz^a$ on $\tilde{S}'$ with a co-frame matrix 
\begin{equation}
\label{sigmaasbeh}
(\sigma^{AB}\,_a) 
 = 
\left( \begin{array}{ccc}
1 & \sigma^{00}\,_2 & \sigma^{00}\,_3\\
0 & \sigma^{01}\,_2 & \sigma^{01}\,_3\\
0 & 0 & 1\\
\end{array} \right)
=
\left( \begin{array}{ccc}
1 & O(u^3) & O(u^2)\\
0 & u + O(u^3) & O(u^2)\\
0 & 0 & 1\\
\end{array} \right)
\,\,\,\,\,\,\,\mbox{as}\,\,\,\,\,\,\,
u \rightarrow 0.
\end{equation}

On $\hat{S}'$ there exist unique, holomorphic vector fields
$e_{AB}$ which satisfy 
\[
< \sigma^{AB}, e_{EF}>\, = h^{AB}\,_{EF}.
\]
If we write 
$e_{AB} = e^a\,_{AB}\,\partial_{z^a}$, the properties noted above imply
for the frame coefficients 
\begin{equation}
\label{easbeh}
(e^a\,_{AB}) =
 \left( \begin{array}{ccc}
1 & e^1\,_{01} & e^1\,_{11}\\
0 & e^2\,_{01} & e^2\,_{11}\\
0 & 0 & 1\\
\end{array} \right)
=
 \left( \begin{array}{ccc}
1 & O(u^2) & O(u^2)\\
0 & \frac{1}{2\,u} + O(u) & O(u)\\
0 & 0 & 1\\
\end{array} \right)
\,\,\,\,\,\,\,\mbox{as}\,\,\,\,\,\,\,
u \rightarrow 0.
\end{equation}
In the following we shall write 
\begin{equation}
\label{frameregsing}
e^a\,_{AB} = e^{*a}\,_{AB} + \hat{e}^a\,_{AB},
\end{equation}
with singular part 
\begin{equation}
\label{framesing}
e^{*a}\,_{AB} = 
  \delta^a_1\,\epsilon_A\,^0\,\epsilon_B\,^0
+ \delta^a_2\,\frac{1}{u}\,\epsilon_{(A}\,^0\,\epsilon_{B)}\,^1
+ \delta^a_3\,\epsilon_A\,^1\,\epsilon_B\,^1,
\end{equation}
and holomorphic functions $\hat{e}^a\,_{AB}$ on $\hat{S}$
 which satisfy
\begin{equation}
\label{ehatOu} 
\hat{e}^a\,_{AB} = O(u) 
\quad \mbox{as} \quad u \rightarrow 0.
\end{equation}

We define connections coefficients on $\hat{S}'$ by writing
$\omega^A\,_B = \Gamma_{CD}\,^A\,_B\,\sigma^{CD}$ with
\[
\Gamma_{CD\,AB} \equiv \,<\omega_{AB},e_{CD}>,
\] 
so that $\Gamma_{CD\,AB} = \Gamma_{(CD)\,(AB)}$. The
definition of the frame then implies
\[
\Gamma_{00\,AB} = 0 
\quad \mbox{on} \quad \hat{S}
\quad \mbox{and} \quad 
\Gamma_{11\,AB} = 0 
\quad \mbox{on} \quad U_0, 
\]
and it follows from the
discussion above that  
\begin{equation}
\label{connregsing}
\Gamma_{ABCD} = \Gamma^*_{ABCD} + \hat{\Gamma}_{ABCD},
\end{equation}
with singular part  
\begin{equation}
\label{connsing}
\Gamma^*_{AB\,CD} = - \frac{1}{u}\,
\epsilon_{(A}\,^0\,\epsilon_{B)}\,^1\,\epsilon_C\,^0\,\epsilon_D\,^0,
\end{equation}
and holomorphic functions $\hat{\Gamma}_{ABCD}$ on $\hat{S}$
which satisfy 
\begin{equation}
\label{GammaOu} 
\hat{\Gamma}_{ABCD} = O(u) 
\quad \mbox{as} \quad u \rightarrow 0.
\end{equation}

The singular parts are `universal' in the sense that their expressions
only depend on the construction of $\hat{S}$ and not on properties of the
metric. If the latter is flat the functions $\hat{e}^a\,_{AB}$ and
$\hat{\Gamma}_{ABCD}$ vanish on $\hat{S}$. 
With the frame and the connection coefficients so defined we have the
spin frame calculus in its standard form.  The expressions
above imply for any holomorphic spinor valued function $\psi_{A \ldots C}$
that $D_{00}\,\psi_{A \ldots C}$ and $D_{11}\,\psi_{A \ldots C}$
extend to $\hat{S}$ as holomorphic functions so that
\[
D_{00}\,\psi_{A \ldots C} = \partial_u\,\psi_{A \ldots C}
\,\,\,\mbox{on}\,\,\,\hat{S} \quad \mbox{and}\quad
D_{11}\,\psi_{A \ldots C} = \partial_w\,\psi_{A \ldots C}
\,\,\, \mbox{on} \,\,\, U_0.
\]

\subsection{Tensoriality and expansion type}

A holomorphic function on $SL(S_c)$ induces a holomorphic
function on $\hat{S}$ which can be considered as a holomorphic function
of the coordinates $z^a$. While these coordinates are holomorphic on the 
submanifold $\hat{S}$ of $SL(S_c)$, the induced map 
$\pi$ of $\hat{S}$ into $S_c$ is singular on $U_0$. As a consequence, not
every holomorphic function of the $z^a$ can arise as a pull-back to 
$\hat{S}$ of a holomorphic function on $SL(S_c)$. The latter must have a
special type of expansion in terms of the $z^a$ which reflects
the particular relation between the `angular' coordinate $v$ the `radial'
coordinate $u$. The following notion will be important for our
discussion.

{\bf Definition}:
{\it A holomorphic function $f$ on $\hat{S}$
will be said to be of $v$-finite expansion type $k_f$, with $k_f$ an
integer, if it has in terms of the coordinates $u$, $v$, and $w$ a Taylor
expansion at the origin of the form
\[
f = \sum_{p = 0}^{\infty} \sum_{m = 0}^{\infty} \sum_{n = 0}^{2\,m + k_f}
f_{m, n, p}\,u^m\,v^n\,w^p,
\]
where it is assumed that $f_{m, n, p} = 0$ if $2\,m + k_f < 0$}.

We note that the construction of $\hat{S}$ does not distinguish the 
set $I = \pi^{-1}(i)$ from the sets $\pi^{-1}(\gamma(w))$.
Correspondingly, the Taylor expansions of the function $f$ above at 
points $(0, 0, w_0)$ with
$w_0$ close to
$0$  have the same structure with respect to $u$ and $v$.

\begin{lemma}
\label{tensorexptype}
Let $\phi_{A_1 \ldots A_j}$ be a holomorphic, symmetric, spinor-valued
function on $SL(S_c)$. Then the restrictions of its essential components  
$\phi_k = \phi_{(A_1 \ldots A_j)_k}$, $0 \le k \le j$, to $\hat{S}$
satisfy 
\begin{equation}
\label{genregcond}
\partial_v\,\phi_k = (j - k)\,\phi_{k + 1},\quad k = 0, \ldots , j,
\quad \mbox{on} \quad U_0,
\end{equation}
(where we set $\phi_{j+1} = 0$) and $\phi_k$ is of expansion type
$j - k$.
\end{lemma}

\begin{proof}: In the following we consider $\hat{S}$ as a submanifold
of $SL(S_c)$. 
The tensorial transformation law of $\phi$ under the action of the
$1$-parameter subgroup (\ref{thegenerator}) with generator
$\alpha^A\,_B = \epsilon_1\,^A\,\epsilon_B\,^0$ implies
\[
Z_{\alpha}\,\phi_k = (j - k)\,\phi_{k + 1}
\quad \mbox{for} \quad 0 \le k \le j
\quad \mbox{on} \quad SL(S_c),
\]
and thus (\ref{genregcond}) because $Z_{\alpha} = \partial_v$ on $U_0$.
From the relations above follows in particular that
\begin{equation}
\label{tenslaw}
Z_{\alpha}^{j - k + 1}\,\phi_k = 0  \quad \mbox{on} \quad SL(S_c).
\end{equation}
A general horizontal vector field $H_x$ has with $Z_{\alpha}$ the
commutator 
\[
[Z_{\alpha}, H_x] = H_{\alpha \cdot x},
\]
where $\alpha$ acts on $x^{AB} = x^{(AB)}$ according to the induced action by
\[
x^{AB} \rightarrow (\alpha \cdot x)^{AB} = 
\alpha^A\,_C\,x^{CB} + \alpha^B\,_C\,x^{AC} = 2\,\epsilon_1^{(A}\,x^{B)0}.
\]
With $x^{AB} = \epsilon_0\,^A\,\epsilon_0\,^B$, so that $H_x = H_{00}$, it
follows
\[
[Z_{\alpha}, H_{00}] = 2\,H_{01},\,\,\,\,\,\,\,
[Z_{\alpha}, H_{01}] = H_{11},\,\,\,\,\,\,\,
[Z_{\alpha}, H_{11}] = 0.
\]
By induction this gives the operator equations
\[
Z^n_{\alpha}\,H_{00} = n\,(n - 1)\,H_{11}\,Z^{n - 2}_{\alpha}
+ 2\,n\,H_{01}\,Z^{n - 1}_{\alpha} + H_{00}\,Z^{n}_{\alpha},\,\,\,\,\,\,\,
n \ge 1,
\]
and, more generally, 
\[
Z^n_{\alpha}\,H^m_{00} = a_{n,m}\,H^m_{11}\,Z^{n - 2m}_{\alpha}
+ \sum_{l = 0}^{2\,m - 1} A_{n, m, l}\,Z^{n - l}_{\alpha} +
H^m_{00}\,Z^{n}_{\alpha},\,\,\,\,\,\,\, m,\,n \ge 1,
\]
where the $a_{n,m}$ are real coefficients, the $A_{n, m, l}$ denote
operators which are sums of products of horizontal vector
fields, and the terms in which $Z_{\alpha}$ formally appears with 
negative exponent are assumed to vanish. 
With (\ref{tenslaw}) this implies
\[
Z^n_{\alpha}\,H^m_{00} \,\phi_k = 0
\quad\mbox{for}\quad n > 2\,m + j - k
\quad \mbox{on} \quad SL(S_c).
\]
The results follows because
$Z^n_{\alpha}\,H^m_{00} \,\phi_k =
\partial^n_v\,\partial^m_u\,\phi_k$ at points of $U_0$. 
\end{proof}

\subsection{The null data on $W_0$}

We shall derive an expansion of the restriction of the
essential component $s_0$ of the Ricci spinor to the hypersurface
$W_0$, i.e.  
\[
s_0(u, v) = s_{(ABCD)_0}|_{W_0},
\]
in terms of quantities
on the base space  $S_c$.  Consider the normal frame $c_{AB}$  on 
$S_c$ near $i$ which agrees at $i$ with the frame associated with
$\delta^*$ and denote by
\[
{\cal D}^*_n \equiv \{D^*_{(A_1B_1} \ldots D^*_{A_{p} B_{p}}
s^*_{ABCD)}(i),\,\,\,p = 0, 1,  2, \ldots\},
\] 
the corresponding null data of $h$ in the frame $c_{AB}$. 
Choose a fixed value of $v$ and consider $s = s(v)$ as in
(\ref{thegenerator}). 
The vector $H_{00}(\delta^* \cdot s)$ then projects onto the null vector
$s^A\,_0\,s^B\,_0\,c_{AB}$ at $i$.
Since $c_{AB}$ is a normal frame near $i$, the null vector field
$s^A\,_0\,s^B\,_0\,c_{AB}$ is tangent to a null geodesic $\eta = \eta(u, v)$
on ${\cal N}_i$ with affine parameter $u$ with $u = 0$ at $i$
and the integral curve of $H_{00}$
through $\delta^* \cdot s$ projects onto this null geodesic.
It follows from this with the explicit expression for $s = s(v)$ that 
\begin{equation}
\label{nulldataonU0}
s_0(u, v) 
= s^A\,_0(v)\,s^B\,_0(v)\,s^C\,_0(v)\,s^D\,_0(v)
\,s^*_{ABCD}|_{\eta(u, v)}
\end{equation}
\[
=
\sum_{m = 0}^{\infty}
\frac{1}{m\,!}\,u^m
\,s^{A_1}\,_0(v)\,s^{B_1}\,_0(v)\ldots\,s^{D}\,_0(v)\,
D^*_{(A_1B_1} \ldots D^*_{A_m B_m}\,s^*_{ABCD)}(i)
\]
\[
= \sum_{m = 0}^{\infty}\sum_{n = 0}^{2\,m + 4}
\,\psi_{m, n}\,\,u^m\,\,v^n,
\]
with
\[
\psi_{m, n} = \frac{1}{m\,!}\,{2\,m + 4 \choose n}
\,D^*_{(A_1B_1} \ldots D^*_{A_m B_m}\,s^*_{ABCD)_n}(i),
\quad 0 \le n \le 2\,m + 4.
\]
This formula shows us how to determine the function $s_0(u, v)$
from the null data ${\cal D}^*_n$ and
vice versa.  We note that the expansion above is consistent with $s_0$ 
being of $v$-finite expansion type $4$. We shall refer to (\ref{nulldataonU0})
as the {\it null data on $W_0$} in our gauge.

\section{The conformal static vacuum field equations on $\hat{S}$}
\label{confstatvaconShat}

With the frame $e_{AB}$ and the connection coefficients $\gamma_{ABCD}$
on $\hat{S}$ we have the standard frame calculus available.
Given the fields $\zeta$, $s$, $s_{ABCD}$, we define on $\hat{S}$ the
quantities 

\[
t_{AB}\,^{EF}\,_{CD}\,e^a\,_{EF} \equiv 
(\Gamma_{AB}\,^{EF}\,_{CD} - \Gamma_{CD}\,^{EF}\,_{AB})\,e^a\,_{EF}
- e^a\,_{CD,b}\,e^b\,_{AB} + e^a\,_{AB,b}\,e^b\,_{CD} ,
\] 

\[
R_{ABCDEF} \equiv r_{ABCDEF} 
- \frac{1}{2}\,\left\{s_{ABCE}\,\epsilon_{DF} + s_{ABDF}\,\epsilon_{CE}
\right\},
\]
with 
\[
r_{ABCDEF} \equiv
e_{CD}(\Gamma_{EFAB}) - e_{EF}(\Gamma_{CDAB})
\]
\[
+ \Gamma_{EF}\,^K\,_C\,\Gamma_{KDAB}
+ \Gamma_{EF}\,^K\,_D\,\Gamma_{CKAB}
- \Gamma_{CD}\,^K\,_E\,\Gamma_{KFAB}
\]
\[
- \Gamma_{CD}\,^K\,_F\,\Gamma_{EKAB}
+ \Gamma_{EF}\,^K\,_B\,\Gamma_{CDAK}
- \Gamma_{CD}\,^K\,_B\,\Gamma_{EFAK}
\]
\[
- t_{CD}\,^{GH}\,_{EF}\,\Gamma_{GHAB},
\]

\[
\Sigma_{AB} \equiv D_{AB}\,\zeta - \zeta_{AB}, 
\]

\[
\Sigma_{ABCD} \equiv D_{AB}\,\zeta_{CD} - s\,h_{ABCD}
+ \zeta\,(1 - \mu\,\zeta)\,s_{ABCD}, 
\]

\[
S_{AB} \equiv D_{AB}\,s + (1 - \mu\,\zeta)\,s_{ABCD}\,\zeta^{CD},
\]

\[
H_{ABCD} \equiv D_A\,^E s_{BCDE} -
\frac{2\,\mu}{1 - \mu\,\zeta}\,s_{E(BCD}\,\zeta_{A)}\,^E.
\]

In terms of the fields on the left hand side, which have been introduced
as labels for the equations as well as for the discussion of their
interdependencies, the conformal static vacuum equations  read 
\[
t_{AB}\,^{EF}\,_{CD}\,e^a\,_{EF} = 0,\quad \quad
R_{ABCDEF} = 0,
\]
\[
 \Sigma_{AB} = 0, 
\]
\[
\Sigma_{ABCD} = 0, \quad \quad 
S_{AB} = 0, \quad \quad
H_{ABCD} = 0.
\]
The first equation is Cartan's first structural equation with the
requirement that the (metric) connection be torsion free 
($t_{AB}\,^{EF}\,_{CD}$ being the torsion tensor). The second equation is
Cartan's second structural equation with the requirement that the Ricci
tensor coincides with the trace free tensor
$s_{ab}$. The third equation defines
$\zeta_{AB}$, the remaining equations have been considered before. 

To discuss these equations in detail we need to write them out in our
gauge, observing in particular the nature of the singularities in
(\ref{frameregsing}) and  (\ref{connregsing}).

The equations $t_{AB}\,^{EF}\,_{00}\,e^a\,_{EF} = 0$: 

\[
\partial_u\hat{e}^1\,_{01} + \frac{1}{u}\,\hat{e}^1\,_{01}
= - 2\,\hat{\Gamma}_{0101}
+ 2\,\hat{\Gamma}_{0100}\,\hat{e}^1\,_{01},
\]
\[
\partial_u\hat{e}^2\,_{01} + \frac{1}{u}\,\hat{e}^2\,_{01}
= 
\frac{1}{u}\,\hat{\Gamma}_{0100}
+ 2\,\hat{\Gamma}_{0100}\,\hat{e}^2\,_{01},
\]
\[
\partial_u\hat{e}^1\,_{11}
= - 2\,\hat{\Gamma}_{1101} 
+ 2\,\hat{\Gamma}_{1100}\,\hat{e}^1\,_{01},
\]
\[
\partial_u\hat{e}^2\,_{11}
= \frac{1}{u}\hat{\Gamma}_{1100}
+ 2\,\hat{\Gamma}_{1100}\,\hat{e}^2\,_{01}.
\]

The equations $R_{AB00EF} = 0$:

\[
\partial_u\hat{\Gamma}_{0100} 
+ \frac{2}{u}\,\hat{\Gamma}_{0100}
- 2\,\hat{\Gamma}^2_{0100}
= \frac{1}{2}\,s_{0},
\]
\[
\partial_u\hat{\Gamma}_{0101} + \frac{1}{u}\,\hat{\Gamma}_{0101}
- 2\,\hat{\Gamma}_{0100}\,\hat{\Gamma}_{0101}
= \frac{1}{2}\,s_{1},
\]
\[
\partial_u\hat{\Gamma}_{0111} + \frac{1}{u}\,\hat{\Gamma}_{0111}
- 2\,\hat{\Gamma}_{0100}\,\hat{\Gamma}_{0111}
= \frac{1}{2}\,s_{2},
\]
\[
\partial_u\hat{\Gamma}_{1100} + \frac{1}{u}\,\hat{\Gamma}_{1100} 
- 2\,\hat{\Gamma}_{1100}\,\hat{\Gamma}_{0100}
= s_{1},
\]
\[
\partial_u\hat{\Gamma}_{1101} 
- 2\,\hat{\Gamma}_{1100}\,\hat{\Gamma}_{0101}
= s_{2},
\]
\[
\partial_u\hat{\Gamma}_{1111} 
- 2\,\hat{\Gamma}_{1100}\,\hat{\Gamma}_{0111}
= s_{3}.
\]

The equations $\Sigma_{00} = 0$, $\Sigma_{00CD} = 0$, $S_{00} = 0$:

\[
0 = \partial_u\zeta - \zeta_{00},
\]
\[
0 = \partial_u\,\zeta_{00} + \zeta\,(1 - \mu\,\zeta)\,s_{0}, 
\]
\[
0 = \partial_u\,\zeta_{01} + \zeta\,(1 - \mu\,\zeta)\,s_{1}, 
\]
\[
0 = \partial_u\,\zeta_{11} - s + \zeta\,(1 - \mu\,\zeta)\,s_{2}, 
\]
\[
0 = \partial_us + (1 - \mu\,\zeta)\,
(s_{0}\,\zeta_{11} - 2\,s_{1}\,\zeta_{01} + s_{2}\,\zeta_{00}). 
\]

The equations $- H_{0(BCD)_k} = 0$ in the order $k = 0, 1, 2, 3$: 

\[
\partial_u\,s_1 - \frac{1}{2\,u}(\partial_v\,s_0 - 4\,s_1)
- \hat{e}^1\,_{01}\partial_u\,s_0 
- \hat{e}^2\,_{01}\partial_v\,s_0 
\]
\[
= - 4\,\hat{\Gamma}_{0101}\,s_0
+ 4\,\hat{\Gamma}_{0100}\,s_1
- \frac{2\,\mu}{(1 - \mu\,\zeta)}\,
\left\{s_0\,\zeta_{01} - s_1\,\zeta_{00}\right\},
\]

\[
\partial_u\,s_2 
- \frac{1}{2\,u}(\partial_v\,s_1 - 3\,s_2) 
- \hat{e}^1\,_{01}\partial_u\,s_1 
- \hat{e}^2\,_{01}\partial_v\,s_1 
\]
\[
= - \hat{\Gamma}_{0111}\,s_0
- 2\,\hat{\Gamma}_{0101}\,s_1
+ 3\,\hat{\Gamma}_{0100}\,s_2
- \frac{\mu}{2\,(1 - \mu\,\zeta)}\,
\left\{s_0\,\zeta_{11} + 2\,s_1\,\zeta_{01}
+ 3\,s_2\,\zeta_{00}\right\},
\]

\[
\partial_u\,s_3 
- \frac{1}{2\,u}(\partial_v\,s_2 - 2\,s_3) 
- \hat{e}^1\,_{01}\partial_u\,s_2 
- \hat{e}^2\,_{01}\partial_v\,s_2 
\]
\[
= - 2\hat{\Gamma}_{0111}\,s_1
+ 2\,\hat{\Gamma}_{0100}\,s_3
- \frac{\mu}{(1 - \mu\,\zeta)}\,\left\{s_1\,\zeta_{11}
+ s_3\,\zeta_{00}\right\},
\]

\[
\partial_u\,s_4
- \frac{1}{2\,u}(\partial_v\,s_3 - s_4) 
- \hat{e}^1\,_{01}\partial_u\,s_3 
- \hat{e}^2\,_{01}\partial_v\,s_3 
\]
\[
= - 3\,\hat{\Gamma}_{0111}\,s_2
+ 2\,\hat{\Gamma}_{0101}\,s_3
+ \hat{\Gamma}_{0100}\,s_4
- \frac{\mu}{2\,(1 - \mu\,\zeta)}\,
\left\{3\,s_2\,\zeta_{11} - 2\,s_3\,\zeta_{01}
- s_4\,\zeta_{00}\right\}.
\]

\vspace{.5cm}

These equations, referred to as the {\it $\partial_u$-equations}, will be read as a system of PDE's for the set of
functions
\[
\hat{e}^1\,_{01},\,\,\,\hat{e}^2\,_{01},\,\,\,
\hat{e}^1\,_{11},\,\,\,\hat{e}^2\,_{11},\,\,\,
\hat{\Gamma}_{01AB},\,\,\,\hat{\Gamma}_{11AB},\,\,\,
\zeta,\,\,\,\zeta_{AB},\,\,\,s,\,\,\,s_1,\,\,\,s_2,\,\,\,s_3,\,\,\,s_4,
\]
which comprises all the unknowns with the exception of $s_0$.
The following features of them will be important.

All $\partial_u$-equations are {\it interior
equations on the hypersurfaces $\{w = w_0\}$} in the sense that only
derivatives in the directions of $u$ and $v$ are involved. 

The equations are singular with
terms $u^{-1}$ occuring in various places. It will be seen later that
these terms come with the `right' signs to possess (unique)  solutions 
which are holomorphic in $u$, $v$ and $w$.
Remarkably, the equations for the $s_k$ ensure regular solution to have
the correct tensorial behaviour by the occurrence of terms $u^{-1}$ with
factors $\partial_v\,s_k - (4 - k)\,s_{k+1}$. By Lemma \ref{tensorexptype}
we know that they have to vanish $U_0$.

The system splits into a hierarchy of subsystems, with
\[
t_{01}\,^{EF}\,_{00}\,e^2\,_{EF} = 0, \quad  
R_{000001} = 0, 
\]
being the first subsystem,
\[
t_{01}\,^{EF}\,_{00}\,e^1\,_{EF} = 0, \quad
R_{010001} = 0, \quad
\Sigma_{00} = 0, \quad
\Sigma_{0000} = 0, \quad \Sigma_{0001} = 0, \quad
H_{0000} = 0,
\]
being the second subsystem, and so on. The hierarchy has the following
property. If $s_0$ is given on $\{w = w_0\}$, the first subsystem reduces
to  singular system of ODE's. Given its solution, the second subsystem
also reduces to a system of ODE's (with coefficients which are 
calculated from the functions known so far by operation interior to $\{w
= w_0\}$), and so on. Thus, given $s_0$ and the appropriate  initial data
on
$U_0 \cap \{w = w_0\}$, all unknowns can be determined on $\{w = w_0\}$ by
solving a sequence of systems of ODE's in the independent variable $u$. 

The functions $\hat{e}^a\,_{AB}$ and $\hat{\Gamma}_{ABCD}$ vanish on $U_0$
by our gauge conditions. Therefore only initial data for $\zeta$,
$\zeta_{AB}$,
$s$, and $s_k$ need to be determined on $U_0$ and the function $s_0$ needs
to be provided on $\{w = w_0\}$. Since $s_0$ will be prescribed on $W_0$
as our initial datum, an equation is needed to determine its evolution off
$W_0$. For this purpose we will consider the following equations.

The equations $H_{1(BCD)_k} = 0$ in the order $k = 0, 1, 2, 3$:

\begin{equation}
\label{dws0}
\partial_w\,s_0 
- \frac{1}{2\,u}(\partial_v\,s_1 - 3\,s_2)
+ \hat{e}^1\,_{11}\partial_u\,s_0 
+ \hat{e}^2\,_{11}\partial_v\,s_0 
- \hat{e}^1\,_{01}\partial_u\,s_1 
- \hat{e}^2\,_{01}\partial_v\,s_1 
\end{equation}
\[
= - (\hat{\Gamma}_{0111} - 4\,\hat{\Gamma}_{1101})\,s_0
- (2\,\hat{\Gamma}_{0101} + 4\,\hat{\Gamma}_{1100})\,s_1
+ 3\,\hat{\Gamma}_{0100}\,s_2
+ \frac{2\,\mu}{(1 - \mu\,\zeta)}\,
\frac{1}{4}\,\left\{s_0\,\zeta_{11} + 2\,s_1\,\zeta_{01}
- 3\,s_2\,\zeta_{00}\right\},
\]
\[
\partial_w\,s_1 
- \frac{1}{2\,u}(\partial_v\,s_2 - 2\,s_3)
+ \hat{e}^1\,_{11}\partial_u\,s_1 
+ \hat{e}^2\,_{11}\partial_v\,s_1 
- \hat{e}^1\,_{01}\partial_u\,s_2 
- \hat{e}^2\,_{01}\partial_v\,s_2 
\]
\[
= \hat{\Gamma}_{1111}\,s_0
- (2\,\hat{\Gamma}_{0111} - 2\,\hat{\Gamma}_{1101})\,s_1
- 3\,\hat{\Gamma}_{1100}\,s_2
+ 2\,\hat{\Gamma}_{0100}\,s_3
+ \frac{2\,\mu}{(1 - \mu\,\zeta)}\,
\frac{1}{2}\,\left\{s_1\,\zeta_{11} - s_3\,\zeta_{00}\right\},
\]
\[
\partial_w\,s_2 
- \frac{1}{2\,u}(\partial_v\,s_3 - 2\,s_4)
+ \hat{e}^1\,_{11}\partial_u\,s_2 
+ \hat{e}^2\,_{11}\partial_v\,s_2 
- \hat{e}^1\,_{01}\partial_u\,s_3 
- \hat{e}^2\,_{01}\partial_v\,s_3 
\]
\[
= 2\,\hat{\Gamma}_{1111}\,s_1
- 3\,\hat{\Gamma}_{0111}\,s_2
- (2\,\hat{\Gamma}_{1100} - 2\,\hat{\Gamma}_{0101})\,s_3
+ \hat{\Gamma}_{0100}\,s_4
+ \frac{2\,\mu}{(1 - \mu\,\zeta)}\,
\frac{1}{4}\,\left\{3\,s_2\,\zeta_{11} - 2\,s_3\,\zeta_{01}
- s_4\,\zeta_{00} \right\},
\]

\[
\partial_w\,s_3 
- \frac{1}{2\,u}\partial_v\,s_4
+ \hat{e}^1\,_{11}\partial_u\,s_3 
+ \hat{e}^2\,_{11}\partial_v\,s_3 
- \hat{e}^1\,_{01}\partial_u\,s_4 
- \hat{e}^2\,_{01}\partial_v\,s_4 
\]
\[
= 3\,\hat{\Gamma}_{1111}\,s_2
- (4\,\hat{\Gamma}_{0111} + 2\,\hat{\Gamma}_{1101})\,s_3
- (\hat{\Gamma}_{1100} - 4\,\hat{\Gamma}_{0101})\,s_4
+ \frac{2\,\mu}{(1 - \mu\,\zeta)}\,
\left\{s_3\,\zeta_{11} - s_4\,\zeta_{01}\right\}.
\]

All singular terms cancel in the equations $0 = H_{0(BCD)_{k+1}} +
H_{1(BCD)_k}$, which are given in the order $k = 0, 1, 2$
by 

\begin{equation}
\label{regdws0}
\partial_w\,s_0 - \partial_u\,s_2 
+ \hat{e}^1\,_{11}\partial_u\,s_0
+ \hat{e}^2\,_{11}\partial_v\,s_0 
\end{equation}
\[
=  4\,\hat{\Gamma}_{1101}\,s_0
- 4\,\hat{\Gamma}_{1100}\,s_1
+ \frac{\mu}{(1 - \mu\,\zeta)}\,
\left\{s_0\,\zeta_{11} + 2\,s_1\,\zeta_{01}
- 3\,s_2\,\zeta_{00}\right\},
\]

\[
\partial_w\,s_1 - \partial_u\,s_3 
+ \hat{e}^1\,_{11}\partial_u\,s_1
+ \hat{e}^2\,_{11}\partial_v\,s_1 
\]
\[
= \hat{\Gamma}_{1111}\,s_0
+ 2\,\hat{\Gamma}_{1101}\,s_1
- 3\,\hat{\Gamma}_{1100}\,s_2
- \frac{2\,\mu}{(1 - \mu\,\zeta)}\,
\left\{s_1\,\zeta_{11} - s_3\,\zeta_{00}\right\},
\]

\[
\partial_w\,s_2 
- \partial_u\,s_4 
+ \hat{e}^1\,_{11}\partial_u\,s_2
+ \hat{e}^2\,_{11}\partial_v\,s_2 
\]
\[
= 2\,\hat{\Gamma}_{1111}\,s_1
- 2\,\hat{\Gamma}_{1100}\,s_3
+ \frac{\mu}{(1 - \mu\,\zeta)}\,
\left\{3\,s_2\,\zeta_{11} - 2\,s_3\,\zeta_{01}
- s_4\,\zeta_{00}\right\}.
\]

We can consider (\ref{dws0}) or (\ref{regdws0}) as equation prescribing the
propagation of $s_0$ transverse to the hypersurfaces $\{w = const.\}$.

\vspace{.3cm}

Because $\Gamma_{11CD} = 0$ on $U_0$,
the equations $\Sigma_{11} = 0$, $\Sigma_{11CD} = 0$, $S_{11} = 0$
reduce on $U_0$ to the ODE's
\[
\partial_w\,\zeta = \zeta_{11}, \quad \quad
\partial_w\,\zeta_{CD} = s\,h_{11CD}
- \zeta\,(1 - \mu\,\zeta)\,s_{11CD}, \quad \quad
\partial_w\,s = - (1 - \mu\,\zeta) \,s_{11CD}\,\zeta^{CD}. 
\]
By (\ref{sigmaati}), (\ref{sati}) we must impose  
\[
\zeta = 0,\,\,\,\,\zeta_{AB} = 0,\,\,\,\,s(i) = - 2
\quad\mbox{on}\quad I = \{u = 0, w = 0\}.
\]
This implies with the equations above 
\begin{equation}
\label{zetazeta01011onU0}
\zeta = 0,\,\,\,\,\zeta_{01} = 0,\,\,\,\,\zeta_{11} = 0
\quad\mbox{on}\quad U_0 = \{u = 0\}.
\end{equation}
To determine $\zeta$, $\zeta_{AB}$, and $s$ on $U_0$ it remains to solve
on $U_0$ the equations
\begin{equation}
\label{dwzeta00s}
\partial_w\,\zeta_{00} = s,\quad
\partial_w\,s = - s_4\,\zeta_{00}. 
\end{equation}

The tensorial properties of $\zeta_{AB}$ and $s$ imply with
(\ref{zetazeta01011onU0}) that
\begin{equation}
\label{zeta00shatvdep}
\partial^n_v\,\zeta_{00} = 0,\,\,\,\,\,
\partial^n_v\,\hat{s} = 0
\quad\mbox{on}\quad U_0
\quad\mbox{for}\quad n \ge 1.
\end{equation}
Later it will be important for that these relations can in
fact be deduced from (\ref{zetazeta01011onU0}), (\ref{dwzeta00s}), 
(\ref{Rictensorial}), and the initial
conditions on $I$.

To ensure the tensor relations for the $s_k$ and thus the existence of
regular solutions to the equation for the $s_k$, we determine the
initial data for $s_1, \ldots, s_4$ on $U_0$ by imposing the conditions 
\begin{equation}
\label{Rictensorial}
\partial_v\,s_k = (4 - k)\,s_{k + 1}, \,\,\,\,\,\,k = 0, \ldots, 3,
\quad\mbox{on}\quad U_0.
\end{equation}
They imply recursively the expressions
\[
\partial_v^n\,\partial_w^p\,s_k = \frac{(4 - k)!}{4!}
\partial_v^{k + n}\,\partial_w^p\,s_{0}, \,\,\,\,\,\,k = 0, \ldots 4,
\,\,\,\,\,\,n,\,p
\ge 0
\quad\mbox{at}\quad \{u = 0, v = 0, w = 0\}.
\]

\subsection{Calculating the formal expansion}

Since the equations are overdetermined there are various ways to determine a
formal expansion of the solution. It will follow from Lemma
\ref{solvesallequonShat} that the expansion obtained by the following procedure
will lead to a formal solution of the full system of equations. A solution
obtained by any other procedure with this property will thus have to coincide with
the present one. 
It will be convenient to replace $s$ by the unknown
\[
\hat{s} = 2 + s.
\]
For certain discussions it is useful to write 
\[
s_k = s^*_k + \hat{s}_k
\quad\mbox{with}\quad
\partial_us^*_k = 0
\quad\mbox{and}\quad
s^*_k|_{u = 0} = s_k|_{u = 0}
\quad\mbox{so that}\quad
\hat{s}_k = O(u)
\quad\mbox{as}\quad u \rightarrow 0.
\]
By (\ref{Rictensorial}) we can then assume that 
\[
\partial_v\,s^*_k = (4 - k)\,s^*_{k + 1},
\]
and the $\partial_u$-equations for the $\hat{s}_k$ can be written in the form
\[
0 = - H_{0\,(BCD)_k}
=
\partial_u\,\hat{s}_{k + 1} 
+ \frac{4 - k}{2\,u}\,\hat{s}_{k + 1} 
- \frac{1}{2\,u}\,\partial_v\,\hat{s}_k 
+ \hat{e}^a\,_{01}\,\partial_a(s^*_k + \hat{s}_k)
\]
\[
+ terms\,\,of \,\,zeroth\,\,order,
\]
so that the coefficient $(4 - k)/2$ of the singular term
$u^{-1}\,\hat{s}_{k + 1}$ is positive and the term 
$u^{-1}\,\partial_v\,\hat{s}_k$, which involves the unknown $\hat{s}_k$ 
determined in an earlier step of the integration procedure, creates no problem
because $\hat{s}_k = 0$ on $U_0$. Writing
\[
x = (\hat{e}^a\,_{AB},\, \hat{\Gamma}_{ABCD},\, \zeta,\,
\zeta_{AB}, \,\hat{s},\, s_1,\,s_2,\,s_3,\,s_4),
\] 
so that the full set of unknowns are given by $x$ and $s_0$, we proceed as
follows.

On $W_0$ we prescribe $s_0$ as given in (\ref{nulldataonU0}) with the null
data ${\cal D}^*_n$ satisfying the reality conditions and the estimates
(\ref{symmest}). By (\ref{Rictensorial}) all components of $x$ can be
determined on $I$.

We successivly integrate the subsystems in the hierachy of
$\partial_u$-equations to determine all components of $x$ on $W_0$. These will be
holomorphic in $u$ and $v$ and unique, because the relevant operators in the
singular equations are of the form  $\partial_uf + c\,u^{-1}\,f$ with non-negative
constants $c$ (a proof of this statement follows from the derivation of the estimates
discussed below). 

The equation $H_{0100} + H_{1000} = 0$ is used to determine 
$\partial_w s_0$ from the fields $x$ and $s_0$ on $W_0$ as a holomorphic function of $u$
and $v$.  

Applying the operator $\partial_w$ formally to the $\partial_u$-equations,
one obtains equations for $\partial_wx$ on $W_0$ which can be solved with the
initial data on $\{w = 0, u = 0\}$ which are obtained by using (\ref{dwzeta00s}) and
by applying $\partial_w$ to (\ref{Rictensorial}).
Applying $\partial_w$ to the equation $H_{0100} + H_{1000} = 0$, one obtains
$\partial^{2}s_0$ on $W_0$. 

Repeating these steps by applying successively the operator $\partial^p_w$,
$p = 2, 3, \ldots$, one gets an sequence of functions $\partial^p_wx$,
$\partial^p_ws_0$ on $W_0$, which are holomorphic in $u$ and $v$. 

Expanding the functions so obtained at $u = 0$, $v = 0$ we get the following
result.

\begin{lemma}
\label{exunformexp}
The procedure described above determines at the point 
$O = (u = 0, v = 0, w = 0)$ from the data $s_0$, given  on $W_0$ according
to (\ref{nulldataonU0}), a unique sequence of expansion coefficients 
\[
\partial^m_u\,\partial^n_v\,\partial^p_w\,f(O),
\quad \quad m, n, p = 0, 1, 2, \ldots ,
\]
where $f$ stands for any of the functions
\[
\hat{e}^a\,_{AB},\, \hat{\Gamma}_{ABCD},\, \zeta,\,
\zeta_{AB}, \,\hat{s},\, s_j.
\]
If the corresponding Taylor series are absolutly convergent in some
neighbourhood
$P$ of
$O$, they define a solution to the $\partial_u$-equations and to the equation
$H_{1000} = 0$ on $P$ which satisfies on $P \cap U_0$
equations (\ref{Rictensorial}) and 
$\Sigma_{11} = 0$, $\Sigma_{11CD} = 0$, $S_{11} = 0$.
\end{lemma}

By Lemma \ref{tensorexptype} all spinor-valued functions should have a
specific $v$-finite expansion type. The following result will be important for
our convergence proof. 

\begin{lemma}
\label{formexptype}
If the data $s_0$ are given on $W_0$ as in (\ref{nulldataonU0}), the formal
expansions of the fields obtained in Lemma \ref{exunformexp} correspond to ones
of functions of $v$-finite expansion types given by
\[
k_{s_j} = 4 - j,\,\,\,\,\,
k_{\zeta_i} = 2 - i,\,\,\,\,\,
k_{\zeta} = 0,\,\,\,\,\,
k_s = k_{\hat{s}} \le 2,
\]
\[
k_{\hat{e}^1_{AB}} = - A - B,\,\,\,\,\,
k_{\hat{e}^2_{AB}} = 3 - A - B
\quad \mbox{for} \quad AB = 01, \,10
\,\,\,\mbox{or}\,\,\, 11.
\]
\[
k_{\hat{\Gamma}_{01AB}} = 2 - A - B,\,\,\,\,
k_{\hat{\Gamma}_{11AB}} = 1 - A - B
\quad \mbox{for} \quad A, B = 0
\,\,\,\mbox{or}\,\,\,1.
\]
\end{lemma}

\begin{remark}
The scalar functions $s$, $\hat{s}$ should have expansion type
$k_s = k_{\hat{s}} = 0$. As  pointed out below, this does not follow with
the simple arguments used here. Since it will not be important
for the following discussions, we shall make no effort to retrieve this
information from the equations. 
\end{remark}

\begin{proof} We note the following properties of $v$-finite expansion types:

For given integer $k$ the functions of expansion type $k$ form a complex
vector space which comprises the functions of expansion type $\le k$.

If the functions $f$ and $g$ have expansion type $k_f$ and $k_g$
respectively, their product $f\,g$ has expansion type 
$k_{f g} = k_f + k_g$. 

If $f$ has expansion type $k_f$, the function $\partial_u f$ has
expansion type $k_f + 2$. Conversely, if $\partial_u f$ has expansion
type $k_f + 2$ and if the function independent of
$u$ which agrees on $U_0$ with $f$ has expansion type $k_f$ (for instance
if $f|_{u = 0} = 0$), then $f$ has expansion type $k_f$.

If $f$ has expansion type $k_f$ and $f|_{u = 0} = 0$ then
$\frac{1}{u}\,f$ has expansion type $k_f + 2$.

If $f$ has expansion type $k_f$, the function $\partial_v f$ has
expansion type $k_f - 1$.

If $f$ has an expansion type, the function $\partial_w f$ has
the same expansion type.

Applying these rules one can check that the expansion types 
listed above are consistent with the $\partial_u$-equations, the
equation $H_{1000} + H_{0100} = 0$ and the equations $S_{11} = 0$, 
$\Sigma_{1100} = 0$ used on $U_0$ in the sense that all terms in the
equations have the same expansion types.

Assuming the given expansion types for the $s_k$, the  
$\partial_u$-equations for the $\hat{\Gamma}_{ABCD}$ imply at lowest
order in $u$ that in general the  $k_{\hat{\Gamma}_{ABCD}}$ must take the
values given above. It follows then from the  
$\partial_u$-equations for the $\hat{e}^a_{AB}$ at lowest
order in $u$ that the  $k_{\hat{e}^a_{AB}}$ must take in general the values
above. The remaining $\partial_u$-equations then imply at lowest order the
other expansion types. 

With these observations the Lemma follows from our procedure by a
straightforward though lengthy induction argument. We do not write out the
details. \end{proof}

The equation
\[
0 = S_{00} = \partial_u\,s + (1 - \mu\,\zeta) \,s_{00CD}\,\zeta^{CD} ,
\]
should imply more precisely $k_s = 0$, because the expansion type of the tensorial
component
$s_{00CD}\,\zeta^{CD}$ should be $2$.  The contraction of the spinor fields 
on the right hand side implies cancellations which lower the
expansion types of the contracted quantities on the right hand side. These
cancellations cannot be controlled in the explicit expression 
\[
0 = \partial_us + (1 - \mu\,\zeta)\,
(s_{0}\,\zeta_{11} - 2\,s_{1}\,\zeta_{01} + s_{2}\,\zeta_{00}),
\]
by the simple rules given above, they only suggest an expansion type $k_s \le
2$. Fortunately, this does not prevent us from determining the other expansion
types. In the equation 
\[
0 = \Sigma_{0011} = \partial_u\,\zeta_{11} - s
+ \zeta\,(1 - \mu\,\zeta)\,s_{0011},
\]
$s$ is added to a field of expansion type $2$ and the equation
\[
0 = S_{11} = \partial_w\,s + s_{11CD}\,\zeta^{CD}
= \partial_w\,s + s_{1111}\,\zeta_{00}
\quad\mbox{on}\quad U_0,
\]
is consistent with $k_s \le 2$. No further equation involving $s$ is needed
in the convergence proof.

\subsection{The complete set of equations on $\hat{S}$}

Because only a certain subset of the system of equations has been
used to determine the formal expansions of the fields, it remains to be shown that the latter
define in fact a formal solution to the complete system of conformal static 
vacuum field equations. To simplify stating the following result it will be assumed in
this subsection that the formal expansions for
\[
\hat{e}^a\,_{AB},\, \hat{\Gamma}_{ABCD},\, \zeta,\,
\zeta_{AB}, \,\hat{s},\,s_j,
\] 
determined in Lemma \ref{exunformexp} define in fact absolutely convergent series on an open neighbourhood of the
point $O$, which we assume to coincide with $\hat{S}$. There will arise no
problem from this assumption because the following two Lemmas will not be 
used in the derivation of the estimates in the next section. 

\begin{lemma}
\label{equonU0}
With the assumptions above the corresponding functions 
\[
e^a\,_{AB},\, \Gamma_{ABCD},\, \zeta,\,
\zeta_{AB}, \,s,\,s_j,
\] 
satisfy the complete set of the conformal vacuum field equations on the set
$\,U_0$ in the sense that the fields 
\[
t_{AB}\,^{EF}\,_{CD},\,\,\,\,
R_{ABCDEF},\,\,\,\,
\Sigma_{AB},\,\,\,\, 
\Sigma_{ABCD},\,\,\,\, 
S_{AB},\,\,\,\,
H_{ABCD},
\]
calculated from these functions on $\hat{S} \setminus U_0$ have vanishing
limit as $u \rightarrow 0$.
\end{lemma}

\begin{proof}
Because of the equations solved already and the symmetries involved, we only
need to examine  the behaviour of the fields 
\[
t_{11}\,^{EF}\,_{01},\,\,\,\,
R_{AB0111},\,\,\,\,
\Sigma_{01},\,\,\,\, 
\Sigma_{01CD},\,\,\,\,
S_{01},\,\,\,\,
H_{1(BCD)_k},\,\,\,\,k = 1, 2, 3,
\]
near $U_0$.

With (\ref{frameregsing}), (\ref{framesing}), (\ref{connregsing}),
(\ref{connsing}) the $\partial_u$-equations imply for the frame and the
dual frame coefficients the slightly stronger results
(\ref{sigmaasbeh}), (\ref{easbeh}). A direct calculation gives then
\[
t_{01}\,^{EF}\,_{11}
= 2\,\Gamma_{01}\,^{(E}\,_{1}\,\epsilon_{1}\,^{F)}
- \Gamma_{11}\,^{(E}\,_{0}\,\epsilon_{1}\,^{F)}
- \Gamma_{11}\,^{(E}\,_{1}\,\epsilon_{0}\,^{F)}
- \sigma^{EF}\,_a\,
(e^a\,_{11, c}\,e^c\,_{01} - e^a\,_{01, c}\,e^c\,_{11}) = O(u),
\]
as $u \rightarrow 0$. 

With the particular result
\[
t_{01}\,^{01}\,_{11}
= \Gamma_{0111}
- \frac{1}{2}\,e^2\,_{11, 2}
 - \frac{1}{2\,u}\,e^1\,_{11} + O(u^2) = O(u),
\]
follows
\[
R_{000111} = 
\Gamma_{1100,1}\,e^1\,_{01} 
+ \Gamma_{1100,2}\,e^2\,_{01} 
- \Gamma_{0100,1}\,e^1\,_{11}
- \Gamma_{0100,2}\,e^2\,_{11}
- \Gamma_{0100,3}
\]
\[
- \Gamma_{1100}\,\Gamma_{1100}
+ 2\,\Gamma_{0100}\,(\Gamma_{1101} - \Gamma_{0111})
- t_{01}\,^{01}\,_{11}\,\Gamma_{0100}
- t_{01}\,^{11}\,_{11}\,\Gamma_{1100}
- \frac{1}{2}\,s_{0011}
\]
\[
= 
\frac{1}{2\,u}\,(\Gamma_{1100,2}
- 2\,\Gamma_{1101} + 3\,\Gamma_{0111}
- \frac{1}{2}\,e^2\,_{11, 2}
- \frac{3}{2\,u}\,e^1\,_{11})
- \frac{1}{2}\,s_{0011} + O(u) \rightarrow
\]
\[
\frac{1}{2}\left(\partial_v\,\partial_u\,\Gamma_{1100}
- 2\,\partial_u\,\Gamma_{1101} + 3\,\partial_u\,\Gamma_{0111}
- \frac{1}{2}\,\partial_v\,\partial_u\,e^2\,_{11}
- \frac{3}{4}\,\partial^2_u\,e^1\,_{11}
- s_{0011}\right) = 0
\,\,\, \mbox{as} \,\,\, u \rightarrow 0,
\]
where the $\partial_u$-equations and the relation
$\partial_v\,s_1 = 3\,s_{2}$ on $U_0$ are used to calculate the limit.
Similarly,
\[
R_{010111}
= 
\frac{1}{2\,u}\,\Gamma_{1101,2}
- \frac{1}{2\,u}\,\Gamma_{1111}
- \frac{1}{2}\,s_{0111} + O(u)
\]
\[
\rightarrow \frac{1}{2}\left( 
\partial_v\,\partial_u\,\Gamma_{1101}
- \partial_u\,\Gamma_{1111}
- \,s_{0111} \right) = 0
\quad \mbox{as} \quad u \rightarrow 0,
\]
where the $\partial_u$-equations and the relation
$\partial_v\,s_2 = 2\,s_3$ on $U_0$ are used,
\[
R_{110111} = 
\frac{1}{2\,u}\,\Gamma_{1111,2} 
- \frac{1}{2}\,s_{1111} + O(u)
\rightarrow 
\frac{1}{2}\left(\partial_v\,\partial_u\,\Gamma_{1111} 
- s_{1111}\right) = 0
\quad \mbox{as} \quad u \rightarrow 0,
\]
where the $\partial_u$-equations and the relation
$\partial_v\,s_3 = s_4$ on $U_0$ are used.

By (\ref{zetazeta01011onU0}) and the remark following (\ref{zeta00shatvdep})
we know that $\zeta = 0$, $\zeta_{01} = 0$,
$\zeta_{11} = 0$, $\partial_v\,\zeta_{00} = 0$, $\partial_v\,s = 0$ on 
$U_0$. The $\partial_u$-equations and (\ref{Rictensorial})
imply
\[
\Sigma_{01} 
= \frac{1}{2\,u}\partial_v\,\zeta - \zeta_{01} + O(u)
\rightarrow \frac{1}{2}\partial_v\,\zeta_{00} - \zeta_{01} = 0, 
\]
\[
\Sigma_{0100}
= \frac{1}{2\,u}\,(\partial_v\,\zeta_{00}
- 2\,\zeta_{01}) + O(u)
\rightarrow
\frac{1}{2}\,(\partial_v\,\partial_u\,\zeta_{00} -
2\,\partial_u\,\zeta_{01}) = 0,
\]
\[
\Sigma_{0101} =
\frac{1}{2\,u}\,(\partial_v\,\zeta_{01} - \zeta_{11})
+ \frac{1}{2}\,s + O(u)
\rightarrow 
\frac{1}{2}\,(\partial_v\,\partial_u\,\zeta_{01} - \partial_u\,\zeta_{11}
+ s) = 0, 
\]
\[
\Sigma_{0111} 
= \frac{1}{2\,u}\,\partial_v\,\zeta_{11} + O(u) 
\rightarrow \frac{1}{2}\,\partial_v\,\partial_u\,\zeta_{11} = 0.
\]
\[
S_{01}
= \frac{1}{2\,u}\,\partial_v\,\,s  
+ s_{0111}\,\zeta_{00} + O(u)
\rightarrow \frac{1}{2}\,(\partial_v\,\partial_u\,s 
+ 2\,s_{0111}\,\zeta_{00}) =
\frac{1}{2}\,\partial_v\,(\partial_u\,s 
+ s_{0011}\,\zeta_{00}) = 0,
\]
 as $u \rightarrow 0$.

With our assumptions (and formally setting $s_5 = 0$) we get for $k = 0,
\ldots, 3$ 
 \[
\gamma_k \equiv \lim_{u \to 0}\,(-2\,H_{0(ABC)_k}) =
(6 - k)\,\partial_u\,s_{k + 1} - \partial_v\,\partial_u\,s_k
- (4 - k)\,\mu\,s_{k + 1}\,\zeta_{00},
\] 
\[
\beta_k \equiv \lim_{u \to 0}\,(-2\,H_{1(ABC)_k}) =
2\,\partial_w\,s_{k} - \partial_v\,\partial_u\,s_{k + 1}
+ (3 - k)\,\,\partial_u\,s_{k + 2}
- (3 - k)\,\mu\,s_{k + 2}\,\zeta_{00}.
\] 

The expected tensorial nature of $s_{ABCD}$ and $H_{ABCD}$ (cf. Lemma
\ref{tensorexptype}) would imply 
\[
4\,\beta_1 = \partial_v\,\beta_0 - \partial_v\,\gamma_1 
+ 2\,\gamma_2, \quad \quad
12\,\beta_2 = \partial_v^2\,\beta_0 - \partial_v^2\,\gamma_1 
- 2\,\partial_v\,\gamma_2 + 4\,\gamma_3,
\]
\[
24\,\beta_3 = \partial_v^3\,\beta_0 - \partial_v^3\,\gamma_1 
- 2\,\partial_v^2\,\gamma_2 - 8\,\partial_v\,\gamma_3
\quad \mbox{on} \quad U_0.
\]
It turns out that these relations can in fact be verified by a direct
calculation with the expressions for $\gamma_k$, $\beta_k$ obtained above.
Because the equations used to establish Lemma \ref{exunformexp} imply 
$\gamma_k = 0$, $\beta_0 = 0$, it follows that $\beta_1 = \beta_2 = \beta_3 =
0$ so that in fact $H_{ABCD} \rightarrow 0$ as $u \rightarrow 0$.
\end{proof}

\vspace{.3cm}
We can now prove the desired result.
\begin{lemma}
\label{solvesallequonShat}
The functions 
\[
e^a\,_{AB},\, \Gamma_{ABCD},\, \zeta,\,
\zeta_{AB}, \,s,\,s_j,
\] 
corresponding to the expansions determined in Lemma \ref{exunformexp}
satisfy the complete set of conformal vacuum field equations on the set
$\hat{S}$.
\end{lemma}

\begin{proof}
It needs to be shown that the {\it zero quantities}
\[
t_{01}\,^{EF}\,_{11},\,\,\,\,
R_{AB0111},\,\,\,\,
\Sigma_{01},\,\,\,\,
\Sigma_{11},\,\,\,\,
\Sigma_{01CD},\,\,\,\,
\Sigma_{11CD},\,\,\,\,
S_{01},\,\,\,\,
S_{11},\,\,\,\,
H_{1ABCD},
\]
vanish on $\hat{S}$. For this purpose we shall derive a system of {\it
subsidiary equations} for these fields.

Given the fields
\[
e^a\,_{AB},\, \Gamma_{ABCD},\, \zeta,\,\zeta_{AB}, \,s,\, s_{ABCD},
\] 
we have the 1-forms $\sigma^{AB}$ dual to 
$e_{AB}$ and the connection form
$\omega^A\,_B = \Gamma_{CD}\,^A\,_B\,\sigma^{CD}$.

To derive the subsidiary system we consider the torsion form 
\[
\Theta^{AB} =
\frac{1}{2}\,t_{CD}\,^{AB}\,_{EF}\,\sigma^{CD}\wedge\sigma^{EF},
\] 
and the form
\[
\Omega^{*A}\,_B \equiv \Omega^A\,_B - \hat{\Omega}^A\,_B  = 
\frac{1}{2}\,R^A\,_{BCDEF}\,\sigma^{CD} \wedge \sigma^{EF},
\]
obtained as difference of the curvature form 
\[
\Omega^A\,_B  = \frac{1}{2}\,r^A\,_{BCDEF}\,\sigma^{CD}
\wedge \sigma^{EF},
\]
and the form 
\[
\hat{\Omega}^A\,_B  = \frac{1}{2}\,s^A\,_{BCE}\,\sigma^C\,_F
\wedge \sigma^{EF}.
\]
The following general relations will be used:
The identity $\sigma^a \wedge \sigma^b \wedge \sigma^c
= \epsilon^{abc}\,\nu$ with
$\nu = \frac{1}{3!}\,\epsilon_{def}\,
\sigma^d \wedge \sigma^e \wedge \sigma^f$,
which holds in $3$-dimensional spaces. In space spinor form it takes the form
\[
\sigma^{AB} \wedge \sigma^{CD} \wedge \sigma^{EF}
= \epsilon^{AB\,CD\,EF}\,\nu
\,\,\,\,\,\mbox{with}\,\,\,\,\,
\epsilon^{AB\,CD\,EF} = \frac{i}{\sqrt{2}}
\,(\epsilon^{AC}\,\epsilon^{BF}\,\epsilon^{DE}
- \epsilon^{AE}\,\epsilon^{BD}\,\epsilon^{FC}),
\]
which implies 
\[
\sigma^{AB} \wedge \sigma^C\,_D \wedge \sigma^{ED} =
- i\,\sqrt{2}\,\epsilon^{A(C}\,\epsilon^{E)B}\,\nu
= i\,\sqrt{2}\,h^{AB\,CE}\,\nu,
\]
and thus
\[
\hat{\Omega}^A\,_B \wedge \sigma^{BD} =
\frac{1}{2}\,s^A\,_{BCE}\,
\sigma^{BD} \wedge \sigma^C\,_F \wedge \sigma^{EF} = 0.
\]
The equations 
\[
i_H\,(\alpha \wedge \beta) =
i_H\,\alpha \wedge \beta + (-1)^k \alpha \wedge i_H\,\beta,
\,\,\,\,\,\,\,\,\,\,\,\,
{\cal L}_H\,\alpha = (d \circ i_H + i_H \circ d)\,\alpha,
\]
which holds for arbitrary vector field $H$,  $k$-form $\alpha$, and $j$-form
$\beta$. 
Finally, we note that in the presence of torsion the Ricci identity for a
spinor field $\iota_{E \ldots H}$ of degree $m$ reads
\[
(D_{AB}\,D_{CD} - D_{CD}\,D_{AB})\,\iota_{E F \ldots H}
= - \,\,\iota_{L F \ldots H}\,\,r^L\,_{E\,AB\,CD} 
- \,\,\iota_{E L \ldots H}\,\,r^L\,_{F\,AB\,CD}
- \dots 
\]
\[
- \,\,\iota_{E F \ldots L}\,\,r^L\,_{H\,AB\,CD} 
-  t_{AB}\,^{KL}\,_{CD}\,D_{KL}\,\iota_{E F\ldots H}. 
\]

\vspace{.3cm}

We shall derive now the subsidiary equations.
The fields $\Theta^{AB}$ and $\Omega^A\,_B$
satisfy the first structural equation
\[
d\,\sigma^{AB} = - \omega^A\,_C \wedge \sigma^{CB}
- \omega^B\,_C \wedge \sigma^{AC} + \Theta^{AB},
\]
and the second structural equation
\[
d\,\omega^A\,_B = - \omega^A\,_C \wedge \omega^C\,_B + \Omega^A\,_B,
\]
respectively. These equations imply
\[
d\,\Theta^{AB} 
= 2\,\Omega^{(A}\,_C \wedge \sigma^{B)C}
- 2\,\omega^{(A}\,_C \wedge \Theta^{B)C}
= 2\,\Omega^{*(A}\,_C \wedge \sigma^{B)C}
- 2\,\omega^{(A}\,_C \wedge \Theta^{B)C}.
\]
We set $H = e_{00}$ and observe
that the gauge conditions and the $\partial_u$-equations imply
\[
i_H\,\sigma^{AB} = \epsilon_0\,^A\,\epsilon_0\,^B = h_{00}\,^{AB},
\,\,\,\,\,\,\,\,\,\,\,\,  
i_H\,\omega^A\,_B = 0,
\,\,\,\,\,\,\,\,\,\,\,\,
i_H\,\Theta^{AB} = 0,
\,\,\,\,\,\,\,\,\,\,\,\,  
i_H\,\Omega^{*A}\,_B = 0.
\]
It follows that
\[
{\cal L}_H\,\Theta^{AB} = (d \circ i_H + i_H \circ d)\,\Theta^{AB}
= 2\,\Omega^{*(A}\,_0 \,\epsilon_0\,^{B)},
\]
and thus 
\[
{\cal L}_H<\Theta^{AB}, e_{01} \wedge e_{11}>\, =
\]
\[
2<\Omega^{*(A}\,_0, e_{01} \wedge e_{11}>\epsilon_0\,^{B)}
+ <\Theta^{AB}, [H, e_{01}] \wedge e_{11}>
+ <\Theta^{AB}, e_{01} \wedge [H, e_{11}]>.
\]

The first structural equation, the gauge conditions, and the
$\partial_u$-equations imply
\[
0 = \,<\Theta^{EF}, H \wedge e_{CD}>e_{EF} =
- \Gamma_{CD}\,^{EF}\,_{00}\,e_{EF} - [H, e_{CD}],
\] 
whence 
\[
[H, e_{CD}] = - 2\,\Gamma_{CD01}\,e_{00}
+ 2\,\Gamma_{CD00}\,e_{01}.
\]
This implies 
\[
{\cal L}_H<\Theta^{AB}, e_{01} \wedge e_{11}>\, =
2\,\Gamma_{0100}<\Theta^{AB}, e_{01} \wedge e_{11}>
+ \,2<\Omega^{*(A}\,_0, e_{01} \wedge e_{11}>\epsilon_0\,^{B)},
\]
i.e.
\begin{equation}
\label{Tsubs}
(\partial_u + \frac{1}{u})\,t_{01}\,^{AB}\,_{11} =
2\,\hat{\Gamma}_{0100}\,\,t_{01}\,^{AB}\,_{11}
+ 2\,R^{(A}\,_{0\,01\,11}\,\epsilon_0\,^{B)}.
\end{equation}

\vspace{.5cm}

With the first structural equation we obtain
\[
d\,\hat{\Omega}_{AB} - \omega^H\,_A \wedge \hat{\Omega}_{HB}
- \omega^H\,_B \wedge \hat{\Omega}_{AH}
= \frac{1}{2}\,
D_{GH}\,s_{ABCD}\,\sigma^{GH} \wedge \sigma^C\,_F \wedge \sigma^{DF}
= \frac{i}{\sqrt{2}}\,H^E\,_{ABE}\,\,\nu,
\]
and from the second structural equation we get
\[
d\,\Omega_{AB} - \omega^H\,_A \wedge \Omega_{HB}
- \omega^H\,_B \wedge \Omega_{AH} = 0,  
\]
which give together 
\[
d\,\Omega^*_{AB} - \omega^H\,_A \wedge \Omega^*_{HB}
- \omega^H\,_B \wedge \Omega^*_{AH}
= -  \frac{i}{\sqrt{2}}\,H^E\,_{ABE}\,\,\nu,
\]
and thus, with the equations above,
\begin{equation}
\label{Rsubs}
(\partial_u + \frac{1}{u})\,R_{AB\,01\,11} =
2\,\hat{\Gamma}_{0100}\,R_{AB\,01\,11}
+ \frac{1}{2}\,H_{1AB0}.
\end{equation}

\vspace{.3cm}

The identity
\[
D_{AB}\,\Sigma_{CD} - D_{CD}\,\Sigma_{AB} = 
t_{AB}\,^{EF}\,_{CD}\,D_{EF}\zeta 
+ \Sigma_{CDAB} - \Sigma_{ABCD},
\]
gives with the gauge conditions and the $\partial_u$-equations
\begin{equation}
\label{sigmaABsubs}
\partial_u\,\Sigma_{CD}
+ \frac{2}{u}\,\epsilon_{(C}\,^0\,\epsilon_{D)}\,^1\,\Sigma_{01} 
= 2\,\hat{\Gamma}_{CD00}\,\Sigma_{01} + \Sigma_{CD00}.
\end{equation}
The identity
\[
D_{AB}\,\Sigma_{CDEF} - D_{CD}\,\Sigma_{ABEF} = 
- 2\,\zeta_{K(E}\,R^K\,_{F)ABCD}
+ t_{AB}\,^{GH}\,_{CD}\,D_{GH}\zeta_{EF} 
\]
\[
+ S_{CD}\,h_{ABEF} - S_{AB}\,h_{CDEF}
\]
\[
+ (1 - 2\,\mu\,\zeta)\,
(\Sigma_{AB}\,s_{CDEF} - \Sigma_{CD}\,s_{ABEF})
+ \zeta\,(1 - \mu\,\zeta)\,
(\epsilon_{CA}\,H_{BDEF} + \epsilon_{DB}\,H_{CAEF}),
\]
implies with the gauge conditions and the $\partial_u$-equations
\begin{equation}
\label{sigmaABCDsubs}
\partial_u\,\Sigma_{CDEF} 
+ \frac{2}{u}\,\epsilon_{(C}\,^0\,\epsilon_{D)}\,^1\,\Sigma_{01EF}
= 
\end{equation}
\[
2\,\hat{\Gamma}_{CD00}\,\Sigma_{01EF}
+ S_{CD}\,h_{00EF} 
- (1 - 2\,\mu\,\zeta)\,\Sigma_{CD}\,s_{00EF}
+ \zeta\,(1 - \mu\,\zeta)\,\epsilon_{D0}\,H_{C0EF}.
\]

The identity
\[
D_{AB}\,S_{CD} - D_{CD}\,S_{AB} = 
t_{AB}\,^{EF}\,_{CD}\,D_{EF}s 
- \mu\,\{\Sigma_{AB}\,s_{CDEF} - \Sigma_{CD}\,s_{ABEF}\}\,\zeta^{EF}
\]
\[
(1 - \mu\,\zeta)\,\left\{ \Sigma_{AB}\,^{EF}\,s_{CDEF} -
\Sigma_{CD}\,^{EF}\,\,s_{ABEF}
+ (\epsilon_{CA}\,H_{BDEF} 
+ \epsilon_{DB}\,H_{CAEF})\,\zeta^{EF}
\right\},
\]
implies with the gauge conditions and the $\partial_u$-equations
\begin{equation}
\label{SABsubs}
\partial_u\,S_{CD} 
+ \frac{2}{u}\,\epsilon_{(C}\,^0\,\epsilon_{D)}\,^1\,S_{01} 
=
\end{equation}
\[
2\,\hat{\Gamma}_{CD00}\,S_{01} 
+ \mu\,\,\Sigma_{CD}\,s_{00EF}\,\zeta^{EF}
- (1 - \mu\,\zeta)\,\left\{
\Sigma_{CD}\,^{EF}\,\,s_{00EF}  
- \epsilon_{D0}\,H_{C0EF}\,\zeta^{EF}
\right\}.
\]

Finally we have the identity
\begin{equation}
\label{AddHident}
2\,D^{EF}\,H_{EFAB} = - 4\,s_{K(BGH}\,R^K\,_{A)}\,^{EG}\,_E\,^H
+ t^E\,_F\,^{KL}\,_{EH}\,D_{KL}\,s_{AB}\,^{FH}
\end{equation}
\[
- \frac{4\,\mu}{1 - \mu\,\zeta}\,s_{H(ABF}\,\Sigma^{EF}\,_{G)}\,^H
- \frac{2\,\mu^2}{(1 - 2\,\mu)^2}\,\Sigma^{EF}\,s_{H(ABF}\,\zeta_{E)}\,^H
\]
\[
+ \frac{\mu}{1 - \mu\,\zeta}\,
\left\{
2\,H_{EHAB}\,\zeta^{EH} - 2\,H^E\,_{EH(A}\,\zeta_{B)}\,^H
\right\},
\]
where the right hand side is a linear function of the zero
quantities. The gauge conditions and the equations $H_{0ABC} = 0$, 
$H_{1000} = 0$ imply for the left hand side
\begin{equation}
\label{BddHident}
D^{EF}\,H_{EFAB} 
= \partial_u\,H_{11AB} 
+ \frac{1}{u}\left\{H_{11AB} + H_{110(A}\,\epsilon_{B)}\,^0
\right\}
- (\frac{1}{2\,u}\,\partial_v + \hat{e}^a\,_{01}\partial_{z^a})
\,H_{10AB}
\end{equation}
\[
- 2\,\hat{\Gamma}_{0100}\,H_{11AB}
- \hat{\Gamma}_{010A}\,H_{110B}
- \hat{\Gamma}_{010B}\,H_{110A}
+ \hat{\Gamma}_{011A}\,H_{100B}
+ \hat{\Gamma}_{011B}\,H_{100A}
+ \hat{\Gamma}_{1100}\,H_{10AB}.
\]  

Equations (\ref{Tsubs}), (\ref{Rsubs}), (\ref{sigmaABsubs}),
(\ref{sigmaABCDsubs}), (\ref{SABsubs}), and 
equation (\ref{AddHident}) with (\ref{BddHident}) observed on the left hand side
provide the system of subsidiary equations.
Note that the right hand side of this system is a linear function of
the zero quantities.
It implies with Lemma \ref{equonU0} that all zero quantities vanish
on $\hat{S}$.
\end{proof}

\vspace{.3cm}

If the series considered in Lemma \ref{exunformexp} are absolutely convergent
it thus follows from Lemma \ref{solvesallequonShat} that they define in fact a
solution to the complete set of static conformal vacuum field equations on 
$\hat{S}$.

\section{Convergence of the formal expansion}
\label{converge}

Let there be given a given sequence 
\[
\hat{{\cal D}}_{n} =
\{\psi_{A_2B_2A_1B_1},\,\,\psi_{A_3B_3A_2B_2A_1B_1},\,\,
\psi_{A_4B_4A_3B_3A_2B_2A_1B_1},\,\,\ldots\},
\]
of totally symmetric spinors as in Lemma \ref{realformalexpansion} and 
set in the expansion (\ref{nulldataonU0}) of $s_0(u, v)$ 
\[
D^*_{(A_1 B_1}\,\ldots\,D^*_{A_m B_m}\,s^*_{A B C D)}(i)
= \psi_{A_1B_1 \ldots A_mB_m A B C D},
\quad m \ge 0.
\]
Observing the estimates (\ref{symmest}), one finds
as a necessary condition for the function $s_0$ on $W_0$ to determine an
analytic solution to the conformal static vacuum field equations that its
non-vanishing Taylor coefficients at the point $O$ satisfy estimates of the
form
\begin{equation}
\label{basest}
|\partial^m_u\,\partial^n_v\,s_0(O)|  
= m!\,n!\,|\psi_{m, n}| 
\le {2m + 4 \choose n}\,
m!\,n!\,M\,r_1^{-m}, \quad
m \ge 0, \quad 0 \le n \le 2\,m + 4.
\end{equation}
A slightly different type of estimate will be more convenient for us.

\begin{lemma}
\label{convest}
For given $ \rho_0 \in  \mathbb{R}$, $0 < \rho_0 \le e^2$, there exist 
positive constants $r_0$, $c_0$ so that (\ref{basest})
implies estimates of the form 
\begin{equation}
\label{formconvest}
|\partial^m_u\,\partial^n_v\,s_0(O)| \le c_0
\,\frac{m!\,n!\,r_0^m\,\rho_0^n}{(1 + m)^2\,(1 + n)^2},
\quad
m \ge 0, \quad 0 \le n \le 2\,m + 4.
\end{equation}
\end{lemma}

\begin{proof}
With $r_0 = 4\,e^6\,r_1^{-1}\,\rho_0^{-2}$ and 
$c_0 = 16\,M\,e^8\,\rho_0^{-4}$, the estimate 
$1 \le {2\,m + 4 \choose n} \le 2^{2 m + 4}$, which follows from the
binomial law 
$(1 + x)^{2 m + 4} = \sum_{n = 0}^{2 m + 4} {2\,m + 4 \choose n}\,x^n$,
and the estimate $e^x \ge 1 + x$, which holds for $x \ge 0$, we get
\[
{2m + 4 \choose n}\,m!\,n!\,M\,r_1^{-m} \le
16\,M\,m!\,n!\,(4\,r_1^{-1})^{m} 
= c_0\,m!\,n!\,\frac{r_0^m}{(e^m)^2}\,\frac{\rho_0^n}{(e^n)^2}
\,\left(\frac{\rho_0}{e^2}\right)^{2\,m + 4 - n}
\]
\[
\le c_0\,m!\,n!\,\frac{r_0^m}{(1 + m)^2}\,\frac{\rho_0^n}{(1 + n)^2},
\quad
m \ge 0, \quad 0 \le n \le 2\,m + 4.
\]
\end{proof}

The following Lemma provides our main estimates.
\begin{lemma}
\label{convergence}
Suppose $s_0 = s_0(u, v)$ is a holomorphic function defined on some open
neighbourhood $U$ of $O = \{u = 0, v = 0, w = 0\}$ in $W_0 = \{w = 0\}$ 
which has an expansion of the form
\[
s_0(u, v) = \sum_{m = 0}^{\infty} \sum_{n = 0}^{2m + 4}
\psi_{m, n}\,u^m\,v^n,
\]
so that its Taylor coefficients at the point $O$ satisfy estimates of the
type (\ref{formconvest}) with some positive constants $c^*_0$, $r_0$, 
and $\rho_0 < 1/2$.
Then there exist positive constants $r \ge r_0$, $\rho$, $c_{\hat{e}^a_{AB}}$,
$c_{\hat{\Gamma}_{ABCD}}$, $c_{\zeta}$, $c_{\zeta_i}$, $c_{\hat{s}}$, $c_k$
so that the expansion coefficients determined from $s_0$ in 
Lemma \ref{exunformexp} satisfy
for $m, n, p = 0, 1, 2, \ldots$
\begin{equation}
\label{skest}
|\partial^m_u\,\partial^n_v\,\partial^p_w\,s_k(O)|
\le c_k\,
\frac{r^{m + p}\,(m + p)!\,\rho^n\,n!}
{(m + 1)^2\,(n + 1)^2\,(p + 1)^2},
\end{equation}
and
\begin{equation}
\label{fest}
|\partial^m_u\,\partial^n_v\,\partial^p_w\,f(O)|
\le c_{f}\,
\frac{r^{m + p - 1}\,(m + p)!\,\rho^n\,n!}
{(m + 1)^2\,(n + 1)^2\,(p + 1)^2},
\end{equation}
where $f$ stands for any of the functions
$\hat{e}^a_{AB},\,\,\hat{\Gamma}_{ABCD},\,\,
\zeta, \,\, \zeta_i, \,\, \hat{s}$.
\end{lemma}

\begin{remark}
Observing the $v$-finite expansion types discussed in Lemma \ref{formexptype},
we can replace the right hand sides in the estimates above by zero if $n$ is
large enough relative to $m$. This will not be pointed out at each step and
for convenience the estimates will be written as above. The expansion types
obtained in Lemma \ref{formexptype} will become important and will be observed,
however, when we derive the estimates.
\end{remark}

We shall make use of arguments discussed in  
\cite{shinbrot: welland}. The following four Lemmas are essentially given in
that article.

\begin{lemma} 
\label{basicestimate}
For any non-negative integer $n$ there is a positive constant $C$
independent of $n$ so that
\[
\sum_{k = 0}^n \frac{1}{(k + 1)^2(n - k + 1)^2}
\le C\,\frac{1}{(n + 1)^2}.
\]
\end{lemma}

\begin{proof} Denoting by $[n/2]$ the largest integer $\le n /2$, we get
with $C = \sum_{k = 0}^{\infty} \frac{8}{(k + 1)^2}$
\[
\sum_{k = 0}^n \frac{1}{(k + 1)^2(n - k + 1)^2}
\le \sum_{k = 0}^{[n/2]} \frac{2}{(k + 1)^2(n - k + 1)^2}
\]
\[
\le \sum_{k = 0}^{[n/2]} \frac{2}{(k + 1)^2([n/2] + 1)^2}
\le C\,\frac{1}{(n + 1)^2}.
\]
\end{proof}

In the following $C$ will always denote the constant above.

\begin{lemma} 
\label{binomestimate}
For any integers $m$, $n$, $k$, $j$ with 
$0 \le k \le m$, and $0 \le j \le n$ resp. $0 \le j \le n - 1$ holds
\[
{m \choose k}\,{n \choose j} \le {m + n \choose k + j}
\quad\quad  \mbox{resp.} \quad
{m \choose k}\,{n - 1 \choose j} \le {m + n \choose k + j}.
\]
\end{lemma}

\begin{proof} This follows by induction, using the 
general formula
${n + 1 \choose j} = {n \choose j} + {n \choose j - 1}$,
or by expanding
$(x + y)^{m + n} = (x + y)^{m}\,(x + y)^{n}$,
using the binomial law
$(x + y)^{p} = \sum_{j = 0}^p {p \choose j} \,x^j\,y^{p - j}$. 
\end{proof}

If $f$ is holomorphic on the polydisk $P = \{(u, v,
w,)\in  \mathbb{C}^3|\,\,\, |u| \le 1/r_1, |v| \le 1 /r_2, |w| \le 1/r_3\}$,
with some $r_1, r_2, r_3 > 0$, one has
the Cauchy estimates
\begin{equation}
\label{Cauchyestimate}
|\partial^m_u\,\partial^n_v\,\partial^p_w\,f(O)| \le
r_1^m\,\,r_2^n\,\,r_3^p\,\,m!\,\,n!\,\,p!\,\,\sup_P |f|,
\quad\quad m, n, p = 0, 1, 2, \ldots
\end{equation}
where $O$ denotes the origin $u = 0$, $v = 0$, $w = 0$. We need a slight
modification of this.

\begin{lemma}
\label{estimatetype}
If $f$ is holomorphic near $O$, there exist positive constants $c$, $r_0$,
$\rho_0$ so that 
\[
|\partial^m_u\,\partial^n_v\,\partial^p_w\,f(O)| \le
c\,\frac{r^{m + p}\,(m + p)!\,\rho^n\,n!}{(m + 1)^2\,(n + 1)^2\,(p +
1)^2},
\quad\quad m, n, p = 0, 1, 2, \ldots
\]
for any $r \ge r_0$, $\rho \ge \rho_0$. If in addition 
$f(0, v, 0) = 0$, the constants can be chosen such that
\[
|\partial^m_u\,\partial^n_v\,\partial^p_w\,f(O)| \le
c\,\frac{r^{m + p - 1}\,(m + p)!\,\rho^n\,n!}
{(m + 1)^2\,(n + 1)^2\,(p + 1)^2},
\quad\quad m, n, p = 0, 1, 2, \ldots
\]
for any $r \ge r_0$, $\rho \ge \rho_0$.
\end{lemma}

\begin{proof} Choosing an estimate of the type (\ref{Cauchyestimate})
with $r_1 = r_3$ and setting $c = \alpha\,\sup_P |f|$, 
$r_0 = e^2\,r_1 = e^2\,r_3$, $\rho_0 = e^2\,r_2$ with some $\alpha > 0$, one
gets from (\ref{Cauchyestimate})
\[
|\partial^m_u\,\partial^n_v\,\partial^p_w\,f(O)| \le
c\,\alpha^{-1} r_0^{m + p}\,(m + p)!\,\rho_0^n\,n!\,e^{- 2(m + n + p)}
\le
c\,\alpha^{-1} \frac{r_0^{m + p}\,(m + p)!\,\rho_0^n\,n!}
{(m + 1)^2\,(n + 1)^2\,(p + 1)^2}.
\]
With $\alpha = 1$ the monotonicity of
$x \rightarrow x^q$, $q \ge 0$, $x > 0$ implies the first estimate. 
With $\alpha = r_0$ the estimate above implies 
\[
|\partial^m_u\,\partial^n_v\,\partial^p_w\,f(O)| 
\le
c\, \frac{r_0^{m + p - 1}\,(m + p)!\,\rho_0^n\,n!}
{(m + 1)^2\,(n + 1)^2\,(p + 1)^2}.
\]
If $f(0, v, 0) = 0$, then
$\partial^0_u\,\partial^n_v\,\partial^0_w\,f(O) = 0$ for $n \in N_0$ and the
last relation remains true for $m + p = 0$, i.e. $m = 0$ and $p = 0$, if
$r_0$ and
$\rho _0$ are replaced by $r \ge r_0$ and $\rho \ge \rho _0$.
If $m + p > 0$ the result follows as above.
\end{proof}

\begin{lemma}
\label{prodestimate}
Let $m$, $n$, $p$ be non-negative integers and $f_i$, $i = 1, \ldots, N$,  be
smooth complex valued functions of $u$, $v$, $w$ on some neighbourhood $U$
of $O$ whose derivatives satisfy on $U$ (resp. at a given point $p \in
U$) estimates of the form 
\[
|\partial^j_u\,\partial^k_v\,\partial^l_w\,f_i|
\le c_i\,\frac{r^{j + l + q_i}\,(j + l)\,!\,\rho^k\,k\,!}
{(j + 1)^2\,(k + 1)^2\,(l + 1)^2}
\quad\mbox{for}\quad
0 \le j \le m,\,\,0 \le k \le n,\,\,0 \le l \le p,
\]   
with some positive constants $c_i$, $r$, $\rho$ and some fixed integers
$q_i$ (independent of $j, k, l$). Then one has on $U$ (resp. at $p$) 
the estimates
\begin{equation}
\label{prodest}
|\partial^{m}_u\,\partial^{n}_v\,\partial^{p}_w\,(f_1 \cdot \ldots \cdot
f_N)|
\le C^{3\,(N - 1)}\,
\, c_1 \cdot \ldots \cdot c_N\,\,\frac{r^{m + p + q_1 + \ldots + q_N}\,
(m + p)\,!\,\rho^{n}\,n\,!} {(m + 1)^2\,(n + 1)^2\,(p + 1)^2}.
\end{equation}
\end{lemma}

\begin{remark}
\label{prodrems} 
(i) Lemma \ref{prodestimate} remains obviously true if
$m$, $n$, $p$ are replaced in (\ref{prodest})  by integers  $m'$, $n'$, $p'$
with $0 \le m' \le m$, $0 \le n' \le n$, $0 \le p' \le p$.

(ii) By the argument given below the factor $C^{3\,(N - 1)}$ in (\ref{prodest})
can be replaced by $C^{(3 - r)\,(N - 1)}$ 
if $r$ of the integers $m$, $n$, $p$ vanish.
\end{remark}

\begin{proof} We prove the case $N = 2$. The general case then
follows with the first of Remarks \ref{prodrems}  by an induction argument.
With the estimates
above and Lemmas (\ref{basicestimate}) and (\ref{binomestimate}) we get on $U$
(resp. at $p$)

\[
|\partial^{m}_u\,\partial^{n}_v\,\partial^{p}_w\,(f_1\,f_2)| 
\le
\sum_{j = 0}^{m}\,\sum_{k = 0}^{n}\,\sum_{l = 0}^{p}
{m \choose j}\,{n \choose k}\,{p \choose l}
|\partial^j_u\,\partial^k_v\,\partial^l_w\,f_1|
|\partial^{m - j}_u\,\partial^{n - k}_v\,\partial^{p - l}_w\,f_2| 
\]
\[
\le \sum_{j = 0}^m\,\sum_{k = 0}^n\,\sum_{l = 0}^p
{m \choose j}\,{n \choose k}\,{p \choose l}
\frac{c_1\,r^{j + l + q_1}\,(j + l)\,!\,\rho^k\,k\,!}
{(j + 1)^2\,(k + 1)^2\,(l + 1)^2}
\frac{c_2\,r^{m - j + p - l + q_2}\,(m - j + p - l)\,!\,
\rho^{n - k}\,(n - k)\,!}
{(m - j + 1)^2\,(n - k + 1)^2\,(p - l + 1)^2}  
\]
\[
\le \sum_{j = 0}^m\,\sum_{k = 0}^n\,\sum_{l = 0}^p
\frac{{m \choose j}\,{p \choose l}}{{m + p \choose j + l}}
\frac{c_1\,c_2\,r^{m + p + q_1 + q_2}\,(m + p)\,!\,\rho^n\,n\,!}
{(j + 1)^2\,(k + 1)^2\,(l + 1)^2\,
(m - j + 1)^2\,(n - k + 1)^2\,(p - l + 1)^2} 
\]
\[
\le C^{3}\,c_1\,c_2\,\frac{r^{m + p + q_1 + q_2}\,(m +
p)\,!\,\rho^n\,n\,!} {(m + 1)^2\,(n + 1)^2\,(p + 1)^2}. 
\]   
\end{proof}

We are now able to proof our main estimates.

{\bf Proof of Lemma} \ref{convergence}.
The proof will essentially be given by induction with respect to $m$ and
$p$, following the procedure which led to Lemma \ref{exunformexp}.
It is easy to see that the constants can be chosen to satisfy the estimates at
lowest order. Leaving the choice of the constants open, we will derive from
the induction hypothesis for the derivatives of the next order estimates of the form
\[
|\partial^m_u\,\partial^n_v\,\partial^p_w\,s_k(O)|
\le c_k\,
\frac{r^{m + p}\,(m + p)!\,\rho^n\,n!}
{(m + 1)^2\,(n + 1)^2\,(p + 1)^2}\,A_{s_k},
\]
\[
|\partial^m_u\,\partial^n_v\,\partial^p_w\,f(O)|
\le c_{f}\,
\frac{r^{m + p - 1}\,(m + p)!\,\rho^n\,n!}
{(m + 1)^2\,(n + 1)^2\,(p + 1)^2}\,A_f,
\]
with certain constants $A_{s_k}$, $A_f$ which depend on $m$, $n$,
$p$ and the constants $c_k$, $c_f$, $r$, and $\rho$. Sometimes superscripts 
will indicate to which order of differentiability particular
constants $A_{s_k}$, $A_f$ refer. Occasionally we will have to make
assumptions on $r$ to proceed with the induction step.
We shall collect these conditions and the constants $A_{s_k}$, $A_f$, or
estimates for them, and at the end it will be shown that the constants 
$c_k$, $c_f$, $r$, and $\rho$ can be adjusted so that all conditions are
satisfied and $A_{s_k} \le 1$, $A_f \le 1$. This will complete the induction
proof. 

In the following it is understood that as above a function in a modulus sign
is evaluated at the origin $O$. The symbol $x$ will stand for any of the
fields
\[
\hat{e}^a\,_{AB},\, \hat{\Gamma}_{ABCD},\, \zeta,\,
\zeta_0,\,\zeta_1,\,\zeta_2, \,\hat{s},\, 
s_1,\,s_2,\,s_3,\,s_4.
\]

For the quantities which are known to vanish at $I$ the estimates are correct
for $m = 0$, $p = 0$. Since we consider
$\hat{s}$ as an unknown and $s(0) = - 2$ as part of the equations, we
thus only need to discuss the $s_k$. They are given on $I$ by
\[
s_k = \frac{(4 - k)!}{4!}\,\partial^k_v\,s_0.
\]
It thus follows by our assumptions
\[
|\partial^0_u\,\partial^n_v\,\partial^0_w\,s_k|
=  |\frac{(4 - k)!}{4!}\partial^{k + n}_v\,s_0|
\]
\[
\le  
\left\{
\begin{array}{c}
\frac{(4 - k)!}{4!}\,c_0\,
\frac{\rho^{n + k}\,(n + k)!}{(n + k + 1)^2} 
\quad\mbox{for}\quad n \le 4 - k \\
0 \quad\mbox{for}\quad n > 4 - k\\  
\end{array}
\right\}
= c_k\,\frac{\rho^n\,n!}{(n + 1)^2}\,A^{m = 0, p = 0}_{s_k},
\]
with
\[
A^{m = 0, p = 0}_{s_k} = \frac{c_0}{c_k}\,\rho^k\,h_{k,n}
\le \frac{c_0}{c_k}\,\rho^k,
\]
because
\[
h_{k,n} \equiv \left\{
\begin{array}{c}
\frac{(4 - k)!}{4!}\,
\frac{(n + k)!}{n!} 
\frac{(n + 1)^2}{(n + k + 1)^2} 
\quad\mbox{for}\quad n \le 4 - k \\
0 \quad\mbox{for}\quad n > 4 - k\\  
\end{array}
\right\} \le 1.
\]

\vspace{.2cm}

We should study now under which conditions on the constants it can be shown
by induction with respect to $m$ that the quantities
$|\partial^m_u\,\partial^n_v\,\partial^0_w\,x|$, $n \in N_0$, satisfy the
estimates  given in the Lemma. We shall skip the details of this step,
because the arguments used here are similar to those used to discuss the
quantities
$|\partial^m_u\,\partial^n_v\,\partial^p_w\,x|$ for general $p$ and the
requirements obtained in that case are in fact stronger that those
obtained for $p = 0$.

It will be assumed now that $p \ge 1$, that the estimates  
for $|\partial^m_u\,\partial^n_v\,\partial^l_w\,x|$ given in the
Lemma hold true for $m, n \in N_0$, $0 \le l \le p - 1$, and try to
determine  conditions so that the induction step $p - 1  \rightarrow p$ can
be performed.

\vspace{.2cm} 

By taking formal derivatives of the equation 
\[
0 = H_{0100} + H_{1000}, 
\]
we get with our assumptions
\[
|\partial^m_u\,\partial^n_v\,\partial^p_w\,s_0| \le
|\partial^{m + 1}_u\,\partial^n_v\,\partial^{p - 1}_w\,s_2|
+ |\partial^m_u\,\partial^n_v\,
\partial^{p - 1}_w\,(\hat{e}^1\,_{11}\,\partial_u\,s_0)| 
\]
\[
+ |\partial^m_u\,\partial^n_v\,
\partial^{p - 1}_w\,(\hat{e}^2\,_{11}\,\partial_v\,s_0)| 
+ 4\,|\partial^m_u\,\partial^n_v\,\partial^{p - 1}_w\,
(\hat{\Gamma}_{1101}\,s_0  + \hat{\Gamma}_{1100}\,s_1)|
\]
\[
+ \mu\,|\partial^m_u\,\partial^n_v\,\partial^{p - 1}_w\,
(\frac{1}{1 -
\mu\,\zeta}\,
\left\{s_0\,\zeta_{2} + 2\,s_1\,\zeta_{1}
- 3\,s_2\,\zeta_{0}\right\})|.
\]
For the first term on the right hand side follows immediately
\[
|\partial^{m + 1}_u\,\partial^n_v\,\partial^{p - 1}_w\,s_2|
\le c_2\,\frac{r^{m + p}\,(m + p)!\,\rho^n\,n!}
{(m + 2)^2 (n + 1)^2 p^2}.
\]
A slight variation of the calculations in the proof Lemma
\ref{prodestimate} gives 
\[
|\partial^m_u\,\partial^n_v\,
\partial^{p - 1}_w\,(\hat{e}^1\,_{11}\,\partial_u\,s_0)| \le
\]
\[
\sum_{j = 0}^{m}\,\sum_{k = 0}^{n}\,\sum_{l = 0}^{p - 1}
{m \choose j}\,{n \choose k}\,{p - 1 \choose l}
|\partial^j_u\,\partial^k_v\,\partial^l_w\,\hat{e}^1\,_{11}|
|\partial^{m - j + 1}_u\,\partial^{n - k}_v\,\partial^{p - l - 1}_w\,s_0|  \le
\]
\[
 \sum_{j = 0}^m\,\sum_{k = 0}^n\,\sum_{l = 0}^{p - 1}
\frac{{m \choose j}\,{p - 1 \choose l}}{{m + p \choose j + l}}
\frac{c_{\hat{e}^1\,_{11}}\,c_0\,\,r^{m + p - 1}\,(m +
p)\,!\,\rho^n\,n\,!} {(j + 1)^2\,(k + 1)^2\,(l + 1)^2\,
(m - j + 2)^2\,(n - k + 1)^2\,(p - l)^2} 
\]
\[
\le C^{3}\,c_{\hat{e}^1\,_{11}}\,\,c_0\,\frac{r^{m + p - 1}\,(m +
p)\,!\,\rho^n\,n\,!} {(m + 2)^2\,(n + 1)^2\,p^2}, 
\]   
where the sum over $j$ has been extended in the last step to $m + 1$.

Similarly one gets
\[
 |\partial^m_u\,\partial^n_v\,
\partial^{p - 1}_w\,(\hat{e}^2\,_{11}\,\partial_v\,s_0)| \le
\]
\[
\sum_{j = 0}^{m}\,\sum_{k = 0}^{n}\,\sum_{l = 0}^{p - 1}
{m \choose j}\,{n \choose k}\,{p - 1 \choose l}
|\partial^j_u\,\partial^k_v\,\partial^l_w\,\hat{e}^2\,_{11}|
|\partial^{m - j}_u\,\partial^{n - k + 1}_v\,\partial^{p - l - 1}_w\,s_0| \le
\]
\[
\sum_{j = 0}^m\,\sum_{k = 0}^n\,\sum_{l = 0}^{p - 1}
\frac{{m \choose j}\,{n \choose k}\,{p - 1 \choose l}}
{{m + p - 1 \choose j + l}\,{n + 1 \choose k}}
\frac{c_{\hat{e}^2\,_{11}}\,c_0\,\,r^{m + p - 2}\,(m +
p - 1)\,!\,\rho^{n + 1}\,(n + 1)\,!} {(j + 1)^2\,(k + 1)^2\,(l + 1)^2\,
(m - j + 1)^2\,(n - k + 2)^2\,(p - l)^2} \le
\]
\[
C^{3}\,c_{\hat{e}^2\,_{11}}\,\,c_0\,\frac{r^{m + p - 2}\,(m +
p - 1)\,!\,\rho^{n + 1}\,(n + 1)\,!} {(m + 1)^2\,(n + 2)^2\,p^2}, 
\]   
where the sum over $k$ has been extended in the last step to $n + 1$.

We emphasis here again an observation which is important
for us. By Lemma \ref{formexptype} the terms
$\partial^j_u\,\partial^k_v\,\partial^l_w\,\hat{e}^2\,_{11}$ and 
$\partial^{m - j}_u\,\partial^{n - k + 1}_v\,\partial^{p - l - 1}_w\,s_0$
in the second line vanish if
$k > 2\,j + 1$ and $n - k + 1 > 2\,(m - j) + 4$ respectively. This
implies, consistent with Lemma \ref{formexptype}, that the term on the left
hand side vanishes if $n > 2\,m + 4$. When we estimate the expression in the
last line above we can thus assume without error that $n \le
2\,m + 4$.

Lemma \ref{prodestimate} implies immediately
\[
4\,|\partial^m_u\,\partial^n_v\,\partial^{p - 1}_w\,
(\hat{\Gamma}_{1101}\,s_0  + \hat{\Gamma}_{1100}\,s_1)|
\le 4\,C^{3}\,(c_0\,c_{\hat{\Gamma}_{1101}} + c_1\,c_{\hat{\Gamma}_{1100}})\,
\frac{r^{m + p - 2}\,(m +
p - 1)\,!\,\rho^{n}\,n\,!} {(m + 1)^2\,(n + 1)^2\,p^2},
\]
and, observing that $\zeta(O) = 0$, 
\[
\mu\,|\partial^m_u\,\partial^n_v\,\partial^{p - 1}_w\,
(\frac{1}{1 - \mu\,\zeta}\,
\left\{s_0\,\zeta_{2} + 2\,s_1\,\zeta_{1}
- 3\,s_2\,\zeta_{0}\right\})|
\]
\[
\le \mu\,\sum_{l = 0}^{\infty}|\partial^m_u\,\partial^n_v\,\partial^{p -
1}_w\, ( (\mu\,\zeta)^l
\left\{s_0\,\zeta_{2} + 2\,s_1\,\zeta_{1}
- 3\,s_2\,\zeta_{0}\right\})|
\]
\[
\le \mu\,\sum_{l = 0}^{\infty}\,\mu^l\,c_{\zeta}^l\,C^{3\,(l + 1)}\,
(c_0\,c_{\zeta_{2}} + 2\,c_1\,c_{\zeta_{1}}
+ 3\,c_2\,c_{\zeta_{0}})
\frac{r^{m + p - l - 2}\,(m +
p - 1)\,!\,\rho^{n}\,n\,!} {(m + 1)^2\,(n + 1)^2\,p^2}
\]
\[
= \frac{\mu}{1 - \frac{\mu\,c_{\zeta}\,C^3}{r}}\,
C^3\,(c_0\,c_{\zeta_{2}} + 2\,c_1\,c_{\zeta_{1}}
+ 3\,c_2\,c_{\zeta_{0}})
\frac{r^{m + p - 2}\,(m +
p - 1)\,!\,\rho^{n}\,n\,!} {(m + 1)^2\,(n + 1)^2\,p^2},
\]
where it is assumed that 
\[
r > \mu\,c_{\zeta}\,C^3.
\]
Together this gives
\[
|\partial^m_u\,\partial^n_v\,\partial^p_w\,s_0|
\le
c_2\,\frac{r^{m + p}\,(m + p)!\,\rho^n\,n!}
{(m + 2)^2 (n + 1)^2 p^2}
\]
\[
+
C^{3}\,c_{\hat{e}^1\,_{11}}\,\,c_0\,\frac{r^{m + p - 1}\,(m +
p)\,!\,\rho^n\,n\,!} {(m + 2)^2\,(n + 1)^2\,p^2}
\]
\[
+
C^{3}\,c_{\hat{e}^2\,_{11}}\,\,c_0\,\frac{r^{m + p - 2}\,(m +
p - 1)\,!\,\rho^{n + 1}\,(n + 1)\,!} {(m + 1)^2\,(n + 2)^2\,p^2}
\]
\[
+
 4\,C^{3}\,(c_0\,c_{\hat{\Gamma}_{1101}} + c_1\,c_{\hat{\Gamma}_{1100}})\,
\frac{r^{m + p - 2}\,(m +
p - 1)\,!\,\rho^{n}\,n\,!} {(m + 1)^2\,(n + 1)^2\,p^2}
\]
\[
+
\frac{\mu}{1 - \frac{\mu\,c_{\zeta}\,C^3}{r}}\,
C^3\,(c_0\,c_{\zeta_{2}} + 2\,c_1\,c_{\zeta_{1}}
+ 3\,c_2\,c_{\zeta_{0}})
\frac{r^{m + p - 2}\,(m +
p - 1)\,!\,\rho^{n}\,n\,!} {(m + 1)^2\,(n + 1)^2\,p^2}
\]
\[
\le c_0\,\frac{r^{m + p}\,(m + p)!\,\rho^n\,n!}
{(m + 1)^2 (n + 1)^2 (p + 1)^2}\,A^*_{s_0},
\]
with a factor 
\[
A^*_{s_0} =
\frac{c_2}{c_0}\,\frac{(m + 1)^2\,(p + 1)^2}
{(m + 2)^2 \,\,p^2}
+
\frac{1}{r}\,C^{3}\,c_{\hat{e}^1\,_{11}}\,\frac{(m + 1)^2\,(p +
1)^2\,}  {(m + 2)^2\,\,p^2}
\]
\[
+
\frac{1}{r^2}\,C^{3}\,c_{\hat{e}^2\,_{11}}\,\frac{
\rho\,(n + 1)^3\,(p + 1)^2} {(n + 2)^2\,p^2\,(m + p)}
+
\frac{4}{r^2}\,\,C^{3}\,(c_{\hat{\Gamma}_{1101}} +
\frac{c_1}{c_0}\,c_{\hat{\Gamma}_{1100}})\,
\frac{(p + 1)^2\,} {p^2\,(m + p)}
\]
\[
+
\frac{1}{r^2}\,\frac{\mu}{1 - \frac{\mu\,c_{\zeta}\,C^3}{r}}\,
C^3\,(c_{\zeta_{2}} + 2\,\frac{c_1}{c_0}\,c_{\zeta_{1}}
+ 3\,\frac{c_2}{c_0}\,c_{\zeta_{0}})
\frac{(p + 1)^2} {p^2\,(m + p)}.
\]
Recalling that we can assume $n \le 2\,m + 4$ in the
third term on the right hand side, this finally gives
\[
A^*_{s_0} \le
4\,\frac{c_2}{c_0} +
\frac{4}{r}\,C^{3}\,c_{\hat{e}^1\,_{11}}
+
\frac{20\,\rho}{r^2}\,C^{3}\,c_{\hat{e}^2\,_{11}}
+
\frac{16}{r^2}\,\,C^{3}\,(c_{\hat{\Gamma}_{1101}} +
\frac{c_1}{c_0}\,c_{\hat{\Gamma}_{1100}})
\]
\[
+
\frac{1}{r^2}\,\frac{4\,\mu}{1 - \frac{\mu\,c_{\zeta}\,C^3}{r}}\,
C^3\,(c_{\zeta_{2}} + 2\,\frac{c_1}{c_0}\,c_{\zeta_{1}}
+ 3\,\frac{c_2}{c_0}\,c_{\zeta_{0}}).
\]

We have the relations
\[
s_k = \frac{(4-k)!}{4\,!}\,\partial^k_v\,s_0
\quad \mbox{on} \quad U_0,
\]
the equation $0 = H_{0100} + H_{1000}$ reduces to
\[
\partial_w\,s_0 = \partial_u\,s_2 + 3\,\mu\,s_2\,\zeta_0
\quad \mbox{on} \quad U_0,
\]
and we have seen that
\[
\partial_v\,\zeta_0 = 0
\quad \mbox{on} \quad U_0.
\] 
This implies for $p \ge 1$ the estimates
\[
|\partial^0_u\,\partial^n_v\,\partial^p_w\,s_k|
\le \frac{(4-k)!}{4\,!}\,
(|\partial^1_u\,\partial^{n + k}_v\,\partial^{p - 1}_w\,s_2|
+ 3\,\mu\,
|\partial^0_u\,\partial^{n + k}_v\,\partial^{p - 1}_w\,(s_2\,\zeta_0)|)
\]
\[
\le  
\left\{
\begin{array}{c}
\frac{(4 - k)!}{4!}\,c_2\,
\frac{r^p\,p!\,\rho^{n + k}\,(n + k)!}{4\,p^2\,(n + k + 1)^2} 
\quad\mbox{for}\quad n \le 4 - k \\
0 \quad\mbox{for}\quad n > 4 - k\\  
\end{array}
\right\}
\]
\[
+   
\left\{
\begin{array}{c}
3\,\mu\,\frac{(4 - k)!}{4!}\,
\sum_{l = 0}^{p - 1}{p - 1 \choose l}
|\partial^{n + k}_v\,\partial^l_w\,s_2|\,
|\partial^{p - 1 - l}_w\,\zeta_0|
\quad\mbox{for}\quad n \le 2 - k \\
0 \quad\mbox{for}\quad n > 2 - k\\  
\end{array}
\right\}
\]
\[
\le c_k\,\frac{r^{p}\,p !\,\rho^n\,n!}
{(n + 1)^2 (p + 1)^2}\,A^{m = 0, p \ge 1}_{s_k},
\]
with
\[
A^{m = 0, p \ge 1}_{s_k} = 
\frac{c_2}{c_k}\,\rho^k\,f_{k,n} + \frac{3}{r}\,\mu\,C\,
\frac{c_2\,c_{\zeta_0}}{c_k}\,\rho^k\,g_{k, n}
\le 
\frac{c_2}{c_k}\,\rho^k + \frac{12}{r}\,\mu\,C\,
\frac{c_2\,c_{\zeta_0}}{c_k}\,\rho^k,
\]
because
\[
f_{k,n} \equiv   
\left\{
\begin{array}{c}
\frac{(4 - k)!}{4!}\,
\frac{(n + k)!\,(n + 1)^2\,(p + 1)^2}{n!\,(n + k + 1)^2\,4\,p^2} 
\quad\mbox{for}\quad n \le 4 - k \\
0 \quad\mbox{for}\quad n > 4 - k\\  
\end{array}
\right\} \le 1,
\]
\[
g_{k,n} \equiv   
\left\{
\begin{array}{c}
\frac{(4 - k)!}{4!}\,
\frac{(n + k)!\,(n + 1)^2\,(p + 1)^2}{n!\,(n + k + 1)^2\,p^3} 
\quad\mbox{for}\quad n \le 2 - k \\
0 \quad\mbox{for}\quad n > 2 - k\\  
\end{array}
\right\} \le 4.
\]

\vspace{.5cm}

From the equation $\Sigma_{1100} = 0$, which reads 
\[
\partial_w\,\zeta_0 = - 2 + \hat{s}
\quad \mbox{on} \quad U_0,
\]
follows
\[
|\partial^0_u\,\partial^n_v\,\partial_w\,\zeta_0| = 
|\partial^n_v\,(- 2 + \hat{s})| = 2\,\delta^n_0 \le 
c_{\zeta_0}\,\frac{\rho^n\,n!}{(n + 1)^2}\,A^{m = 0, p = 1}_{\zeta_0},
\]
with
\[
A^{m = 0, p = 1}_{\zeta_0} = \frac{2}{c_{\zeta_0}}.
\]
Furthermore, for $p \ge 2$,
\[
|\partial^0_u\,\partial^n_v\,\partial^p_w\,\zeta_0| = 
|\partial^n_v\,\partial^{p - 1}_w\,\hat{s}| = 
c_{\hat{s}}\,\frac{r^{p - 1}\,(p - 1)!\,\rho^n\,n!}
{(n + 1)^2\,p^2}
\le c_{\zeta_0}\,\frac{r^{p}\,p !\,\rho^n\,n!}
{(n + 1)^2\,(p + 1)^2}\,A^{m = 0, p \ge 2}_{\zeta_0},
\]
with 
\[
A^{m = 0, p \ge 2}_{\zeta_0} =
\frac{1}{r}\,\frac{c_{\hat{s}}}{c_{\zeta_0}}\,
\frac{(p + 1)^2}{p^3} \le \frac{2}{r}\,\frac{c_{\hat{s}}}{c_{\zeta_0}}.
\]

\vspace{.5cm}

The equation $S_{11} = 0$, which reads
\[
\partial_w\,\hat{s} = - s_4\,\zeta_0
\quad \mbox{on} \quad U_0,
\]
implies
\[
|\partial^0_u\,\partial^n_v\,\partial_w\,\hat{s}| = 0
\le c_{\hat{s}}\,\frac{\rho^n\,n!}{(n + 1)^2},
\]
and for $p \ge 2$
\[
|\partial^0_u\,\partial^n_v\,\partial^p_w\,\hat{s}| 
= |\partial^0_u\,\partial^n_v\,\partial^{p - 1}_w\,(s_4\,\zeta_0)|
\le 
C^2\,c_4\,c_{\zeta_0}\frac{r^{p - 2}\,(p - 1) !\,\rho^n\,n!}
{(n + 1)^2\,p^2} 
\]
\[
\le
c_{\hat{s}}\,
\frac{r^{p - 1}\,p !\,\rho^n\,n!}
{(n + 1)^2\,(p + 1)^2}\,A^{m = 0, p \ge 2}_{\hat{s}},
\]
with 
\[
A^{m = 0, p \ge 2}_{\hat{s}} =
\frac{1}{r}\,C^2\,
\frac{c_4\,c_{\zeta_0}}{c_{\hat{s}}}\,\frac{(p + 1)^2}{p^3}
\le \frac{2}{r}\,C^2\,
\frac{c_4\,c_{\zeta_0}}{c_{\hat{s}}}.
\]

\vspace{.2cm}

Having studied the quantities
$|\partial^m_u\,\partial^n_v\,\partial^p_w\,x|$ for $m = 0$,
we shall now derive the conditions which arise from the requirement that we
can obtain the desired estimates for these quantities inductively for all
positive integers $m$. We shall provide detailed arguments only for
some representative $\partial_u$-equations and just state the analogues
results for the remaining equations.

Multiplication of the equation
\[
\partial_u\hat{e}^2\,_{01} + \frac{1}{u}\,\hat{e}^2\,_{01}
= 
\frac{1}{u}\,\hat{\Gamma}_{0100}
+ 2\,\hat{\Gamma}_{0100}\,\hat{e}^2\,_{01},
\]
with $u$ and formal differentiation gives with Lemma \ref{prodest} for 
$m \ge 1$

\[
|\partial^m_u\,\partial^n_v\,\partial^p_w\,\hat{e}^2\,_{01}|
\le 
\frac{1}{m + 1}\,
(|\partial^m_u\,\partial^n_v\,\partial^p_w\,\hat{\Gamma}_{0100}| 
+ 2\,m\,
|\partial^{m - 1}_u\,\partial^n_v\,\partial^p_w\,
(\hat{\Gamma}_{0100}\,\hat{e}^2\,_{01})|)
\]
\[
\le \frac{1}{m + 1}\,\left(c_{\hat{\Gamma}_{0100}}\,
\frac{r^{m + p -1}\,(m + p) !\,\rho^n\,n!}{(m + 1)^2\,(n + 1)^2\,(p + 1)^2}
+ 2\,m\,C^3\,c_{\hat{e}^2\,_{01}}\,c_{\hat{\Gamma}_{0100}}\,
\frac{r^{m + p - 3}\,(m + p - 1) !\,\rho^n\,n!}{m^2\,(n + 1)^2\,(p +
1)^2}\right)
\]
\[
= c_{\hat{e}^2\,_{01}}\,
\frac{r^{m + p -1}\,(m + p) !\,\rho^n\,n!}{(m + 1)^2\,(n + 1)^2\,(p + 1)^2}
\,A^{m \ge 1}_{\hat{e}^2\,_{01}},
\]
with 
\[
A^{m \ge 1}_{\hat{e}^2\,_{01}} =
\frac{c_{\hat{\Gamma}_{0100}}}{c_{\hat{e}^2\,_{01}}}\,\frac{1}{m + 1}
+ \frac{1}{r^2}\,C^3\,c_{\hat{\Gamma}_{0100}}\,
\frac{2\,(m + 1)}{m\,(m + p)}.
\]
Proceeding in a similar way with the equations for the other frame
coefficients one gets for the factors which need to be controlled the estimates
\[
A^{m \ge 1}_{\hat{e}^2_{01}} \le 
\frac{c_{\hat{\Gamma}_{0100}}}{2\,c_{\hat{e}^2_{01}}}
+ \frac{4}{r^2}\,C^3\,c_{\hat{\Gamma}_{0100}},\quad\quad
A^{m \ge 1}_{\hat{e}^2_{11}} \le 
\frac{c_{\hat{\Gamma}_{1100}}}{c_{\hat{e}^2_{11}}}
+ \frac{8}{r^2}\,C^3\,
\frac{c_{\hat{\Gamma}_{1100}}\,c_{\hat{e}^2_{01}}}{c_{\hat{e}^2_{11}}},
\]
\[
A^{m \ge 1}_{\hat{e}^1_{01}} \le 
 \frac{4}{r}\,\frac{c_{\hat{\Gamma}_{0101}}}{c_{\hat{e}^1_{01}}}
+ \frac{4}{r^2}\,C^3\,c_{\hat{\Gamma}_{0100}},\quad\quad
A^{m \ge 1}_{\hat{e}^1_{11}} \le 
\frac{8}{r}\,\frac{c_{\hat{\Gamma}_{1101}}}{c_{\hat{e}^1_{11}}}
+ \frac{8}{r^2}\,C^3\,
\frac{c_{\hat{\Gamma}_{1100}}\,c_{\hat{e}^1_{01}}}{c_{\hat{e}^1_{11}}}.
\]
The same inequalities, with $C^3$ replaced by $C^2$, are obtained in the case
$p = 0$.
In the last two inequalities the occurence of $1/r$ in both terms reflects
the fact that $\hat{e}^1_{01}$ and $\hat{e}^1_{11}$ are both of the order
$O(u^2)$ near $O$.

\vspace{.2cm}

Multiplication of the equation
\[
\partial_u\,\hat{\Gamma}_{0100} 
+ \frac{2}{u}\,\hat{\Gamma}_{0100}
= 2\,\hat{\Gamma}^2_{0100} + \frac{1}{2}\,s_{0},
\]
with $u$ and formal differentiation gives for $m \ge 1$
\[
|\partial^m_u\,\partial^n_v\,\partial^p_w\,\hat{\Gamma}_{0100}| \le
\frac{m}{m + 2}(
2\,|\partial^{m - 1}_u\,\partial^n_v\,\partial^p_w\,\hat{\Gamma}_{0100}| 
+ \frac{1}{2}\,|\partial^{m - 1}_u\,\partial^n_v\,\partial^p_w\,s_0|)
\]
\[
\le
\frac{2\,m}{m + 2}\,C^3\,c_{\hat{\Gamma}_{0100}}^2\,
\frac{r^{m + p - 3}\,(m + p - 1) !\,\rho^n\,n!}
{m^2\,(n + 1)^2\,(p + 1)^2}
+ \frac{m}{2\,(m + 2)}\,c_0\,
\frac{r^{m + p -1}\,(m + p - 1) !\,\rho^n\,n!}{m^2\,(n + 1)^2\,(p + 1)^2}
\]
\[
\le
c_{\hat{\Gamma}_{0100}}\,
\frac{r^{m + p -1}\,(m + p) !\,\rho^n\,n!}{(m + 1)^2\,(n + 1)^2\,(p +
1)^2}\,A^{m \ge 1}_{\hat{\Gamma}_{0100}},
\]
with
\[
A^{m \ge 1}_{\hat{\Gamma}_{0100}} 
= 
\frac{1}{r^2}\,C^3\,c_{\hat{\Gamma}_{0100}}\,
\frac{2\,(m + 1)^2}{m\,(m + 2)\,(m + p)}
+ \frac{c_0}{c_{\hat{\Gamma}_{0100}}} \,
\frac{(m + 1)^2}{2\,m\,(m + 2)\,(m + p)}.
\]
Proceeding in a similar way with the equations for the other connection
coefficients one gets for the factors which need to be controlled 
the estimates
\[
A^{m \ge 1}_{\hat{\Gamma}_{0100}} \le 
\frac{c_0}{c_{\hat{\Gamma}_{0100}}}
+ \frac{4}{r^2}\,C^3\,c_{\hat{\Gamma}_{0100}},\quad\quad
A^{m \ge 1}_{\hat{\Gamma}_{0101}} \le 
\frac{c_1}{c_{\hat{\Gamma}_{0101}}}
+ \frac{4}{r^2}\,C^3\,c_{\hat{\Gamma}_{0100}},
\]
\[
A^{m \ge 1}_{\hat{\Gamma}_{0111}} \le 
\frac{c_2}{c_{\hat{\Gamma}_{0111}}}
+ \frac{4}{r^2}\,C^3\,c_{\hat{\Gamma}_{0100}},\quad\quad
A^{m \ge 1}_{\hat{\Gamma}_{1100}} \le 
\frac{2\,c_1}{c_{\hat{\Gamma}_{1100}}}
+ \frac{4}{r^2}\,C^3\,c_{\hat{\Gamma}_{0100}},
\]
\[
A^{m \ge 1}_{\hat{\Gamma}_{1101}} \le 
\frac{4\,c_2}{c_{\hat{\Gamma}_{1101}}}
+ \frac{8}{r^2}\,C^3\,
\frac{c_{\hat{\Gamma}_{1100}}c_{\hat{\Gamma}_{0101}}}
{c_{\hat{\Gamma}_{1101}}},\quad\quad
A^{m \ge 1}_{\hat{\Gamma}_{1111}} \le 
\frac{4\,c_3}{c_{\hat{\Gamma}_{1111}}}
+ \frac{8}{r^2}\,C^3\,
\frac{c_{\hat{\Gamma}_{1100}}c_{\hat{\Gamma}_{0111}}}
{c_{\hat{\Gamma}_{1111}}},
\]
The same inequalities, with $C^3$ replaced by $C^2$, are obtained in the case
$p = 0$. Being slightly more generous, one gets inequalities which can be
written in the concise form
\[
A^{m \ge 1}_{\hat{\Gamma}_{01AB}} \le 
\frac{c_{A + B}}{c_{\hat{\Gamma}_{01AB}}}
+ \frac{4}{r^2}\,C^3\,c_{\hat{\Gamma}_{0100}},
\quad
A^{m \ge 1}_{\hat{\Gamma}_{11AB}} \le 
\frac{4\,c_{A + B + 1}}{c_{\hat{\Gamma}_{11AB}}}
+ \frac{8}{r^2}\,C^3\,
\frac{c_{\hat{\Gamma}_{1100}}c_{\hat{\Gamma}_{01AB}}}
{c_{\hat{\Gamma}_{11AB}}},
\quad A, B = 0, 1,
\]
where the $c_{A + B}$, $c_{A + B + 1}$ denote for suitable numerical
values of the indices $A$, $B$ the constants $c_0, \ldots, c_4$.

\vspace{.2cm}

The analogous discussion of the equations 
\[
\partial_u\zeta = \zeta_{0},
\]
\[
\partial_u\,\zeta_{0} = - \zeta\,(1 - \mu\,\zeta)\,s_{0}, 
\]
\[
\partial_u\,\zeta_{1} = - \zeta\,(1 - \mu\,\zeta)\,s_{1}, 
\]
\[
\partial_u\,\zeta_{2} = 
- 2 + \hat{s} - \zeta\,(1 - \mu\,\zeta)\,s_{2}, 
\]
\[
\partial_u\,\hat{s} - (1 - \mu\,\zeta)\,
(s_{0}\,\zeta_{11} - 2\,s_{1}\,\zeta_{01} + s_{2}\,\zeta_{00}),
\]
does not require new considerations. For the factors which need to be
controlled we get the estimates
\[
A^{m \ge 1, p \ge 0}_{\zeta} \le \frac{4}{r}\,\frac{c_{\zeta_0}}{c_{\zeta}},
\]
\[
A^{m \ge 1, p \ge 0}_{\zeta_0} \le 
\frac{4}{r}\,C^3\,\frac{c_0\,c_{\zeta}}{c_{\zeta_0}}
+ \frac{4}{r^2}\,\mu\,C^6\,\frac{c_0\,c^2_{\zeta}}{c_{\zeta_0}},\quad\quad
A^{m \ge 1, p \ge 0}_{\zeta_1} \le 
\frac{4}{r}\,C^3\,\frac{c_1\,c_{\zeta}}{c_{\zeta_1}}
+ \frac{4}{r^2}\,\mu\,C^6\,\frac{c_0\,c^2_{\zeta}}{c_{\zeta_1}},
\]
\[
A^{m \ge 1, p \ge 0}_{\zeta_2}  \le
\left\{
\begin{array}{c}
\frac{8}{c_{\zeta_2}} + 
\frac{4}{r}\,
(\frac{c_{\hat{s}}}{c_{\zeta_2}} + C^3\,\frac{c_2\,c_{\zeta}}{c_{\zeta_2}})
+ \frac{4}{r^2}\,\mu\,C^6\,\frac{c_2\,c^2_{\zeta}}{c_{\zeta_2}}
\quad\mbox{for}\quad  m = 1, n = 0, p = 0,\\
\frac{4}{r}\,
(\frac{c_{\hat{s}}}{c_{\zeta_2}} + C^3\,\frac{c_2\,c_{\zeta}}{c_{\zeta_2}})
+ \frac{4}{r^2}\,\mu\,C^6\,\frac{c_2\,c^2_{\zeta}}{c_{\zeta_2}}
 \quad\quad\quad\quad\mbox{otherwise}\\  
\end{array}
\right\},
\]
\[
A^{m \ge 1}_{\hat{s}} \le 
(\frac{4}{r}\,C^3 + \frac{4}{r^2}\,\mu\,C^6\,c_{\zeta})
(\frac{c_0\,c_{\zeta_2}}{c_{\hat{s}}} 
+ 2\,\frac{c_{1}\,c_{\zeta_1}}{c_{\hat{s}}} 
+ \frac{c_{2}\,c_{\zeta_0}}{c_{\hat{s}}}).
\]

\vspace{.2cm}

We consider the $\partial_u$-equations for the curvature component $s_1$.
Multiplication with $2\,u$ gives
\[
2\,u\,\partial_u\,s_1 + 4\,s_1 
= \partial_v\,s_0
+ 2\,u\,\hat{e}^1\,_{01}\partial_us_0  
+ 2\,u\,\hat{e}^2\,_{01}\partial_vs_0 
\]
\[
- 8\,u\,(\hat{\Gamma}_{0101}\,s_0 - \hat{\Gamma}_{0100}\,s_1)
- u\,\frac{4\,\mu}{(1 - \mu\,\zeta)}\,
\left\{s_0\,\zeta_{1} - s_1\,\zeta_{0}\right\},
\]
which implies for $m \ge 1$
\[
|\partial^m_u\,\partial^n_v\,\partial^p_w\,s_1| \le
\frac{1}{2\,m + 4}\,|\partial^{m}_u\,\partial^{n+1}_v\,\partial^p_w\,s_0|
\]
\[
+ \frac{2\,m}{2\,m + 4}\,\left(|\partial^{m-1}_u\,\partial^n_v\,
\partial^{p}_w\,(\hat{e}^1\,_{01}\,\partial_u\,s_0)| 
+ |\partial^{m-1}_u\,\partial^n_v\,
\partial^p_w\,(\hat{e}^2\,_{01}\,\partial_v\,s_0)|\right) 
\]
\[
+ \frac{4\,m}{2\,m +
4}\,\left(2\,\,|\partial^{m-1}_u\,\partial^n_v\,\partial^p_w\,
(\hat{\Gamma}_{0101}\,s_0  - \hat{\Gamma}_{0100}\,s_1)| +
\mu\,|\partial^{m-1}_u\,\partial^n_v\,\partial^p_w\,\{\frac{1}{1 -
\mu\,\zeta}\,(s_0\,\zeta_{1} - s_1\,\zeta_{0})\}|\right).
\]
The terms arising here are estimated in a similar way
as the terms in the curvature equation above. Again the expansion types
allows one assumed that $0 \le n \le 2\,m +
4 - k$. Again $r$ is restricted to values with  
\[
r > \mu\,c_{\zeta}\,C^3.
\]
Proceeding similarly with the other $\partial_u$-equations for the curvature,
the following estimates are obtained for the factors which need to be
controlled.

\[
A^{m \ge 1}_{s_1} \le \frac{c_0}{c_1}\,\rho 
+ \frac{1}{r}\,C^3\,\frac{c_0}{c_1}\,c_{\hat{e}^1_{01}}
+ \frac{8\,\rho}{r^2}\,C^3\,\frac{c_0}{c_1}\,c_{\hat{e}^2_{01}}
+ \frac{8}{r^2}\,C^3\,
(\frac{c_0}{c_1}\,c_{\hat{\Gamma}_{0101}} + c_{\hat{\Gamma}_{0100}})
\]
\[
+ \frac{1}{r^2}\,C^3\,\frac{4\,\mu}{1 - \frac{\mu\,c_{\zeta}\,C^3}{r}}
(\frac{c_0}{c_1}\,c_{\zeta_1} + c_{\zeta_0})
\]

\[
A^{m \ge 1}_{s_2} \le \frac{c_1}{c_2}\,\rho 
+ \frac{1}{r}\,C^3\,\frac{c_1}{c_2}\,c_{\hat{e}^1_{01}}
+ \frac{8\,\rho}{r^2}\,C^3\,\frac{c_1}{c_2}\,c_{\hat{e}^2_{01}}
+ \frac{4}{r^2}\,C^3\,
(\frac{c_0}{c_2}\,c_{\hat{\Gamma}_{0111}}
+ 2\,\frac{c_1}{c_2}\,c_{\hat{\Gamma}_{0101}} + 3\,c_{\hat{\Gamma}_{0100}})
\]
\[
+ \frac{1}{r^2}\,C^3\,\frac{2\,\mu}{1 - \frac{\mu\,c_{\zeta}\,C^3}{r}}
(\frac{c_0}{c_2}\,c_{\zeta_2} + 2\,\frac{c_1}{c_2}\,c_{\zeta_1} +
3\,c_{\zeta_0})
\]

\[
A^{m \ge 1}_{s_3} \le \frac{c_2}{c_3}\,\rho 
+ \frac{1}{r}\,C^3\,\frac{c_2}{c_3}\,c_{\hat{e}^1_{01}}
+ \frac{8\,\rho}{r^2}\,C^3\,\frac{c_2}{c_3}\,c_{\hat{e}^2_{01}}
+ \frac{8}{r^2}\,C^3\,
(\frac{c_1}{c_3}\,c_{\hat{\Gamma}_{0111}}
+ c_{\hat{\Gamma}_{0100}})
\]
\[
+ \frac{1}{r^2}\,C^3\,\frac{4\,\mu}{1 - \frac{\mu\,c_{\zeta}\,C^3}{r}}
(\frac{c_1}{c_3}\,c_{\zeta_2} + c_{\zeta_0})
\]

\[
A^{m \ge 1}_{s_3} \le \frac{c_3}{c_4}\,\rho 
+ \frac{1}{r}\,C^3\,\frac{c_3}{c_4}\,c_{\hat{e}^1_{01}}
+ \frac{8\,\rho}{r^2}\,C^3\,\frac{c_3}{c_4}\,c_{\hat{e}^2_{01}}
+ \frac{4}{r^2}\,C^3\,
(3\,\frac{c_2}{c_4}\,c_{\hat{\Gamma}_{0111}}
+ 2\,\frac{c_3}{c_4}\,c_{\hat{\Gamma}_{0101}} + c_{\hat{\Gamma}_{0100}})
\]
\[
+ \frac{1}{r^2}\,C^3\,\frac{2\,\mu}{1 - \frac{\mu\,c_{\zeta}\,C^3}{r}}
(3\,\frac{c_2}{c_4}\,c_{\zeta_2} + 2\,\frac{c_3}{c_4}\,c_{\zeta_1} +
c_{\zeta_0}).
\]
This gives all the needed information.

\vspace{.5cm}

To arrange now the constants so that the induction argument can
successfully carried out, we proceed as follows. The estimates for the
decisive factors which have been obtained above are of the general form
\[
A \le \alpha + \frac{1}{r}\,\beta + \frac{1}{r^2}\,\gamma,
\]
with $\alpha$, $\beta$, and $\gamma$ depending on all the constants except 
$r$. If
$\beta = 0$ and $\gamma = 0$ it suffices to ensure $\alpha \le 1$.  In the
other cases we require
$\alpha
\le a$ where $a$ is a given constant, $a < 1$, and then choose $r$ large
enough so that $A \le 1$.
A first set of conditions arising this way reads 

\[
\frac{c_k}{c_{k + 1}}\,\rho \le a,\quad
\frac{c_0}{c_k}\,\rho^k \le 1,\quad
\frac{c_2}{c_k}\,\rho^k \le a,\quad
4\,\frac{c_2}{c_0} \le a.
\]
These conditions can be satisfied simultaneously. The first equation implies 
$c_k \ge  (\rho/a)^k\,c_0$. With
\[
c_k = (\frac{\rho}{a})^k\,c^*_0,
\]
where $0 < \rho, \,a < 1$, the first two relations hold true, the
forth relation implies $\rho^2 \le a^3/4$ and with this restriction the third
relation holds as well. We choose 
\[
\rho = \rho_0, \quad
a = (4\,\rho^2_0)^{1/3}.
\]
The conditions
\[
\frac{2}{c_{\zeta_0}} \le 1, \quad\quad\frac{8}{c_{\zeta_2}} \le a,
\]
are met by setting 
\[
c_{\zeta_0} \equiv 2, \quad\quad
c_{\zeta_2} \equiv \frac{8}{a}.
\]
The conditions
\[
\frac{c_{A + B}}{c_{\hat{\Gamma}_{01AB}}} \le a,\quad
\frac{4\,c_{1 + A + B}}{c_{\hat{\Gamma}_{11AB}}} \le a,
\quad\quad A, B = 0, 1,
\]
are then dealt with by setting 
\[
c_{\hat{\Gamma}_{01AB}} \equiv \frac{1}{a}\,c_{A + B},\quad\quad
c_{\hat{\Gamma}_{11AB}} \equiv \frac{1}{a}\,c_{1 + A + B}.
\]
The conditions
\[
\frac{c_{\hat{\Gamma}_{0100}}}{c_{\hat{e}^2_{01}}} \le a,
\quad
\frac{c_{\hat{\Gamma}_{1100}}}{c_{\hat{e}^2_{11}}} \le a,
\]
are satisfied by setting
\[
c_{\hat{e}^2_{01}} \equiv \frac{1}{a}\,c_{\hat{\Gamma}_{0100}}, \quad\quad
c_{\hat{e}^2_{11}} \equiv \frac{1}{a}\,c_{\hat{\Gamma}_{1100}}.
\]
After this we choose some positive constants
\[
\hat{e}^1_{01},\quad \hat{e}^1_{11}, \quad
c_{\zeta}, \quad c_{\zeta_1},\quad c_{\hat{s}}.
\]
That these constants are not further restricted by the procedure
reflects the fact that the corresponding functions vanish to higher order at
$O$. Their choice affects, however, the value of the constant $r$.
After all constants except $r$ have been fixed we can choose $r$ so large
that 
\[
r >  \max\{r_0, \,\,\mu\,c_{\zeta}\,C^3\},
\] 
and that all the $A$'s are $\le 1$. This finishes the induction proof.
$\Box$

\vspace{.2cm}

The following statement of the convergence result, obtained by using the 
$v$-finite expansion types of the various functions,  emphasizes the role of 
$v$ as an angular coordinate.

\begin{lemma}
\label{extconv}
The estimates (\ref{skest}) and (\ref{fest}) for the derivatives of the functions
$s_k$ and $f$ and the expansion types given in Lemma \ref{formexptype} 
imply that the associated Taylor series are
absolutely convergent in the domain $|v| < \frac{1}{\alpha\,\rho}$,
$|u| + |w| < \frac{\alpha^2}{r}$, for any real number $\alpha$, 
$0 < \alpha \le 1$. It follows that the formal expansion determined in Lemma 
\ref{exunformexp} defines indeed a (unique) holomorphic solution to the conformal
static vacuum field equations which induces the datum $s_0$ on $W_0$.
\end{lemma}

\begin{proof}
The estimates (\ref{skest}) and (\ref{fest}) imply
\[
|\partial^m_u\,\partial^n_v\,\partial^p_w\,s_k(O)|
\le \frac{c_k}{\alpha^{4 - k}}\,
\frac{(r/\alpha^2)^{m + p}\,(m + p)!\,(\alpha\,\rho)^n\,n!}
{(m + 1)^2\,(n + 1)^2\,(p + 1)^2}\,\alpha^{4 - k + 2m + 2p - n}
\]
\[
\le \frac{c_k}{\alpha^{4 - k}}\,
\frac{(r/\alpha^2)^{m + p}\,(m + p)!\,(\alpha\,\rho)^n\,n!}
{(m + 1)^2\,(n + 1)^2\,(p + 1)^2}
\quad \mbox{for} \quad n \le 2\,m + 4 - k,\,\,\,\,m, p = 0, 1, 2, \ldots
\]

\[
|\partial^m_u\,\partial^n_v\,\partial^p_w\,f(O)|
\le \frac{c_{f}}{\alpha^{k_f - 2}}\,
\frac{(r/\alpha^2)^{m + p - 1}\,(m + p)!\,(\alpha\,\rho)^n\,n!}
{(m + 1)^2\,(n + 1)^2\,(p + 1)^2}\,\alpha^{k_f + 2m + 2p - n}
\]
\[
\le \frac{c_{f}}{\alpha^{k_f - 2}}\,
\frac{(r/\alpha^2)^{m + p - 1}\,(m + p)!\,(\alpha\,\rho)^n\,n!}
{(m + 1)^2\,(n + 1)^2\,(p + 1)^2}
\quad \mbox{for} \quad n \le 2\,m + k_f,\,\,\,\,m, p = 0, 1, 2, \ldots\,.
\]
Since the other derivatives vanish because of the respective expansion types,
the first assertion is an immediate consequence of the majorizations
(\ref{Amajorant}), (\ref{Bmajorant}). The second assertion then follows with Lemma
\ref{solvesallequonShat}.
\end{proof}

\section{Analyticity at space-like infinity}
\label{analyticati}

Due to our singular gauge the holomorphic solution of the
conformal static field equations obtained in Lemma \ref{extconv} does not
cover a full neighbourhood of the point $i$. To analyse the situation we
study the part of the solution which we have obtained by the convergence
proof in terms of a normal frame based on the frame $c_{AB}$ at $i$
and associated normal coordinates. We write the
geodesic equation 
$D_{\dot{z}}\dot{z} = 0$ for $z^a(s) = (u(s), v, (s), w(s))$ in the
form
\[
\dot{z}^a = m^{AB}\,e^a_{AB} =  m^{AB}\,(e^{*a}_{AB} + \hat{e}^a_{AB}),
\]
\[
\dot{m}^{AB} = - 2\,m^{CD}\Gamma_{CD}\,^{(A}\,_B\,m^{B)E}
= - 2\,m^{CD}\Gamma^*_{CD}\,^{(A}\,_B\,m^{B)E}
- 2\,m^{CD}\hat{\Gamma}_{CD}\,^{(A}\,_B\,m^{B)E},
\]
With the explicit expressions for the singular parts,
the system takes the form
\[
\dot{u} = \,\,\,\,\,m^{00} + m^{AB}\,\hat{e}^1_{AB},
\quad \quad \quad
\dot{m}^{00} =
- 2\,m^{CD}\hat{\Gamma}_{CD}\,^{0}\,_B\,m^{0B},
\quad \quad \quad\quad \quad \quad\,\,\,
\]
\[
\dot{v} = \frac{1}{u}\,m^{01} + m^{AB}\,\hat{e}^2_{AB},
\quad  \quad  \quad
\dot{m}^{01} = - \frac{1}{u}\,m^{01}\,m^{00}
- 2\,m^{CD}\hat{\Gamma}_{CD}\,^{(0}\,_B\,m^{1)B},
\]
\[
\dot{w} = m^{11},
\quad  \quad \quad  \quad \quad \quad \quad\quad \quad\,\,
\dot{m}^{11} =
- \frac{2}{u}\,m^{01}\,m^{01}
- 2\,m^{CD}\hat{\Gamma}_{CD}\,^{1}\,_B\,m^{1B}.
\,\,\, 
\]
These equations have to be solved with the initial conditions 
\begin{equation}
\label{uwindat}
u|_{s = 0} = 0,\,\,\,\,\,
w|_{s = 0} = 0,
\end{equation}
for the curves to start at $i$. An arbitrary value 
\begin{equation}
\label{vindat}
v_0 = v|_{s = 0},
\end{equation}
can be prescribed to determine the $\partial_u$-$\partial_w$-plane over
$i$ in which the tangent vector is lying, and an arbitrary choice of 
\[
m^{AB}|_{s = 0} = m^{AB}_0 = m^{AB}_0\,\epsilon_0\,^A\,\epsilon_0\,^B
+ m^{AB}_0\,\epsilon_1\,^A\,\epsilon_1\,^B,
\,\,\,\,\,\,\,\,\,\,\,\,\,\,\,\,\dot{u}_0 \neq 0,
\]
can be prescribed to specify the tangent vector in the
$\partial_u$-$\partial_w$-plane. Regularity and the equations require  
\begin{equation}
\label{mindat}
m^{00}_0 = \dot{u}|_{s = 0} \equiv \dot{u}_0,\,\,\,\,\,\,
m^{01}_0 = 0,\,\,\,\,\,\,
m^{11}_0 = \dot{w}|_{s = 0} \equiv \dot{w}_0.
\end{equation}
If the frame $e_{AB}$ at a point of $I$ is identified with its projection
into $T_iS_c$, then
\[
m^{AB}_0\,e_{AB} = m^{AB}_0\,s^C\,_A(v_0)\,s^D\,_B(v_0)\,c_{CD}
= m^{*AB}\,c_{AB}, 
\]
holds at $i$ with 
\[
m^{*00} = \dot{u}_0, \quad
m^{*01} = \dot{u}_0\,v_0,\quad
m^{*11} 
= \dot{u}_0\,v^2_0 + \dot{w}_0,
\,\,\,\,\,\,\,\,\,\,\,\,\,\,\,\,\dot{u}_0 \neq 0.
\]
For arbitrarily given $m^{*AB} \in C^3$ with  $m^{*00} \neq 0$ this relation
determines $\dot{u}_0$, $v_0$, $\dot{w}_0$ uniquely.
Using $c_{AB} = \alpha^a\,_{AB}\,c_{\bf a}$, the tangent vectors can be
written $m^{*AB}\,c_{AB} = x^a\,c_{\bf a}$ with
\begin{equation}
\label{xuvwdep}
x^1 = \frac{1}{\sqrt{2}}\,(\dot{w}_0 + (v_0^2 - 1)\,\dot{u}_0),\,\,\,\,\,
x^2 = \frac{i}{\sqrt{2}}\,(\dot{w}_0 + (v_0^2 + 1)\,\dot{u}_0),\,\,\,\,\,
x^3 = \sqrt{2}\,v_0\,\dot{u}_0
\quad \quad \dot{u}_0 \neq 0,
\end{equation}
or, equivalently, 
\begin{equation}
\label{uvwxdep}
\dot{u}_0(x^a) =  -  \frac{x^1 + i\,x^2}{\sqrt{2}},\,\,\,\,\,
v_0(x^a) =  - \frac{x^3}{x^1 + i\,x^2},\,\,\,\,\,\,
\dot{w}_0(x^a) = 
\frac{\delta_{ab}\,x^a\,x^b}{\sqrt{2}\,(x^1 + i\,x^2)},
\quad x^1 + i\,x^2 \neq 0.
\end{equation}
The vectors $x^a c_{\bf a}$ cover all directions at $i$ except those tangent
to the  complex null hyperplane $(c_{\bf 1} + i\,c_{\bf 2})^{\perp} =
\{a(c_{\bf 1} + i\,c_{\bf 2}) + b\,c_{\bf 3}|\,\,a, b \in \mathbb{C}\}$.

To determine the normal frame centered at $i$
and based on the frame $c_{AB}$ at $i$, we write  the equation
$D_{\dot{x}}c_{AB} = 0$ for the normal frame as an equation for the transformation
$t^A\,_B \in SL(2, \mathbb{C})$, which relates the frame
$e_{AB}$ to the normal frame $c_{AB} = t^C\,_A\,t^D\,_B\,e_{CD}$. The
resulting equation
\[
0 = \frac{d}{ds}(t^C\,_A\,t^D\,_B)
+ m^{GH}\,\Gamma_{GH}\,^{CD}\,_{EF}\,t^E\,_A\,t^F\,_B,
\]
can be written in the form $\dot{t}^A\,_B = -
m^{DE}\Gamma_{DE}\,^A\,_C\,t^C\,_B$. Taking into account the
structure of the connection coefficients, this gives  
\begin{equation}
\label{tevol}
\dot{t}^A\,_B =
- \frac{1}{u}\,m^{01}\,\epsilon_1\,^A\,t^0\,_B
- m^{DE}\,\hat{\Gamma}_{DE}\,^A\,_C\,t^C\,_B.
\end{equation}
This equation has to be solved along $z(s)$ with the initial condition
\begin{equation}
\label{tevoldata}
t^A\,_B|_{s = 0} = s^A\,_B(-v_0).
\end{equation}

The initial value problems above make sense because the functions
$\hat{e}^a\,_{AB}$ and $\hat{\Gamma}_{ABCD}$ are,  by Lemma \ref{extconv}, 
holomorphic near  the point $u = 0$,
$v = v_0$, $w = 0$ for any prescribed value of $v_0$.
The singularity of the system at that
particular point requires, however, some attention. 

We prepare the statement and the proof of the existence result, to be given
in Lemma \ref{geodexist}, by casting the system of ODE's into a suitable
form. It will be convenient to make use of the {\it replacements resp.
change of notation}
\begin{equation}
\label{vminusv0etc}
v \rightarrow v_0 + v,\,\,\,
m^{AB} \rightarrow m_0^{AB} + m^{AB},
\end{equation}
so that all unknowns vanish at $s = 0$.
Furthermore, by setting
\[
\tilde{e}^a_{AB}(u, v, w) = \hat{e}^a_{AB}(u, v_0 + v, w),
\quad \quad
\tilde{\Gamma}_{ABCD}(u, v, w) = \hat{\Gamma}_{ABCD}(u, v_0 + v, w),
\]
we define functions $\tilde{e}^a_{AB}$, $\tilde{\Gamma}_{ABCD}$
of the new unknowns which are holomorphic near $u = v = w = 0$. 
The regular equations read with this notation 
\[
\dot{u} = \dot{u}_0 + m^{00} 
+ \dot{w}_0\,\tilde{e}^1_{11}
+ 2\,\tilde{e}^1_{01}\,m^{01}
+ \tilde{e}^1_{11}\,m^{11},
\]

\[
\dot{w} = \dot{w}_0 + m^{11},
\]

\[
\dot{m}^{00} =
- 2\left\{\dot{u}_0\,\dot{w}_0\,\tilde{\Gamma}_{1101}
+  \dot{u}_0\,(2\,\tilde{\Gamma}_{0101}\,m^{01}
+ \tilde{\Gamma}_{1101}\,m^{11})
+ \dot{w}_0\,(\tilde{\Gamma}_{1101}\,m^{00}
+ \tilde{\Gamma}_{1111}\,m^{01})\right.
\]
\[
\left. + 2\tilde{\Gamma}_{0101}\,m^{00}\,m^{01}
+ 2\tilde{\Gamma}_{0111}\,m^{01}\,m^{01}
+ \tilde{\Gamma}_{1101}\,m^{00}\,m^{11}
+ \tilde{\Gamma}_{1111}\,m^{01}\,m^{11}\right\}
\]
The singular equations take the form
\[
u\,\dot{v} = m^{01} + u\,(
\dot{w}_0\,\tilde{e}^2_{AB} 
+ 2\,\tilde{e}^2_{01}\,m^{01}
+ \tilde{e}^2_{11}\, m^{11})
\]

\[
u\,\dot{m}^{01} =  - \dot{u}_0\,m^{01} - m^{00}\,m^{01}
+ u\left\{
\dot{u}_0\,\dot{w}_0\,\tilde{\Gamma}_{1100}
- \dot{w}_0^2\,\tilde{\Gamma}_{1111}
\right.
\]
\[
+ \dot{u}_0\,(2\,\tilde{\Gamma}_{0100}\,m^{01}
+ \tilde{\Gamma}_{1100}\,m^{11})
+ \dot{w}_0\,(\tilde{\Gamma}_{1100}\,m^{00}
- 2\,\tilde{\Gamma}_{0111}\,m^{01}
- 2\,\tilde{\Gamma}_{1111}\,m^{11})
\]
\[
\left. 
+ 2\,\tilde{\Gamma}_{0100}\,m^{00}\,m^{01}
- 2\,\tilde{\Gamma}_{0111}\,m^{01}\,m^{11}
+ \tilde{\Gamma}_{1100}\,m^{00}\,m^{11}
- \tilde{\Gamma}_{1111}\,m^{11}\,m^{11}\right\},
\]

\[
u\,\dot{m}^{11} =
- 2\,m^{01}\,m^{01}
+ 2\,u\left\{\dot{w}_0^2\,\tilde{\Gamma}_{1101}
+ \dot{w}_0\,(2\,\tilde{\Gamma}_{0101}\,m^{01}
+ \tilde{\Gamma}_{1100}\,m^{01}
+ 2\,\tilde{\Gamma}_{1101}\,m^{11})\right.
\]
\[
\left.
+ 2\,\tilde{\Gamma}_{0100}\,m^{01}\,m^{01}
+ 2\,\tilde{\Gamma}_{0101}\,m^{01}\,m^{11}
+ \tilde{\Gamma}_{0100}\,m^{01}\,m^{11}
+ \tilde{\Gamma}_{1101}\,m^{11}\,m^{11}
\right\}.
\]
Finally, equation (\ref{tevol}) reads
\begin{equation}
\label{0dattevol}
\dot{t}^A\,_B =
- \frac{1}{u}\,m^{01}\,\epsilon_1\,^A\,t^0\,_B
- (2\,m^{01}\,\hat{\Gamma}_{01}\,^A\,_C
+ \dot{w}_0\,\hat{\Gamma}_{11}\,^A\,_C
+ m^{11}\,\hat{\Gamma}_{11}\,^A\,_C)
\,t^C\,_B.
\end{equation}

After applying $\partial_s$ resp. $\partial_s^2$ to the geodesic equations
and restricting all equations to $s = 0$ one obtains with the initial
conditions
(\ref{uwindat}), (\ref{vindat}), (\ref{mindat})  the relations
\begin{equation}
\label{1indat}
\dot{v}|_{s = 0} = 0,\,\,\,\,\,\,\,\,
\dot{m}^{AB}|_{s = 0} = 0,\,\,\,\,\,\,\,\,
\ddot{u}|_{s = 0} = 0,
\end{equation}
and, by taking a further derivative,
\[
\partial^3_su(0) =
\dot{u}^2_0\,\dot{w}_0\,
\left\{\partial^2_u\,\hat{e}^1\,_{11}
- 2\,\partial_u\,\hat{\Gamma}_{1101} 
\right\}_{u = 0, v = v_0, w = 0}.
\]
This gives with the $\partial_u$-equations
\begin{equation}
\label{3indat}
\partial^3_su(0) =
- 4\,\dot{u}^2_0\,\dot{w}_0\,(s_2)_{u = 0, v = v_0, w = 0}
= - \frac{1}{3}
\dot{u}^2_0\,\dot{w}_0\,(\partial^2_vs_0)_{u = 0, v = v_0, w = 0},
\end{equation}
which can be determined from the null data.

Because of Lemma \ref{extconv} and the behaviour (\ref{easbeh}), (\ref{GammaOu})
of the metric and the connection coefficients, which follows from the
$\partial_u$-equations,
there exist functions $f$, $g$, $h$, $k$, $l$ which are holomorphic on 
a polycylinder $P_{\epsilon'} = \{x \in \mathbb{C}^6|\,\, |x_j| < \epsilon'\}$
with some $\epsilon' > 0$ so that the equations above can be
written 
\begin{eqnarray}
\label{1modgoedequ}
\dot{u} & = & \dot{u}_0 + m^{00} + u^2\,f,\\
\label{2modgoedequ}
u\,\dot{v} & = & m^{01} + u^2\,g,\\
\label{3modgoedequ}
\dot{w} & = & \dot{w}_0 + m^{11},\\
\label{4modgoedequ}
\dot{m}^{00} & = & u\,h,\\
\label{5modgoedequ}
u\,\dot{m}^{01} & = & - \dot{u}_0\,m^{01} - m^{00}\,m^{01} + u^2\,k,\\
\label{6modgoedequ}
u\,\dot{m}^{11} & = & - 2\,m^{01}\,m^{01} + u^2\,l,
\end{eqnarray}
with $f$, $g$, $h$, $k$, $l$ depending on the $\mathbb{C}^6$-valued function
$z(s)$ comprising our unknowns in the form
\[
z(s) = (z^j(s))_{j = 1, \ldots , 6} = 
(u(s), \,v(s), \,w(s), \,m^{00}(s), \,m^{01}(s), \,m^{11}(s)),
\]
(which agrees after the replacement $v \rightarrow v - v_0$ in the first
$3$ components with the notation introduced earlier).
 
If $F$ stands for any of the functions  $f$, $g$, $h$, $k$,
$l$, then it has on $P_{\epsilon'}$ an absolutely convergent expansion
\[
F = \sum_{\alpha \in N^6} F_{\alpha}\,z^{\alpha},
\]
at $z^j = 0$, where again the multi-index notation is used. If  
$0 < \epsilon < \epsilon'$, there exists thus an $M > 0$ so that 
\[
\sup_{x \in P_{\epsilon}} 
\sum_{\alpha} |F_{\alpha}|\,|z^{\alpha}| \le M.
\]

\begin{lemma}
\label{Festimates}
Let $p \ge0 $ be an integer and $c$ and $t$ real numbers which
satisfy  with the constant $C$ of Lemma \ref{basicestimate}
\begin{equation}
\label{1estimate}
c \ge \frac{M}{C}, \quad \quad
t \ge \max \,\{1, \frac{c\,C}{\epsilon}\}.
\end{equation}
If the derivatives of the functions $z^j(s)$ at $s = 0$ exist and statisfy
the estimates 
\[
|\partial^k_sz^j| \le c\,\frac{t^{k - 1}\,k\,!}{(k + 1)^2},
\quad  k = 1, \ldots , 6, \quad \quad k \le p,
\]
then 
\[
|\partial^p_s F(z(s))|_{s = 0} \le c\,\frac{t^{p}\,p\,!}{(p + 1)^2}.
\]
If, in addition, $u$ satisfies $u(0) = 0$, $\dot{u}(0) = \dot{u}_0$ and  
\[
|\partial^k_su(s)|_{s = 0} \le c\,\frac{t^{k - 2}\,k\,!}{(k + 1)^2},
\quad \quad 2 \le k \le p,
\]
then
\[
|\partial^p_s (u\,F(z(s)))|_{s = 0} \le 
|\dot{u}_0|\,c\,\frac{t^{p - 1}\,p\,!}{p^2}
+ c^2\,C\,\frac{t^{p -  2}\,p\,!}{(p + 1)^2},
\]
for $p \ge 1$, where the second term on the right hand side is to be
dropped if $p < 2$, and
\[
|\partial^p_s (u^2\,F(z(s)))|_{s = 0} \le 
2\,|\dot{u}_0|^2\,c\,\frac{t^{p - 2}\,p\,!}{(p - 1)^2}
+ 4\,|\dot{u}_0|\,c^2\,C\,\frac{t^{p - 3}\,p\,!}{(p + 1)^2}
+ c^3\,C^2\,\frac{t^{p -  4}\,p\,!}{(p + 1)^2},
\]
for $p \ge 2$, where the second term on the right hand side is to be
dropped if $p < 3$ and the third term is to be dropped if $p < 4$.
\end{lemma}

In the following a function in the modulus sign has argument $s = 0$. 

\vspace{.2cm}

\begin{proof}
Observing Lemma \ref{prodestimate} and the subsequent remark, one gets
\[
|\partial^p_s F(z)| \le
\sum_{|\alpha| \le p} |F_{\alpha}|| \partial^p_s z^{\alpha}|
\le 
\sum_{|\alpha| \le p} |F_{\alpha}|\,C^{|\alpha| - 1}\,c^{|\alpha|}
\frac{t^{p - |\alpha|}\,p\,!}{(p + 1)^2}
\]
\[
\le 
\frac{1}{c\,C}
\sum_{|\alpha| \le p} |F_{\alpha}|\,
\left(\frac{c\,C}{t}\right)^{|\alpha|}
\,c\,\frac{t^{p}\,p\,!}{(p + 1)^2}
\le 
\frac{M}{c\,C}\,
\,c\,\frac{t^{p}\,p\,!}{(p + 1)^2}
\le 
c\,\frac{t^{p}\,p\,!}{(p + 1)^2},
\]
by the choice of $c$ and $t$. With Lemma \ref{basicestimate} this gives
\[
|\partial^p_s (u\,F(z))| 
\le
p\,|\dot{u}_0|\,|\partial^{p - 1}_s F(z)|
+ \sum_{j = 2}^p {p \choose j} |\partial^j_su|\,|\partial^{p - j}_s F(z)|
\]
\[
\le p\,|\dot{u}_0|\,c\,\frac{t^{p - 1}\,(p - 1)!}{p^2}
+ \sum_{j = 2}^p {p \choose j} 
c\,\frac{t^{j - 2}\,(j)!}{(j + 1)^2}\,
c\,\frac{t^{p - j}\,(p - j)!}{(p - j + 1)^2}
\le
|\dot{u}_0|\,c\,\frac{t^{p - 1}\,p\,!}{p^2}
+ c^2\,C\,\frac{t^{p -  2}\,p\,!}{(p + 1)^2},
\]
and similarly
\[
|\partial^p_s (u^2\,F(z))| \le 
\sum_{l = 0}^p {p \choose l}
\,\sum_{j = 0}^l {l \choose j}\,|\partial^j_su|\,|\partial^{l - j}_su|\,
|\partial^{p - l}_sF(z)|
\]
\[
= 4\,{p \choose 2}\,|\dot{u}_0|^2\,|\partial^{p - 2}_sF(z)|
+ \sum_{l = 3}^p {p \choose l}\,2\,l\,|\dot{u}_0|
\,|\partial^{l - 1}_su|\,|\partial^{p - l}_sF(z)|
\]
\[
+ \sum_{l = 2}^p {p \choose l}
\,\sum_{j = 2}^{l - 2} {l \choose j}\,|\partial^j_su|\,|\partial^{l -
j}_su|\, |\partial^{p - l}_sF(z)|
\]
\[
\le 2\,|\dot{u}_0|^2\,c\,\frac{t^{p - 2}\,p\,!}{(p - 1)^2}
+ 4\,|\dot{u}_0|\,c^2\,C\,\frac{t^{p - 3}\,p\,!}{(p + 1)^2}
+ c^3\,C^2\,\frac{t^{p -  4}\,p\,!}{(p + 1)^2},
\]
\end{proof}

\begin{lemma}
\label{geodexist}
The requirement that $z(s)$ be a holomorphic solution of equations 
(\ref{1modgoedequ}) - (\ref{6modgoedequ}) near $s = 0$ satisfying $x(0) = 0$ and
$\partial_su(0) = \dot{u}_0 \neq 0$ determines a unique formal expansion of
$z(s)$ at $s = 0$. 
There exist real constants $c$ and $t$ satisfying 
\begin{equation}
\label{2estimate}
c \ge \max\,\{4\,|\dot{u}_0|, 4\,|\dot{w}_0|, \,
|\dot{u}_0|^2\,|\,\dot{w}_0|\,|(\partial_v^2s_0)_{u = 0, v = v_0, w = 0}|,
\,\frac{M}{C}\}, 
\quad \quad
t \ge \max \,\{1, \frac{c\,C}{\epsilon}\},
\end{equation}
with $C$ the constant of Lemma \ref{basicestimate}, so that the Taylor
coefficients of $z(s)$ at $s = 0$ satisfy the estimates
\begin{equation}
\label{xqest}
|\partial^q_sz^j| \le c\,\frac{t^{q - 1}\,q\,!}{(q + 1)^2},
\quad \quad q = 0, 1, 2, \ldots\,,
\end{equation}
and the Taylor coefficients of $u(s)$ at $s = 0$ satisfy in addition the
estimates
\begin{equation}
\label{uqest}
|\partial^{q + 2}_su| \le c\,\frac{t^{q}\,(q+2)!}{(q + 3)^2},
\quad \quad q = 0, 1, 2, \ldots\,.
\end{equation}

It follows that for any given initial data  $\dot{u}_0$, $v_0$, $\dot{w}_0$
with $\dot{u}_0 \neq 0$  there exists a number 
$t = t(\dot{u}_0, v_0,\dot{w}_0)$ and a unique holomorphic solutions 
$z^j(s) = z^j(s, \dot{u}_0, v_0,\dot{w}_0)$ of the initial value problem
for the geodesic equations with initial data as described above which is
defined for $|s| \le 1/t$. The functions 
$z^j(s, \dot{u}_0, v_0,\dot{w}_0)$ are in fact holomorphic
functions of all four variables in a certain domain.
\end{lemma}

\begin{proof}
The existence of a unique formal expansion follows immediately by applying
$\partial^p_s$ for $p = 1, 2, 3, \ldots$ formally to equations
(\ref{1modgoedequ}) - (\ref{6modgoedequ}), restricting to $s = 0$, and observing
$\dot{u}_0 \neq 0$ and the initial data. 

That the estimates (\ref{xqest}) hold for $q = 0, 1$
follows from the initial condition $x(0) = 0$, the equations at $s = 0$ and
our conditions on $c$ and $t$. 
That the estimates (\ref{uqest}) hold for $q = 0, 1$ follows from (\ref{1indat}),
(\ref{3indat}), and our conditions on $c$ and $t$.

Let $p \ge 1$ be an integer. We show that $c$ and $t$ can be chosen such
that the estimates
(\ref{xqest}), (\ref{uqest}) for $q \le p$ imply with the equations the
corresponding estimates for $p +1$.
From equation (\ref{4modgoedequ}) and Lemma \ref{Festimates} (with the
provisos given there not repeated here) follows 
\[
|\partial^{p + 1}_s\,m^{00}| = |\partial^{p}_s\,(u\,h)|
\le
|\dot{u}_0|\,c\,\frac{t^{p - 1}\,p\,!}{p^2}
+ c^2\,C\,\frac{t^{p -  2}\,p\,!}{(p + 1)^2}
\le A_{00}\,c\,\frac{t^p\,(p + 1)\,!}{(p + 2)^2},
\]
with 
\[
A_{00} = 
\frac{1}{t}\,|\dot{u}_0|\,\frac{p\,!}{p^2}
\,\frac{(p + 2)^2}{(p + 1)\,!}   
+ \frac{1}{t^2}\,c\,C\,\frac{p\,!}{(p + 1)^2}
\,\frac{(p + 2)^2}{(p + 1)\,!} \le
\frac{5}{t}\,|\dot{u}_0|  
+ \frac{2}{t^2}\,c\,C.
\]
Similarly one gets from (\ref{1modgoedequ})
\[
|\partial^{p + 2}_s\,u| \le |\partial^{p + 1}_s\,m^{00}| +
|\partial^{p + 1}_s\,(u^2\,f)|
\le A_{m^{00}}\,\,c\,\frac{t^p\,(p + 1)\,!}{(p + 2)^2}
\]
\[
+ 2\,|\dot{u}_0|^2\,c\,\frac{t^{p - 1}\,(p+1)\,!}{p^2}
+ 4\,|\dot{u}_0|\,c^2\,C\,\frac{t^{p - 2}\,(p + 1)\,!}{(p + 2)^2}
+ c^3\,C^2\,\frac{t^{p -  3}\,(p + 1)\,!}{(p + 2)^2}
\le A_u\,c\,\frac{t^{p}\,(p+2)!}{(p + 3)^2},
\]
with 
\[
A_u = A_{00}\,\frac{(p + 1)\,!}{(p + 2)^2}\,\frac{(p + 3)^2}{(p + 2)\,!}
+ \frac{2}{t}\,|\dot{u}_0|^2\,\frac{(p+3)^2}{p^2(p+2)}
+ \frac{4}{t^2}\,|\dot{u}_0|\,c\,C\,\frac{(p + 3)^2}{(p + 2)^3} 
+ \frac{1}{t^3}\,c^2 \,C^2\,\frac{(p + 3)^2}{(p + 2)^3}
\]
\[
\le \frac{3}{t}\,|\dot{u}_0|(1 + 4\,|\dot{u}_0|)
+ \frac{1}{t^2}\,c\,C\,(1 + 4\,|\dot{u}_0|)
+ \frac{1}{t^3}\,c^2\,C^2,
\]
and from (\ref{3modgoedequ})
\[
|\partial^{p + 1}_s\,w| = |\partial^{p}_s\,m^{11}|
\le c\,\frac{t^{p - 1}\,p\,!}{(p + 1)^2} 
\le A_w\,c\,\frac{t^{p}\,(p + 1)\,!}{(p + 2)^2},
\]
with 
\[
A_w = \frac{1}{t}\,\frac{(p + 2)^2}{(p + 1)^3} \le \frac{2}{t}.
\]
Applying $\partial_s^{p + 1}$ to equation (\ref{5modgoedequ}) and observing
the initial conditions, gives at $s = 0$ for $p \ge 1$
\[
(p + 2)\,\dot{u}_0\,\partial_s^{p + 1}m^{01} =
- \sum_{j = 2}^{p + 1}
{p +1 \choose j}\,\partial_s^{j}\,u\,\partial_s^{p + 2 - j}m^{01}
\]
\[
- \sum_{j = 1}^{p}
{p +1 \choose j}\,\partial_s^{j}\,m^{00}\,\partial_s^{p + 1 - j}m^{01} 
+ \partial_s^{p + 1}(u^2\,k),
\]
whence
\[
|\partial_s^{p + 1}m^{01}| \le 
\frac{1}{(p + 2)\,|\dot{u}_0|}
\left\{
\sum_{j = 2}^{p + 1}{p +1 \choose j}\,c^2\,\frac{t^{j - 2}\,j\,!}{(j + 1)^2}
\,\frac{t^{p + 1 - j}\,(p + 2 - j)\,!}{(p + 3 - j)^2} \right.
\]
\[
\left. + \sum_{j = 1}^{p}{p +1 \choose j}\,
\,c^2\,\frac{t^{j - 1}\,j\,!}{(j + 1)^2}
\,\frac{t^{p - j}\,(p + 1 - j)\,!}{(p + 2 - j)^2}
+ |\partial_s^{p + 1}(u^2\,k)|
\right\}
\]
\[
\le
\frac{1}{|\dot{u}_0|}c^2\,C\,t^{p - 1}\,(p+1)\,!\left\{
\frac{1}{(p + 3)^2}
+ \frac{1}{(p + 2)^2}\right\}
+ 
2\,|\dot{u}_0|\,c\,\frac{t^{p - 1}\,(p + 1)\,!}{p^2\,(p + 2)}
\]
\[ 
+ 4\,c^2\,C\,\frac{t^{p - 2}\,(p + 1)\,!}{(p + 2)^3}
+ \frac{1}{|\dot{u}_0|}\,c^3\,C^2\,\frac{t^{p -  3}\,(p + 1)\,!}{(p +
2)^3}
= A_{01}\,\,c\,\frac{t^p\,(p + 1)\,!}{(p + 2)^2},
\]
with 
\[
A_{01} =
\frac{1}{t}\,\left\{
\frac{c\,C}{|\dot{u}_0|}
(1 + \frac{(p + 2)^2}{(p + 3)^2})
+ 2\,|\dot{u}_0|\,\frac{(p + 2)}{p^2}\right\}
+\frac{4\,c\,C}{t^2}\,\frac{1}{p + 2}
+ \frac{c^2\,C^2}{t^3\,|\dot{u}_0|}\,\frac{1}{p + 2}
\]
\[
\le \frac{1}{t}\,\left\{
\frac{2\,c\,C}{|\dot{u}_0|}
+ 4\,|\dot{u}_0|\right\}
+\frac{2\,c\,C}{t^2}
+ \frac{c^2\,C^2}{t^3\,|\dot{u}_0|}.
\]
Similarly we get from (\ref{2modgoedequ})
\[
|\partial_s^{p + 1}v| \le \frac{1}{(p + 1)\,|\dot{u}_0|}
\left\{\sum_{j = 2}^{p + 1} {p + 1 \choose
j}\,|\partial_s^ju|\,|\partial_s^{p + 2 - j}v|
+ |\partial_s^{p + 1}m^{01}| +
| \partial_s^{p + 1}(u^2\,h)|\
\right\}
\]
\[
\le \frac{1}{(p + 1)\,|\dot{u}_0|}
\left\{
\sum_{j = 2}^{p + 1}{p +1 \choose j}\,c^2\,\frac{t^{j - 2}\,j\,!}{(j + 1)^2}
\,\frac{t^{p + 1 - j}\,(p + 2 - j)\,!}{(p + 3 - j)^2}
+ |\partial_s^{p + 1}m^{01}| +
| \partial_s^{p + 1}(u^2\,h)|\
\right\}
\]
\[
\le A_{v}\,\,c\,\frac{t^p\,(p + 1)\,!}{(p + 2)^2},
\]
with
\[
A_v = \frac{A_{01}}{(p + 1)\,|\dot{u}_0|}
+ \frac{1}{t}\,\frac{2\,c\,C}{|\dot{u}_0|}\,\frac{(p + 2)^2}{(p + 3)^2}
+ \frac{2\,|\dot{u}_0|}{t}\,\frac{(p + 2)^2}{p\,(p + 1)}
+ \frac{4\,c\,C}{t^2}\,\frac{1}{p + 1}
+ \frac{c^2\,C^2}{t^3\,|\dot{u}_0|}\,\frac{1}{p + 1}
\]
\[
\le \frac{1}{t}\,\left\{9\,|\dot{u}_0| + 2 + \frac{2\,c\,C}{|\dot{u}_0|}
+ \frac{c\,C}{|\dot{u}_0|^2} \right\} 
+ \frac{c\,C}{t^2}\,\left\{2 + \frac{1}{|\dot{u}_0|}\right\} 
+ \frac{c^2\,C^2}{t^3}\,\left\{\frac{1}{|\dot{u}_0|}
+ \frac{1}{|\dot{u}_0|^2} \right\},
\]
and finally from (\ref{6modgoedequ})
\[
|\partial_s^{p + 1}m^{11}| \le 
\frac{1}{(p + 1)\,|\dot{u}_0|}
\left\{
\sum_{j = 2}^{p + 1}{p +1 \choose j}\,c^2\,\frac{t^{j - 2}\,j\,!}{(j + 1)^2}
\,\frac{t^{p + 1 - j}\,(p + 2 - j)\,!}{(p + 3 - j)^2} \right.
\]
\[
\left. + \sum_{j = 1}^{p}{p +1 \choose j}\,
\,c^2\,\frac{t^{j - 1}\,j\,!}{(j + 1)^2}
\,\frac{t^{p - j}\,(p + 1 - j)\,!}{(p + 2 - j)^2}
+ |\partial_s^{p + 1}(u^2\,l)|
\right\}
\le A_{11}\,\,c\,\frac{t^p\,(p + 1)\,!}{(p + 2)^2},
\]
with
\[
A_{11}
\le \frac{1}{t}\,\left\{18\,|\dot{u}_0| + \frac{2\,c\,C}{|\dot{u}_0|}
\right\} 
+ \frac{2\,c\,C}{t^2} 
+ \frac{c^2\,C^2}{t^3\,|\dot{u}_0|}.
\]
From the estimates for the $A$'s it follows that given a choice of $c$ which
satisfies the first of the estimates (\ref{2estimate}), we can determine $t$
large enough such that the second of the estimates (\ref{2estimate}) and the
conditions
\[
A_u, A_v, A_w, A_{00}, A_{01}, A_{11} \le 1,
\]
are satisfied. With this choice the induction step can be carried out.

\vspace{.2cm}

It follows immediately from estimates (\ref{xqest}) that the Taylor
expansions of the functions $z^j$ at $s = 0$,
$z^j(s) = \sum_{p = 0}^{\infty}\,z^j_p\,s^p$
with $z^j_p = \frac{1}{p!}\,\partial^p_sz^j(0)$, are absolutely convergent
for $|s| < 1/t$. 

The coefficients 
$z^j_p  = z^j_p(\dot{u}_0, v_0, \dot{w}_0)$ depend on $v_0$ via the
expansion coefficients of the functions $\tilde{e}^a_{AB}$, 
$\tilde{\Gamma}_{ABCD}$. This implies a polynomial dependence of the 
$z^j_p$ on $v_0$ due to the $v$-finite expansion types of the
functions $\hat{e}^a_{AB}$, $\hat{\Gamma}_{ABCD}$. The explicit
dependence of the right hand sides of equations 
(\ref{1modgoedequ}) - (\ref{6modgoedequ}) on $\dot{u}_0$ and $\dot{w}_0$ 
alone would lead to a polynomial dependence of the $z^j_p$ on $\dot{u}_0$
and $\dot{w}_0$. The occurence of the factors $u$ on the left hand sides of
equations (\ref{4modgoedequ}) - (\ref{6modgoedequ}) implies, however, that the 
$z^j_p$ are polynomials in $\dot{u}_0$, $v_0$, $\dot{w}_0$
divided by certain powers of $\dot{u}_0$.

The number $t$ which restricts the domain of convergence ensured by our
argument depends via $\epsilon$ and $M$ on $v_0$, and via $c$ and the $A$'s
on $\dot{u}_0$, $1/\dot{u}_0$ and $\dot{w}_0$ with the effect that 
$t \rightarrow \infty$ as $\dot{u}_0 \rightarrow 0$.  
It follows, however, from the form of the estimates (\ref{xqest}) and the way
they have been obtained that  for $(\dot{u}_0, v_0, \dot{w}_0)$ in a
compactly embedded subset $U$ of 
$(\mathbb{C} \setminus \{0\}) \times \mathbb{C} \times \mathbb{C}$ a
common number $t$ can be determined so that the Taylor series will be
absolutely convergent for 
$(s, \dot{u}_0, v_0, \dot{w}_0) \in P_{1/t}(0) \times U$.

If $K$ is compact in $P_{1/t}(0) \times U$, there exists $t' > t$ with 
$K \subset P_{1/t'}(0) \times U$ and it follow from  (\ref{xqest}) that 
the sequence of holomorphic functions $f^j_n = \sum_{p = 0}^{n}\,z^j_p\,s^p$
on $P_{1/t}(0) \times U$ satisfies
\[
\sup_K\,|f^j_n - z^j| \le \sum_{p = n + 1}^{\infty} 
c\,\frac{t^{p - 1}}{(p + 1)^2}\,(\frac{1}{t'})^p
\le \frac{c}{t'}\,\frac{(t /t')^n}{1 - t/t'} \rightarrow 0 
\quad \mbox{as} \quad n \rightarrow \infty,
\]
so that the $f^j_n$ converge uniformly to $z^j$ on $K$.
Standard results on compactly converging sequences of holomorphic functions 
(\cite{range}) then imply that  the $z^j  = z^j(s, \dot{u}_0, v_0,
\dot{w}_0)$ are holomorphic function of all four variables on $P_{1/t}(0)
\times U$. \end{proof}

\begin{lemma}
\label{tsexists}
Along the geodesic corresponding to $s \rightarrow z^j(s, \dot{u}_0, v_0,
\dot{w}_0)$ equations (\ref{0dattevol}) have a unique holomorphic solution
$t^A\,_B(s) = t^A\,_B(s, \dot{u}_0, v_0, \dot{w}_0)$
satisfying the  initial conditions (\ref{tevoldata}). The functions 
$t^A\,_B(s, \dot{u}_0, v_0, \dot{w}_0)$ are holomorphic in all four variables
in the domain where the $z^j(s, \dot{u}_0, v_0, \dot{w}_0)$ are holomorphic.
\end{lemma}

\begin{proof}
By the previous discussion we have $m^{01} = O(s^2)$, $u = O(s)$ with
$\dot{u}_0 \neq 0$ so that $m^{01}/u = O(s)$ as $s \rightarrow 0$.
It follows that equation  (\ref{0dattevol}) is in fact a linear
ODE with holomorphic coefficients  and the Lemma follows from
standard ODE theory.
\end{proof}

For later use we note that (\ref{tevoldata}), (\ref{0dattevol}) imply  as an
immediate consequence that
\begin{equation}
\label{tinversexp}
t^{-1 A}\,_B(s) = s^A\,_B(v_0) + O(|s|^2)
\quad \mbox{as} \quad s \rightarrow 0.
\end{equation}

\vspace{.2cm}

To discuss the transformation to normal coordinates the notation
employed before the transition (\ref{vminusv0etc}) will be used again, so that
\[
s \rightarrow z^a(exp(s\,x^ac_a)) = z^a(s, \dot{u}_0, v_0, \dot{w}_0),
\]
denotes in the coordinates $z^1 = u$, $z^2 = v$, $z^3 = w$ the geodesic
which has at $s = 0$ the tangent vector $x^ac_{\bf a}$ with 
$x^a = x^a(\dot{u}_0, v_0, \dot{w}_0)$ at $i$. We note that by the
discussion above 
\begin{equation}
\label{uvwsexp}
u(s, \dot{u}_0, v_0, \dot{w}_0) = \dot{u}_0\,s + O(|s|^3),\,
v(s, \dot{u}_0, v_0, \dot{w}_0) = v_0 + O(|s|^2),\,
w(s, \dot{u}_0, v_0, \dot{w}_0) = \dot{w}_0\,s + O(|s|^3).
\end{equation}

In terms of the map
(\ref{uvwxdep}) the transformation of the normal coordinates $x^c$ centered at
$i$ and based on the frame $c_{\bf a}$ at $i$ into the coordinates $z^a$ is the
given by
\begin{equation}
\label{xtoztransf}
x^a \rightarrow 
z^a(x^c) = z^a(1, \dot{u}_0(x^c), v_0(x^c), \dot{w}_0(x^c)),
\end{equation}
for small enough $|x^a|$ with $x^1 + i\,x^2 \neq 0$. The geodesics being 
given in normal coordinates by the curves $s \rightarrow s\,x^a$, this
implies
\[
s\,x^a \rightarrow z^a(1, \dot{u}_0(s\,x^c), v_0(s\,x^c),
\dot{w}_0(s\,x^c)) = z^a(s, \dot{u}_0(x^c), v_0(x^c), \dot{w}_0(x^c)).
\]
We use the relation on the right hand side to derive a convenient
expression for the map (\ref{xtoztransf}). 
Observing that
\[
\dot{u}_0(s\,x^c) = s\,\dot{u}_0(x^c),\,\,\,\,\, 
v_0(s\,x^c) = v_0(x^c),\,\,\,\,\, 
\dot{w}_0(s\,x^c) = s\,\dot{w}_0(x^c), \quad s \in C,
\]
by (\ref{uvwxdep}), we write $x^a = s\,x_*^a$ with $s$ chosen so
that $\dot{u}_0(x^c_*) = 1$, whence $\dot{u}_0(x^c) = s$, and get with the
relation above the map (\ref{xtoztransf}) in the form
\[
z^a(x^c) =
z^a(1, \dot{u}_0(x^c), v_0(x^c), \dot{w}_0(x^c))
= z^a(s, \dot{u}_0(x^c_*), v_0(x^c_*), \dot{w}_0(x^c_*))
\]
\[
= z^a(\dot{u}_0(x^c), 1, v_0(x^c), \frac{\dot{w}_0(x^c)}{\dot{u}_0(x^c)}).
\]
With (\ref{uvwsexp}) this gives,
as $|x| \equiv \sqrt{\delta_{ab}\,\bar{x}^a\,x^b}
\rightarrow 0$, $x^1 + i\,x^2 \neq 0$,
\begin{equation}
\label{uvwxexp}
u(x^c) = -  \frac{x^1 + i\,x^2}{\sqrt{2}} + O(|x|^3),\,\,\,\,\,
v(x^c) = - \frac{x^3}{x^1 + i\,x^2} + O(|x|^2),
\end{equation}
\begin{equation}
\label{wexp}
w(x^c)
= \frac{1}{\sqrt{2}}\,(x^1 - i\,x^2 + \frac{(x^3)^2}{x^1 + i\,x^2})
+ O(|x|^3)
 = \frac{\delta_{ab}\,x^a\,x^b}{\sqrt{2}\,(x^1 + i\,x^2)} + O(|x|^3).
\end{equation}
In the flat case the order symbols must be omitted in these expressions.

With (\ref{sigmaasbeh}), (\ref{tinversexp}) and 
\[
du =  -  \frac{1}{\sqrt{2}}(dx^1 + i\,dx^2) + O(|x|^2), \quad
dv = \frac{dx^3}{\sqrt{2}\,\,u} 
+ \frac{v}{\sqrt{2}\,\,u}(dx^1 + i\,dx^2) + O(|x|),
\]
\[
dw = \frac{1}{\sqrt{2}}\,(dx^1 - i\,dx^2 
- 2\,v\,dx^3
- v^2\,(dx^1 + i\,dx^3)) +
O(|x|^2),
\]
one gets for the forms $\chi^{AB}= \chi^{AB}\,_c\,dx^c$ 
dual to the normal frame $c_{AB}$ indeed
\[
\chi^{AB}(x^c) = t^{-1A}\,_C\,t^{-1B}\,_D\,
(\sigma^{CD}\,_1\,du
+ \sigma^{CD}\,_2\,dv
+ \sigma^{CD}\,_3\,dw)
= (\alpha^{AB}\,_a + \hat{\chi}^{AB}\,_a)\,dx^a,
\]
with some functions $\hat{\chi}^{AB}\,_a(x^c)$ which satisfies 
$\hat{\chi}^{AB}\,_a = O(|x|^2)$ as $|x| \rightarrow 0$. Correspondingly,
the coefficients $c^a_{AB} = \,<dx^a,\,c_{AB}>$ of the normal frame in
the normal coordinates satisfy 
\[
c^a\,_{AB}(x^c) = \alpha^a\,_{AB} + \hat{c}^a\,_{AB}(x^c),
\]
with holomorphic functions $\hat{c}^a\,_{AB}(x^c)$  which satisfy
$\hat{c}^a\,_{AB}(x^c) = O(|x|^2)$ as $|x| \rightarrow 0$

Since the three 1-forms $\alpha^{AB}\,_a\,dx^a$ are linearly independent
this shows that for small $|x^c|$ the coordinate transformation $x^a
\rightarrow z^a(x^c)$, where defined, is non-degenerate and the forms
$\chi^{AB}$ behave as required by normal forms in normal coordinates. The
relations (\ref{normalformchar}), which characterize coefficients of normal
forms in normal coordinates, are a consequence of the equations
satisfied by $z^a(s)$ and $t^A\,_B(s)$. All the tensor fields which enter
the conformal static vacuum field equations can now be expressed in term
of the coordinates $x^c$ and the frame field $c_{AB}$. 

All ingredients are now available to derive our main result. 

{\bf Proof of Theorem \ref{mainres} }
The coordinates $x^c$ cover a certain (connected) domain $U$ in 
$\mathbb{C}^3$
on which the frame vector fields
$c^a\,_{AB}\,\partial/\partial_{x^c}$ exist, are linearly
independent and holomorphic and where the other tensor fields expressed in
terms of the $x^a$ and $c_{AB}$ are holomorphic. It follows from Lemmas
\ref{extconv}, \ref{geodexist}, and \ref{tsexists} that given any initial data 
$\dot{u}_0$, $v_0$, $\dot{w}_0$ with 
$\dot{u}_0 \neq 0$, there exists a solution $z^a(s, \dot{u}_0, v_0,
\dot{w}_0)$ of the geodesics equations on the solution provided by
Lemma \ref{extconv} which is defined for $|s| \le 1/t$ with some $t >
0$. The dicussion above shows, however, that $t$ will become large 
 if $|v_0|$ becomes large or $|\dot{u}_0|$ becomes very
small.  This implies that the domain $U$ will not contain the 
hypersurface $x^1 + i\,x^2 = 0$ but its boundary will become tangent to this
hypersurface at $x^a = 0$.
From the estimates obtained so far it cannot be concluded that the
coordinates extend holomorphically to a domain 
containing an open neighbourhood of the origin. 

To analyse this question, we make use of the remaining gauge freedom to
perform with some $\,t^A\,_B \in SU(2)$
a rotation $\delta^* \rightarrow \delta^* \cdot t$ of the spin
frame and the associated rotation  
\[
c_{AB} \rightarrow c^t_{AB} = t^C\,_A\,t^D\,_B\,c_{CD}
\]
of the frame $c_{AB}$ at $i$ on which the construction of the submanifold 
$\hat{S}$ and the related gauge is based. Starting with
these frames at $i$ all the previous constructions and
derivations can be repeated.

Let $u'$, $v'$, $w'$ and $e^t_{AB}$ denote the analogues in the new
gauge of the coordinates $u$, $v$, $w$ and the frame $e_{AB}$. 
The sets $\{w = 0\}$ and $\{w' = 0\}$ are then both to be thought of as
lift of the set ${\cal N}_i$ to the bundle of spin frames, the coordinates
$u$ and $u'$ can both be interpreted as affine parameters on the null
generators of ${\cal N}_i$ which vanish at $i$, the coordinates $v$, $v'$
both label these null generators, and the frame vectors $e_{00}$ and
$e^t_{00}$ can be identified with auto-parallel vector fields tangent to the
null generators.

If $v$ and $v'$ then label the same generator $\eta$ of ${\cal N}_i$, a
relation
\[
s^C\,_0(v')\,s^D\,_0(v')\,t^E\,_C\,t^F\,_D\,c_{EF} = 
e^t_{00} = f^2\,e_{00} = f^2\,s^C\,_0(v)\,s^D\,_0(v)\,c_{CD},
\]
must hold at $i$ with some $f \neq 0$ and $e^t_{00} = f^2\,e_{00}$ must 
hold in fact along $\eta$, with $f$ constant along $\eta$ because 
$e^t_{00}$ and $e_{00}$ are auto-parallel. Absorbing the undetermined sign
in $f$, this leads to  
\[
t^E\,_C\,s^C\,_0(v') = f\,s^E\,_0(v).
\]
With 
\begin{equation}
\label{tbasicgaugetransf}
(t^A\,_B) =
\left( \begin{array}{cc}
a & - \bar{c} \\
c & \bar{a}
\end{array} \right)
\quad \mbox{where} \quad a, c \,\in C, \,\,\,\,\,|a|^2 = |c|^2 = 1,
\end{equation}
this gives
\[
v' = \frac{-c + a\,v}{\bar{a} + \bar{c}\,v},\,\,\,\,\,
f = \frac{1}{\bar{a} + \bar{c}\,v},
\quad \mbox{resp.} \quad
v = \frac{c + \bar{a}\,v'}{a - \bar{c}\,v'},\,\,\,\,\,
f = a - \bar{c}\,v'.
\]
Moreover, the relations  
\[
<du, e_{00}>\, = 1 = \,<du', e^t_{00}>\,= \,<du', f^2\,e_{00}>,
\]
imply for the affine parameters satisfy along $\eta$
\[
u = f^2\,u',
\]
so that $\eta(u', v') = \eta(u, v)$ holds with these relations. 
We note that choices of $t^A\,_B$ with $c \neq 0$ can 
supply new information, because then $v \rightarrow \infty$ as 
$v' \rightarrow a/\bar{c}$ so that the singular generator of the
$c_{AB}$-gauge, about whose neighbourhood we need
information, is then contained in the regular domain of the
$c^t_{AB}$-gauge.

For the null datum in the new gauge one gets with (\ref{nulldataonU0})
\[
s^t_0(u', v') = s^A\,_0(v')\,\ldots\,s^C\,_0(v')\,
t^E\,_A\,\ldots\,t^H\,_D\,s^*_{E\,\ldots\,H}|_{\eta(u', v')}
= f^4\,s_0(u, v)
\]
\[
=
\sum_{m = 0}^{\infty}
\frac{1}{m\,!}\,u'^{\,m}\,f^{2\,m + 4}
\,s^{A_1}\,_0(v)\,s^{B_1}\,_0(v)\ldots\,s^{D}\,_0(v)\,
D^*_{(A_1B_1} \ldots D^*_{A_m B_m}\,s^*_{ABCD)}(i)
\]
\[
=
\sum_{m = 0}^{\infty}
\frac{1}{m\,!}\,u'^{\,m}\,f^{2\,m + 4}
\,s^{A_1}\,_0(v')\,s^{B_1}\,_0(v')\ldots\,s^{D}\,_0(v')\,
D^t_{(A_1B_1} \ldots D^t_{A_m B_m}\,s^t_{ABCD)}(i),
\]
and thus 
\begin{equation}
\label{stpsit}
s^t_0(u', v') = \sum_{m = 0}^{\infty}\sum_{n = 0}^{2\,m + 4}
\,\psi^t_{m, n}\,\,u'^{\,m}\,\,v'^{\,n},
\end{equation}
with
\[
D^t_{(A_1B_1} \ldots D^t_{A_m B_m}\,s^t_{ABCD)}(i)
\equiv 
t^{G_1}\,_{A_1}\,t^{H_1}\,_{B_1}\,\ldots \,t^N\,_D\,
D^*_{(G_1H_1} \ldots D^*_{G_m H_m}\,s^*_{LKMN)}(i),
\]
and 
\[
\psi^t_{m, n} = \frac{1}{m\,!}\,{2\,m + 4 \choose n}
\,D^t_{(A_1B_1} \ldots D^t_{A_m B_m}\,s^t_{ABCD)_n}(i)
\]
\[
=\frac{1}{m\,!}\,{2\,m + 4 \choose n} \sum_{j = 0}^{2\,m + 4}
{2\,m + 4 \choose j}
t^{(G_1}\,_{(A_1}\,t^{H_1}\,_{B_1}\,\ldots \,t^{N)_j}\,_{D)_n}\,
D^*_{(G_1H_1} \ldots D^*_{G_m H_m}\,s^*_{LKMN)j}(i)
\]
\[
={2\,m + 4 \choose n} \sum_{j = 0}^{2\,m + 4}
t^{(G_1}\,_{(A_1}\,t^{H_1}\,_{B_1}\,\ldots \,t^{N)_j}\,_{D)_n}\,
\psi_{m, j}.
\]
It is convenient to write this in the form
\begin{equation}
\label{stu'v'exp}
\psi^t_{m, n} = \sum_{j = 0}^{2\,m + 4}
{2\,m + 4 \choose n}^{1/2} \,{2\,m + 4 \choose j}^{- 1/2}\,T_{2\,m +
4}\,^j\,_n(t)\,
\psi_{m, j}
\end{equation}
where the numbers 
\[
T_{2\,m + 4}\,^j\,_n(t) = 
{2\,m + 4 \choose n}^{1/2}\,{2\,m + 4 \choose j}^{1/2}\,
t^{(G_1}\,_{(A_1}\,t^{H_1}\,_{B_1}\,\ldots \,t^{N)_j}\,_{D)_n},
\]
are so defined (\cite{friedrich:pure rad}) that they represent the matrix
elements of a unitary representation of $SU(2)$ and thus satisfy
\[
|T_{2\,m + 4}\,^j\,_n(t)| \le 1, \quad m = 0, 1, 2, \ldots,
\quad 0 \le j, \,n \le 2\,m + 4. 
\]
With the expressions above it is easy to see that the type of the
estimate (\ref{symmest}) and the type of the resulting estimate (\ref{basest})
are preserved under the gauge transformation. With (\ref{stpsit}) and
(\ref{stu'v'exp}) follows from (\ref{basest}) at the point 
$O' = (u' = 0, v' = 0)$
\begin{equation}
\label{tbasest}
|\partial^m_{u'}\,\partial^n_{v'}\,s^t_0(O')|
= m!\,n!\,|\psi^t_{m, n}|
\le m!\,n!\, \sum_{j = 0}^{2\,m + 4}
{2\,m + 4 \choose n}^{1/2} \,{2\,m + 4 \choose j}^{- 1/2}\,
|T_{2\,m + 4}\,^j\,_n(t)|\,
|\psi_{m, j}|
\end{equation}
\[
\le m!\,n!\,\sum_{j = 0}^{2\,m + 4}
{2\,m + 4 \choose n}^{1/2} \,{2\,m + 4 \choose j}^{1/2}\,
\,M\,r_1^{-m}
\le m!\,n!\, {2\,m + 4 \choose n}\,\sum_{j = 0}^{2\,m + 4}
{2\,m + 4 \choose j}\,
\,M\,r_1^{-m} 
\]
\[
= 
m!\,n!\, {2\,m + 4 \choose n}\,
\,M'\,r_t^{-m},
\]
with $M' = 16\,M$ and $r_t = r_1/4$.

\vspace{.5cm}

Assuming now that $c \neq 0$ in (\ref{tbasicgaugetransf}), 
the resulting {\it $c^t_{AB}$-gauge} can be studied from two different points of view:

i) The singular generator of ${\cal N}_i$ in the $c^t_{AB}$-gauge will
coincide with the regular generator of ${\cal N}_i$ on which 
$v = - \bar{a}/\bar{c}$ in the $c_{AB}$-gauge. By starting from the
solution in the $c_{AB}$-gauge, we are thus able to directly determine
near that generator the transformation into the 
$c^t_{AB}$-gauge and to determine the expansion of the solution in the
$c_{AB}$-gauge in terms of the coordinates $u'$, $v'$, $w'$ and the 
frame field $e^t_{AB}$.

ii) Alternatively, with the null data $s^t_0(u', v')$ at hand, 
one can go through the discussions of the previous sections
to show the existence of a solution to the conformal static vacuum
equations in the coordinates $u'$, $v'$, $w'$ pertaining to the
$c^t_{AB}$-gauge. All the observations made above, in particular 
statements about domains of convergence, apply to this
solution as well. Important for us is that this solution 
covers the generator $v' = a/\bar{c}$ near $u' = 0$ and $w' = 0$, which
corresponds to the singular generator in the $c_{AB}$-gauge.

Because the formal expansions of the fields in terms of $u'$, $v'$, $w'$
are uniquely determined by the data $s^t_0(u', v')$, the solutions obtained 
by the two methods are holomorphically related to each
other on certain domains by the gauge transformation obtained in (i). The
solution obtained in (ii) can be expressed in terms of the normal
coordinates $x^a_t$ and the normal frame field $c^t_{AB}$ so that the $x^a_t$
cover an certain domain
$U_t \subset \mathbb{C}^3$ and the frame field $c^t_{AB}$ is non-degenerate and
all our tensor fields expressed in terms of $x^a_t$ and $c^t_{AB}$ are 
holomorphic on $U_t$ as discussed above. It follows then that the solution
in the $c_{AB}$-gauge and the solution in
the $c^t_{AB}$-gauge are related on certain domains by the simple
transformation (cf. (\ref{SU2SO3}))  
\[
x_t^a = t^{-1\,a}\,_b\,x^b,
\quad 
c^t_{AB} = t^C\,_A\,t^D\,_B\,c_{CD}.
\]
Extending this as a coordinate and frame transformation to the
solution obtained in (ii) to express all field in terms $x^a$ and
$c_{AB}$ so that they are defined and holomorphic on $t^{-1}\,U_t$, one
finds that the solution obtained in (ii) and our original solution define in
fact genuine holomorphic extensions of each other because each one 
covers the singular generator of the other one away from the origin in a regular way.

By letting $t^A\,_B$ go through  $SU(2)$ and
observing the corresponding extensions, one obtains in fact a holomorphic
solution to the conformal static vacuum field equations in the normal
coordinates $x^a$ centered at $i$ associated with the frame $\delta^*$ resp.
$c_{AB}$ at ${i}$ on a domain which covers a full
neighbourhood of space-like infinity.
Consider again the solution we obtained in the $c_{AB}$-gauge. From the
discussion above it follows that the domain $U$ in $\mathbb{C}^3$ on which the
solution is holomorphic in the coordinates
$x^a$ covers a (connected) domain $U'$ of the hypersurface $\{x^3 = 0\}$ of
$\mathbb{C}^3$ which has empty intersection with the line 
$\{x^1 + i\,x^3 = 0,\,x^3 = 0\}$
(corresponding to the singular generator of the $c_{AB}$-gauge) and whose
boundary becomes tangent to this line at the origin $x^a = 0$.  
Under the transition
\[
\dot{u}_0 \rightarrow \dot{u}_0, \quad
v_0 \rightarrow e^{i\,\theta/2}\,v_0, \quad
\dot{w}_0 \rightarrow e^{i\,\theta}\,\dot{w}_0,\quad \theta \in R, 
\]
which leaves the quantities $|\dot{u}_0|$, $|v_0|$, $|\dot{w}_0|$ entering
the estimates above invariant, one gets by (\ref{xuvwdep})  
\[
x^1 + i\,x^2 \rightarrow x^1 + i\,x^2, \quad
x^1 - i\,x^2 \rightarrow e^{i\,\theta}\,(x^1 - i\,x^2), \quad
x^3 \rightarrow e^{i\,\theta}\,x^3. 
\]
Thus the set $U'$ can be assumed to be invariant under this
transformation.

Consider now the $c^{t^*}_{AB}$-gauge where the special
transformation $t^{*\,A}\,_B$ is given by
(\ref{tbasicgaugetransf}) with $a = 0$, $c = 1$. Let $U'_{t^*}$ denote 
the subset of the hypersurface $\{x^3_{t^*} = 0\}$ in $\mathbb{C}^3$
analogous to $U'$. It has empty insection with the line 
$\{x^1_{t^*} + i\,x^2_{t^*} = 0, \,x^3_{t^*} = 0\}$ but its boundary becomes
tangent to it at
$x^a_{t^*} = 0$. It holds
\[
c^{t^*}_{00} = c_{11}, \,\,\,\, c^{t^*}_{01} = - c_{01}, 
\,\,\,\,c^{t^*}_{11} = c_{00} 
\quad \mbox{at} \quad i,
\]
and the corresponding normal coordinates are related by 
\[
x^1_{t^*} = - x^1, \,\,\,\, x^2_{t^*} = x^2, \,\,\,\, x^3_{t^*} = - x^3.
\]
The holomorphic transformation 
$\{x^3_{t^*} = 0\} \ni (x^1_{t^*}, x^2_{t^*}) \rightarrow (- x^1, x^2)
\in \{x^3 = 0\}$ maps $U'_{t^*}$ onto a subset of 
$\mathbb{C}^2 \sim \mathbb{C}^2 \times \{0\}$,
denoted by
$t^{*\,-1}\,U'_{t^*}$, which has non-empty intersection with $U'$. After the
transformation above the two solutions coincide on 
$t^{*\,-1}\,U'_{t^*} \cap U'$. 

On the other hand, the image of the
$c^{t^*}_{AB}$-regular line 
$\{x^1_{t^*} - i\,x^2_{t^*} = 0, \,x^3_{t^*} = 0\}\,\cap \,U'_{t^*}$ 
under this transformation contains the intersection of a
neighbourhood of the origin with the singular line
$\{x^1 - i\,x^2 = 0, \,x^3 = 0, \,x^a \neq 0\}$ of the $c_{AB}$-gauge. In
fact, the set
$t^{*\,-1}\,U'_{t^*} \,\cup \,U'$, which admits a holomorphic extension of
our solution in the coordinates $x^a$ and the frame $c_{AB}$, contains a
punctured neighbourhood of the origin. As we have seen above, the
field $c_{AB}$ on this neighbourhood extends continously to the origin.

Let now $x^a_* \neq 0$ be an arbitrary point in $\mathbb{C}^3$. We want to show that
the solution extends in the coordinates $x^a$ to a domain which
covers the set $s\,x^a_*$ for $0 < |s| < \epsilon$ for some 
$\epsilon > 0$. Since $x^a_* = y^a + i\,z^a$ with $y^a, \,z^a \in \mathbb{R}^3$
there is a vector $u^a \in \mathbb{R}^3$ of unit length and orthogonal to $x^a$
with respect to the standard product $u \cdot x = \delta_{ab}\,u^a\,x^b$. 
Consider the $c^t_{AB}$-gauges with $t^A\,_B \in SU(2)$ so
that $u^a\,_t = t^{-1\,a}\,_b\,u^b = \delta^a\,_3$.
It follows then that $x^a_{* t}  = t^{-1\,a}\,_b\,x^b_* \in \{x^3_t = 0\}$ 
and by the preceeding observation $t^A\,_B$ can in fact be chosen such 
that there exist an $\epsilon > 0$ so that
the points $s\,x^a_{* t}$ with $0 < |s| < \epsilon$ are covered by $U'_t$.
Transforming back we find that the set $U \in \mathbb{C}^3$ covered by the
coordinates $x^a$ can be extended so that the points 
$s\,x^a_*$ with $0 < |s| < \epsilon$ are covered by $U$ and all field are
holomorphic on $U$ in the coordinates $x^a$.  It follows that 
we can assume $U$ to contain a punctured neighbourhood of the origin in which
the solution is holomorphic in the normal coordinates $x^a$ and the normal
frame $c_{AB}$. Since holomorphic functions in more than one dimensions cannot
have isolated singularities  (\cite{fritzsche:grauert:2002}) 
the solution is then in fact holomorphic on a full neighbourhood of the origin
$x^a = 0$, which represents the point $i$.

\vspace{.3cm}

By Lemma \ref{realformalexpansion} the exact sets of
equations argument determines from null data satisfying the reality conditions  a formal expansion of
the solution with expansion coefficients satisfying the reality conditions.
By the various uniqueness statements obtained in the Lemmas this expansion must
coincide with the expansion in normal coordinates of the solution obtained
above. This implies the existence of a
$3$-dimensional real slice on which the tensor fields satisfy the reality
conditions. It is obtained by requiring 
the coordinates $x^a$ to assume values in $\mathbb{R}^3$. 
$\Box$

\section{Acknowledgements}
 I would like to thank Piotr Chru\'sciel for
discussions. A substantial part of this work has been done during the
programme ``Global Problems in Mathematical Relativity'' at the 
Isaac Newton Institute, Cambridge. I am grateful to the Newton Institute for
hospitality and financial support.

}


\begin{thebibliography}{9}


\bibitem{baeckdahl:herberthson}
T. B\"ackdahl, M. Herberthson.
\newblock Static axisymmetric space-times with prescribed multipole
moments.
\newblock {Class. Quantum Grav.} 22 (2005) 1607 - 1621.

\bibitem{beig:simon}
R. Beig, W. Simon.
\newblock Proof of a multipole conjecture due to Geroch.
\newblock { Comm. Math. Phys.} 78 (1980) 75 - 82.

\bibitem{chrusciel:delay:2002}
P. T. Chru\'sciel, E. Delay.
\newblock Existence of non-trivial, vacuum, asymptotically simple
spacetimes.
\newblock {\em Class. Quantum Grav.}, 19 (2002) L 71 - L 79.
\newblock Erratum
\newblock {\em Class. Quantum Grav.}, 19 (2002) 3389.

\bibitem{chrusciel:delay:2003}
P. T. Chru\'sciel, E. Delay.
\newblock On mapping properties of the general relativistic constraints
operator in weighted function spaces, with application.
\newblock {\it M\'{e}m. Soc. Math. France} submitted.
\newblock http://xxx.lanl.gov/abs/gr-qc/0301073

\bibitem{corvino}
J. Corvino.
\newblock Scalar curvature deformation and a gluing
construction for the Einstein constraint equations.
\newblock { Comm. Math. Phys.} 214 (2000) 137--189. 

\bibitem{corvino:2005}
J. Corvino.
\newblock On the Existence and Stability of the Penrose Compactification.
\newblock Submitted for publication.

\bibitem{corvino:schoen}
J. Corvino, R. Schoen
\newblock On the Asymptotics for the Vacuum Einstein Constraint
Equations.
\newblock  http://xxx.lanl.gov/abs/gr-qc/0301071

\bibitem{dieudonne:IV}
J. Dieudonn\'e.
\newblock {\em Treatise on Analysis}, Vol IV.
\newblock Academic Press, New York, 1974.

\bibitem{friedrich:1981}
H. Friedrich.
\newblock On the regular and the asymptotic characteristic initial value
problem for Einstein's vacuum field equations. 
\newblock Proceedings of the 3rd Gregynog Relativity
Workshop on  Gravitational Radiation Theory
\newblock MPI-PEA/Astro 204 (1979) 137--160 and
\newblock { Proc. Roy. Soc.}, 375 (1981) 169--184.
\newblock The asymptotic characteristic initial value problem for Einstein's
vacuum field equations as an initial value problem for a first-order
quasilinear symmetric hyperbolic system. 
\newblock { Proc. Roy. Soc.}, A 378 (1981) 401--421.

\bibitem{friedrich:1982}
H. Friedrich.
\newblock On the Existence of Analytic Null Asymptotically Flat
Solutions of Einstein's Vacuum Field Equations.
\newblock { Proc. Roy. Soc.\ Lond.\ A} 381 (1982) 361--371.

\bibitem{friedrich:pure rad}
H. Friedrich.
\newblock On purely radiative space-times.
\newblock { Comm. Math. Phys.} 103 (1986) 35--65.

\bibitem{friedrich:static}
H. Friedrich.
\newblock On static and radiative space-times.
\newblock { Comm. Math. Phys.}, 119 (1988) 51--73.

\bibitem{friedrich:i-null}
H. Friedrich.
\newblock Gravitational fields near space-like and null
infinity.
\newblock { J. Geom. Phys.}  24 (1998)  83--163.

\bibitem{friedrich:cargese}
H. Friedrich.
\newblock Smoothness at null infinity and the structure of initial data.
\newblock In: P. T. Chru\'sciel, H. Friedrich (eds.):
{\it The Einstein equations and the large scale behaviour of gravitational fields.}
\newblock Birkh\"auser, Basel, 2004.

\bibitem{fritzsche:grauert:2002}
K. Fritzsche, H. Grauert.
\newblock From holomorphic functions to complex manifolds.
\newblock Springer, Berlin 2002.


\bibitem{geroch:1970}
R. Geroch
\newblock Multipole moments. I. Flat space.
\newblock {\em J. Math. Phys.} 11 (1970) 1955 - 1961.
\newblock Multipole moments. II. Curved space.
\newblock {\em J. Math. Phys.} 11 (1970) 2580 - 2588.

\bibitem{kennefick:o'murchadha}
D. Kennefick, N. O'Murchadha. 
\newblock Weakly decaying asymptotically flat static and
stationary solutions to the Einstein equations. 
\newblock { Class. Quantum. Grav.} 12 (1995) 149--158.

\bibitem{miao:2003}
P. Miao.
\newblock On Existence of Static Metric Extensions in General Relativity.
\newblock {\em Commun. Math. Phys.} 241 (2003) 27 - 46.


\bibitem{morrey}
C. B. Morrey.
\newblock Multiple integrals in the calculus of variations.
\newblock Springer, Berlin 1966.


\bibitem{mueller zum hagen}
H. M\"uller zum Hagen.
\newblock On the analyticity of stationary vacuum solutions of Einstein's equation.
\newblock {\em Proc. Camb. Phil. Soc.} 68 (1970) 199 - 201.

\bibitem{penrose:rindler:I}
R. Penrose, W. Rindler.
\newblock {\it Spinors and space-time}, Vol. 1 and 2.
\newblock Cambridge University Press, 1984.

\bibitem{range}
R. M. Range.
\newblock Holomorphic Functions and Integral Representations in Several
Complex Variables.
\newblock Springer, Berlin 1986.

\bibitem{reula:1989}
O. Reula.
\newblock On existence and behaviour of asymptotically flat solutions to the
stationary Einstein equations.
\newblock {\em Commun. Math. Phys.} 122 (1989) 615 - 624.

\bibitem{shinbrot: welland}
M. Shinbrot, R. Welland.
\newblock The Cauchy-Kowalewskaya Theorem.
\newblock {\em J. Math. Anal. Appl.} 55 (1976) 757 - 772.

\bibitem{valientekroon:2004a}
J. Valiente Kroon.
\newblock Does asymptotic simplicity allow for radiation near spatial infinity?
\newblock {\em Commun. Math. Phys.} 251 (2004) 211 - 234.

\bibitem{Weyl:1917}
H. Weyl.
\newblock Zur Gravitationstheorie.
\newblock {\em Ann. Phys. Leipzig} 54 (1917) 117
\end{thebibliography}
\end{document}